\documentclass[a4paper,fleqn,usenatbib]{mnras}
\usepackage{newtxtext,newtxmath}
\usepackage[T1]{fontenc}
\usepackage{ae,aecompl}
\usepackage{graphicx}	
\usepackage{amsmath}	
\usepackage{amssymb}	
\usepackage{booktabs,caption,fixltx2e}	
\usepackage{aas_macros} 
\usepackage[flushleft]{threeparttable}
\usepackage[percent]{overpic}	
\usepackage{upgreek}
\usepackage{url}
\usepackage[usenames]{color}

\usepackage{array} 
\newcolumntype{H}{>{\setbox0=\hbox\bgroup}c<{\egroup}@{}} 
\usepackage{threeparttable} 

\title[{\it Kepler} Pulsating Binaries]{Finding binaries from phase modulation of pulsating stars with {\it Kepler}: V. Orbital parameters, with eccentricity and mass-ratio distributions of 341 new binaries}

\author[S. J. Murphy et al.] 
{Simon J. Murphy$^{1,2,\dagger}$, Maxwell Moe$^{3}$, Donald W. Kurtz$^4$, Timothy R. Bedding$^{1,2}$, \and Hiromoto Shibahashi$^5$, Henri M.~J. Boffin$^{6}$\\
\\
$^1$Sydney Institute for Astronomy (SIfA), School of Physics, The University of Sydney, NSW 2006, Australia\\
$^2$Stellar Astrophysics Centre, Department of Physics and Astronomy, Aarhus University, DK-8000 Aarhus C, Denmark\\
$^3$Steward Observatory, University of Arizona, 933 N. Cherry Ave., Tucson, AZ 85721, USA\\
$^4$Jeremiah Horrocks Institute, University of Central Lancashire, Preston, PR1 2HE, UK\\
$^5$Department of Astronomy, The University of Tokyo, Tokyo 113-0033, Japan\\
$^6$ESO, Karl Schwarzschild Strasse 2, 85748 Garching, Germany\vspace{2mm} \\
$^{\dagger}$email: simon.murphy@sydney.edu.au\\
}
 
\begin{document}

\maketitle 

\begin{abstract}
The orbital parameters of binaries at intermediate periods ($10^2$\,--\,$10^3$\,d) are difficult to measure with conventional methods and are very incomplete. We have undertaken a new survey, applying our pulsation timing method to \textit{Kepler} light curves of 2224 main-sequence A/F stars and found 341 non-eclipsing binaries. We calculate the orbital parameters for 317 PB1 systems (single-pulsator binaries) and 24 PB2s (double-pulsators), tripling the number of intermediate-mass binaries with full orbital solutions. The method reaches down to small mass ratios $q \approx 0.02$ and yields a highly homogeneous sample. We parametrize the mass-ratio distribution using both inversion and MCMC forward-modelling techniques, and find it to be skewed towards low-mass companions, peaking at $q \approx 0.2$. 
While solar-type primaries exhibit a brown dwarf desert across short and intermediate periods, we find a small but statistically significant (2.6$\sigma$) population of extreme-mass-ratio companions (\mbox{$q < 0.1$}) to our intermediate-mass primaries.
Across periods of 100\,--\,1500\,d and at $q>0.1$, we measure the binary fraction of current A/F primaries to be 15.4\%\,$\pm$\,1.4\%, though we find that a large fraction of the companions (21\%\,$\pm$\,6\%) are white dwarfs in post-mass-transfer systems with primaries that are now blue stragglers, some of which are the progenitors of Type Ia supernovae, barium stars, symbiotics, and related phenomena.
Excluding these white dwarfs, we determine the binary fraction of original A/F primaries to be 13.9\%\,$\pm$\,2.1\% over the same parameter space.
Combining our measurements with those in the literature, we find the binary fraction across these periods is a constant 5\% for primaries $M_1 < 0.8$\,M$_{\odot}$, but then increases linearly with $\log M_1$, demonstrating that natal discs around more massive protostars \mbox{$M_1 \gtrsim 1$\,M$_{\odot}$} become increasingly more prone to fragmentation.
Finally, we find the eccentricity distribution of the main-sequence pairs to be much less eccentric than the thermal distribution.
\end{abstract}

\begin{keywords}
blue stragglers -- stars\,:\,variables\,:\,$\delta$\,Scuti -- binaries\,:\,general -- stars\,:\,statistics --  -- stars\,:\,oscillations -- stars\,:\,formation \vspace{-6mm}
\end{keywords}


\section{Introduction}
\label{sec:intro}

Binary and multiple systems are so common that they outnumber single stars by at least 2:1 \citep{duchene&kraus2013,guszejnovetal2017}, and even more so at birth \citep{ghezetal1993}. Their influence on star formation and stellar populations lends an importance to their distribution functions that is comparable to that of the stellar initial mass function (IMF). It is therefore no surprise that reviews of these distributions are among the most highly cited papers in astronomy (e.g. \citealt{duquennoy&mayor1991}). Their overarching significance spans the intricacies of star formation \citep{bate&bonnell1997, white&ghez2001} to the circumstances of stellar deaths \citep{narayanetal1992, hillebrandt&niemeyer2000, abbottetal2016}, with clear consequences for stellar population synthesis \citep{zhangetal2005}.

Interaction between binary components can be significant even on the main-sequence and before any mass transfer \citep{zahn1977,demarco&izzard2017}. Tidal effects in binary systems alter not only the orbit but also the stellar structure, which among intermediate-mass stars leads to the development of stratified abundances and chemical peculiarities (i.e. the Am stars; \citealt{abt1967,baglinetal1973}). If the eccentricity is high, tides can also excite oscillations \citep{willems2003,welshetal2011,fuller2017,hambletonetal2017}, revealing information on the stellar structure and thereby extending the utility of binary stars well beyond their fundamental role in the provision of dynamical masses.

Binary stars are expected to derive from two formation processes: fragmentation of molecular cores at separations of several 100s to 1000s of au and fragmentation within protostellar discs on much smaller spatial scales \citep{tohline2002,bate2009,kratter2011,tobinetal2016}.
The orbital distribution of solar-type main-sequence binaries peaks at long periods $P \sim 10^5$\,d \citep{duquennoy&mayor1991,raghavanetal2010}, indicating turbulent fragmentation of molecular clouds is the dominant binary star formation process \citep{offneretal2010}. Moreover, the frequency of very wide companions ($P > 10^6$\,d) to T~Tauri stars is 2\,--\,3 times larger than that observed for solar-type primaries in the field \citep{ghezetal1993,ducheneetal2007,connelleyetal2008b,tobinetal2016},
demonstrating that the solar-type binary fraction was initially much larger as the result of efficient core fragmentation, but then many wide companions were subsequently dynamically disrupted \citep{goodwin&kroupa2005,marks&kroupa2012,thiesetal2015}.
Conversely, companions to more massive stars are skewed toward shorter periods, peaking at $10^3$\,d for early B main-sequence primaries ($M_1 \sim 10$\,M$_{\odot}$, \citealt{abtetal1990,rizzutoetal2013,moe&distefano2017}) and $P<20$\,d for O main-sequence primaries ($M_1 \sim 30$\,M$_{\odot}$, \citealt{sanaetal2012a}).
These observations suggest disc fragmentation plays a more important role in the formation of massive binaries. The physics of binary formation in the intermediate-mass regime ($M_1 \sim 2$\,M$_{\odot}$) is less clear, primarily because the orbital parameters of large numbers of binaries at intermediate periods have previously proven hard to constrain \citep{fuhrmann&chini2012}.  Binaries with intermediate orbital periods are less likely to be dynamically disrupted, and so their statistical distributions and properties directly trace the processes of fragmentation and subsequent accretion in the circumbinary disc.  While binaries with solar-type primaries and intermediate periods exhibit a uniform mass-ratio distribution and a small excess fraction of twin components with $q > 0.9$ \citep{halbwachsetal2003,raghavanetal2010}, binaries with massive primaries and intermediate periods are weighted toward small mass ratios \citep{abtetal1990,rizzutoetal2013,gulliksonetal2016,moe&distefano2017}.  The transition between these two regimes is not well understood.  Precise, bias-corrected measurements of both the binary star fraction and the mass-ratio distribution of intermediate-mass systems across intermediate orbital periods would provide important constraints for models of binary star formation.

Historically, binary systems were discovered by eclipses \citep{goodricke1783}, astrometry \citep{bessell1844}, or spectroscopy \citep{vogel1889, pickering1890}, but in the last few decades new methods have been invented. The discovery of a binary pulsar \citep{hulse&taylor1975} spurred larger pulsar-timing surveys \citep[e.g.][]{manchesteretal1978,lorimeretal2015,lyneetal2017}, ultimately constraining not only the masses of neutron stars \citep{antoniadisetal2016}, but also therewith the equation of state of cold, dense matter \citep{ozel&freire2016}. Large samples of binaries have now been collected from speckle imaging \citep{hartkopfetal1996,davidsonetal2009,tokovinin2012}, adaptive optics \citep{tokovininetal1999,shatsky&tokovinin2002,jansonetal2013,derosaetal2014}, long-baseline interferometry \citep[LBI,][]{rizzutoetal2013}, and common proper motion \citep{abtetal1990,catalanetal2008}, extending sensitivity to long orbital periods ($\gg10^3$\,d).

The greatest leap has been made with the availability of continuous space-based photometry, from MOST \citep{walkeretal2003}, CoRoT \citep{auvergneetal2009} and \textit{Kepler} \citep{boruckietal2010}. In addition to transforming the study of eclipsing binaries \citep{prsaetal2011}, these ultra-precise data have revealed binaries by reflection and mutual irradiation \citep{gropp&prsa2016}, Doppler beaming \citep{bloemenetal2011}, and ellipsoidal variability \citep{welshetal2010}. Eclipse timing variations can also lead to the discovery of non-eclipsing third bodies \citep{conroyetal2014}. More of these photometric discoveries can be expected for nearby stars with the launch of TESS \citep{rickeretal2015}, and a further quantum leap in binary orbital solutions is anticipated from Gaia astrometry \citep{debruijne2012}, particularly when combined with radial velocities from the RAVE survey \citep{zwitter&munari2004,steinmetzetal2006}.

\begin{figure}
\begin{center}
\includegraphics[width=0.48\textwidth]{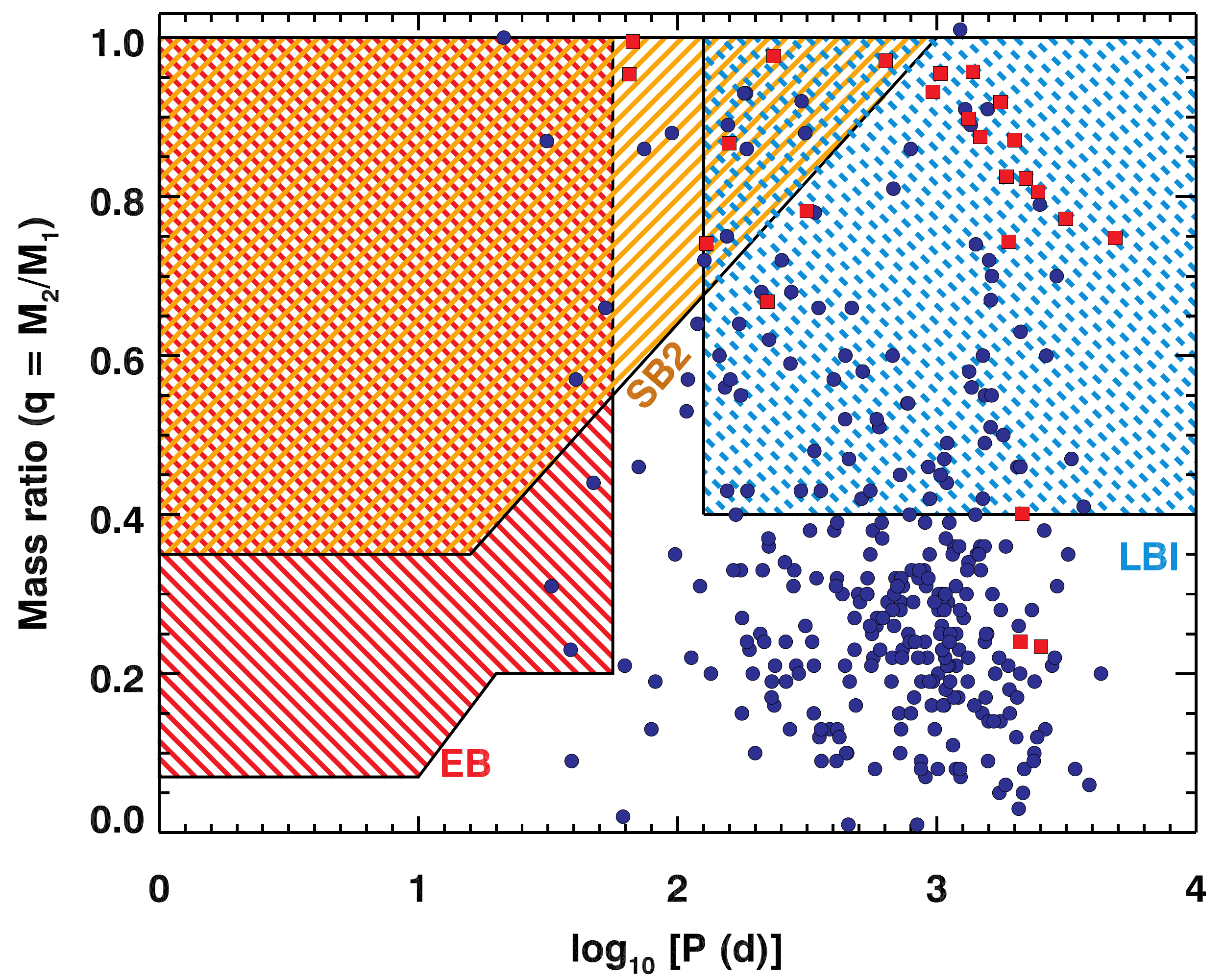}
\caption{The sensitivity of traditional techniques (eclipsing binaries, `EB', red; double-lined spectroscopic binaries, `SB2', orange; long-baseline interferometry, `LBI', cyan) to companions of 2-M$_{\odot}$ stars, as a function of orbital period ($P$) and mass ratio. PB2s (double-pulsator binaries) found by phase modulation in this work have directly measured mass ratios and are shown as red squares. The PB1 (single-pulsator binary) systems are shown as blue circles, for which we assumed orbital inclinations of $i=60^{\circ}$, that is, the median inclination of randomly distributed orbits. Although mass ratios of individual PB1s cannot be measured directly, we can reconstruct the intrinsic mass-ratio distribution with our large, well-characterized sample. Primary masses, $M_1$, come from \citet{huberetal2014}, though the mass ratio is far more sensitive to the measured binary mass function than to $M_1$. Figure adapted from \citet{moe&distefano2017}.}
\label{fig:sensitivity}
\vspace{-3mm}
\end{center}
\end{figure}

The latest innovation in binary star detection using \textit{Kepler}'s exquisite data sets is pulsation timing. The method has previously been applied to intermediate-mass stars \citep[e.g.][]{barnes&moffett1975}, but \textit{Kepler} has prompted a revival of interest leading to exploration of analytical orbital solutions, including in earlier papers of this series \citep{shibahashi&kurtz2012,murphyetal2014,murphy&shibahashi2015,shibahashietal2015}. Pulsation timing can detect companions down to planetary masses \citep{silvottietal2007}, even for main-sequence primaries \citep{murphyetal2016c}. It complements the existing methods well, with sensitivity at intermediate periods of $10^2$\,--\,$10^3$\,d, where radial velocity (RV) amplitudes tail off, eclipses are geometrically unlikely, and LBI lacks the sensitivity to identify binaries with large brightness contrasts, i.e., faint, low-mass companions with $q < 0.3$ \citep{moe&distefano2017}, as Fig.\,\ref{fig:sensitivity} shows.\footnote{In this paper we only discuss {\it orbital} periods, not {\it pulsation} periods, thus $P$ is always the orbital period.}
 Binaries with $P = 100$\,--\,2000\,d will interact via Case C Roche-lobe overflow \citep{lauterborn1970,toonenetal2014} and/or wind accretion with asymptotic giant branch (AGB) donors. Interacting binaries with intermediate-mass AGB donors are the progenitors of barium stars, blue stragglers, symbiotics, AM\,CVn stars, Helium stars, subdwarfs, R\,CrB stars, 1991bg-like Type Ia supernovae and, of course, Type Ia supernovae themselves \citep{boffin&jorissen1988,karakasetal2000,hanetal2002,mikolajewska2007,ruiteretal2009,geller&mathieu2011,zhang&jeffrey2012,claeysetal2014,maozetal2014}.

In this paper we present a catalogue of 341 full orbits from pulsation timing, making the method more successful than even spectroscopy for characterising binary systems with intermediate-mass primaries. A summary of the method is given in Sect.\,\ref{sec:method} and more details can be found in the references therein. Sect.\,\ref{sec:overview} describes the properties of the binaries and the catalogue. We determine the completeness of the method in Sect.\,\ref{sec:completeness}. In Sect.\,\ref{sec:MRD} we use a period--eccentricity relation to distinguish binaries consisting of main-sequence pairs from systems likely containing a white-dwarf companion to a main-sequence A/F primary. We derive the mass-ratio distribution of both populations and calculate the binary fraction of A/F stars at intermediate periods (100\,--\,1500\,d). We discuss the eccentricity distribution of main-sequence pairs in Sect.\,\ref{sec:eccentricity} and present our conclusions in Sect.\,\ref{sec:conclusions}.


\section{Sample selection and methodology}
\label{sec:method}

We have applied the phase modulation (PM) method \citep{murphyetal2014} to all targets in the original \textit{Kepler} field with effective temperatures between 6600\,K and 10\,000\,K, as given in the revised stellar properties catalogue \citep{huberetal2014}. The temperature range was chosen on three criteria. Firstly, we wanted to capture all $\delta$\,Scuti pulsators, since these have been shown to be excellent targets for PM \citep{comptonetal2016}, and because they address a large gap in binary statistics at intermediate stellar masses \citep{moe&distefano2017}. Secondly, the cut-off at 6600\,K avoids the rapidly increasing number of stars without coherent pressure modes (p\:modes) beyond the $\delta$\,Sct instability strip's red edge \citep{dupretetal2005b}. Thirdly, the upper limit of 10\,000\,K avoids the pulsating B stars, whose oscillation frequencies differ from $\delta$\,Sct stars and therefore have different sensitivity to companions. Crucially, this limit avoids subdwarfs, which are often in binaries and can be found by pulsation timing (e.g.\ \citealt{kawaleretal2010,teltingetal2012}), but again have different oscillation frequencies and they are not main-sequence stars. Although many $\delta$\,Sct stars are also $\gamma$\,Dor pulsators, the g\:modes of the latter have not proven to be as useful \citep{comptonetal2016}, so we did not fine-tune our temperature range to include them.

We used \textit{Kepler} long-cadence (LC; 29.45-min sampling) light curves from the multi-scale MAP data pipeline \citep{stumpeetal2014}, and only included targets with an observational timespan exceeding 100\,d. We calculated the discrete Fourier transform of each light curve between 5.0 and 43.9\,d$^{-1}$. The frequency limits generally selected only p\:modes, which, compared to low-frequency g\:modes, have 10\,--\,50 times more oscillation cycles per orbit and therefore allow the binary orbit to be measured more precisely. Since the presence of any unused Fourier peaks (such as the g\:modes) contributes to the phase uncertainty estimates of the useful modes, we applied a high-pass filter to the data to remove the low-frequency content. We ensured that the high-pass filtering did not inadvertently remove any useful oscillation content, and we looked for obvious eclipses before filtering. \citet{murphyetal2016c} have given an example of the filtering process.

The oscillation frequencies of intermediate-mass stars often lie above the LC Nyquist frequency (24.48\,d$^{-1}$). The Nyquist aliases are distinguishable from the real peaks in \textit{Kepler} data because of the correction of the time-stamps to the solar-system barycentre \citep{murphyetal2012b}. A consequence of this correction is that real peaks have higher amplitudes than their aliases, allowing them to be automatically identified. If the real peaks exceeded the {\it sampling} frequency, only their aliases would be detected in the prescribed frequency range. These alias peaks show strong phase modulation at the orbital frequency of the \textit{Kepler} satellite around the Sun and are easily identified \citep{murphyetal2014}. The frequency range was then extended for these stars only, such that the real peaks were included, but without increasing the computation time of the Fourier transforms for the entire \textit{Kepler} sample of 12\,650 stars.

The sample also included many non-pulsating stars, since much of the temperature range lies outside of known instability strips. We classified stars as non-pulsators if their strongest Fourier peak did not exceed 0.02\,mmag. Otherwise, up to nine peaks were used in the PM analysis. Frequency extraction ceased if the peak to be extracted had an amplitude below 0.01\,mmag or less than one tenth of the amplitude of the strongest peak (whichever was the stricter).

The method for detecting binarity is that of \citet{murphyetal2014}, which we briefly summarise here. We subdivide the light curve into short segments and calculate the pulsation phases in those segments. Since our survey used 10-d segments, Fourier peaks closer than 0.1\,d$^{-1}$ are unresolved from each other in those segments, and their phases are not measured independently. Unresolved peaks are excluded from the analysis, on a case-by-case basis, when the time delays of each pulsator are visually inspected. Binary motion imparts correlated phase shifts upon each pulsation mode, which are converted into light arrival-time delays (equations 1--3 of \citealt{murphyetal2014}). We check that the binary motion is consistent among the time-delay series of different modes, and never classify a star as binary that has only one pulsation mode showing phase modulation. We calculate the weighted average of the (up to nine) time-delay series, weighting by the phase uncertainty estimates, and use this weighted average to solve the orbit, starting with a semi-analytic solution \citep{murphy&shibahashi2015} and continuing with an MCMC method \citep{murphyetal2016b}. For PB2s, we identify two sets of time-delay series showing the same orbital motion, where one set is the mirror image of the other, scaled by an amplitude factor that is the binary mass ratio. Further details are given in the references above.

\begin{figure}
\begin{center}
\includegraphics[width=0.48\textwidth]{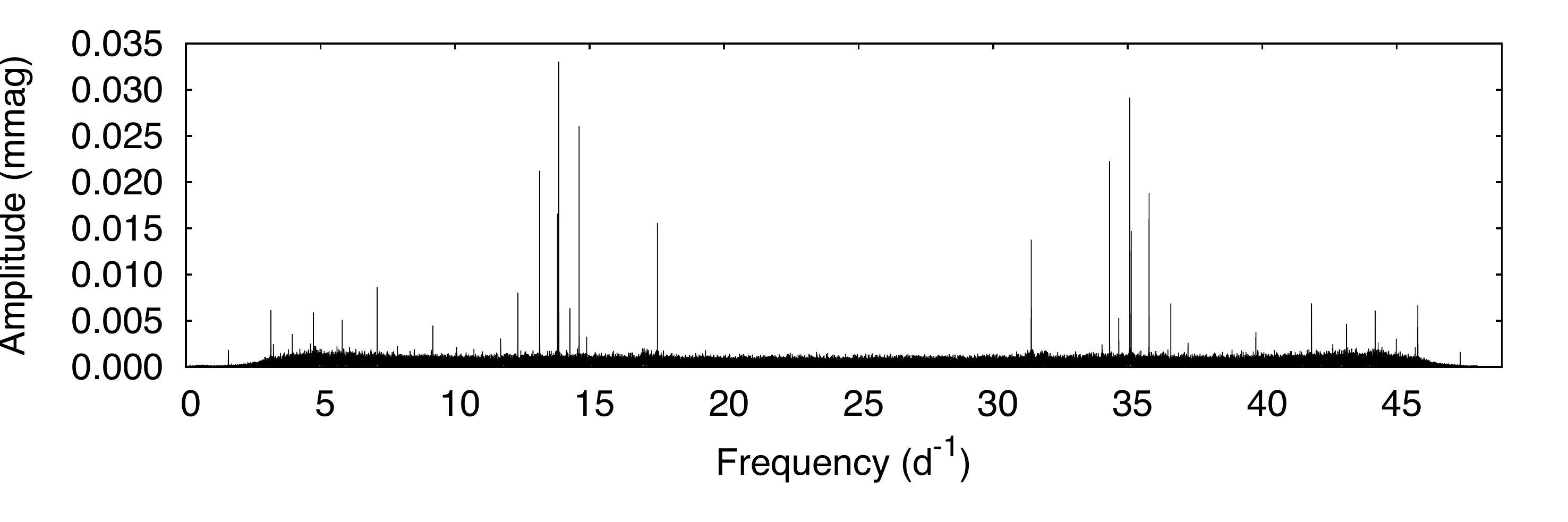}
\includegraphics[width=0.48\textwidth]{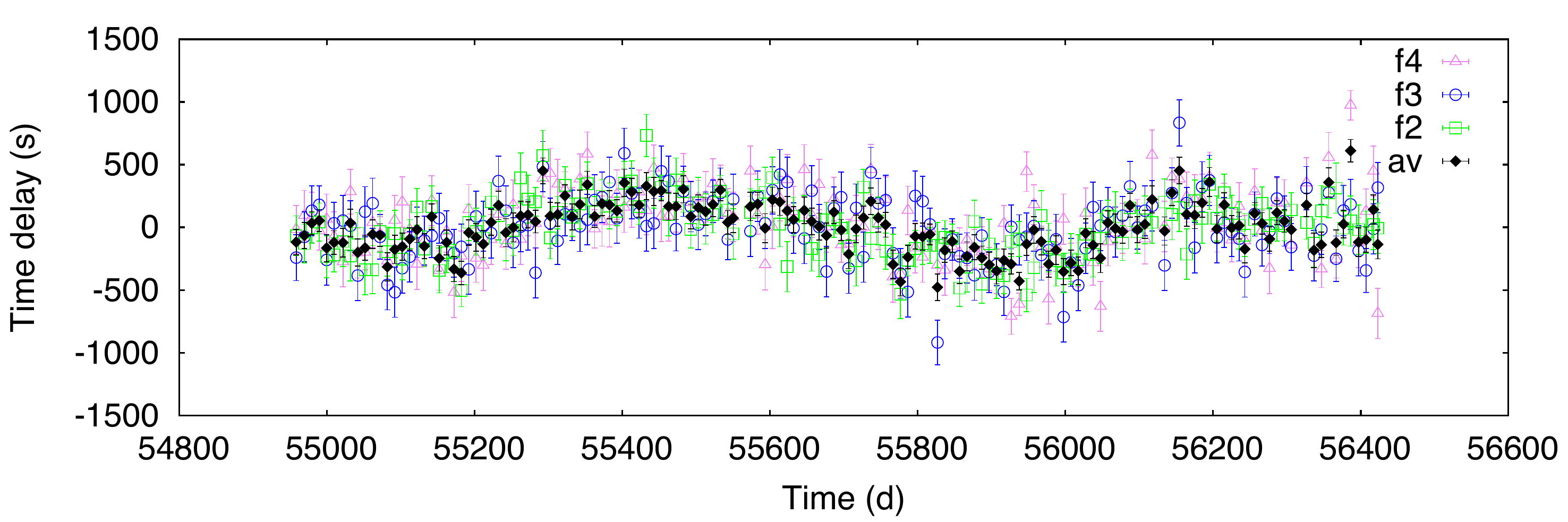}
\includegraphics[width=0.48\textwidth]{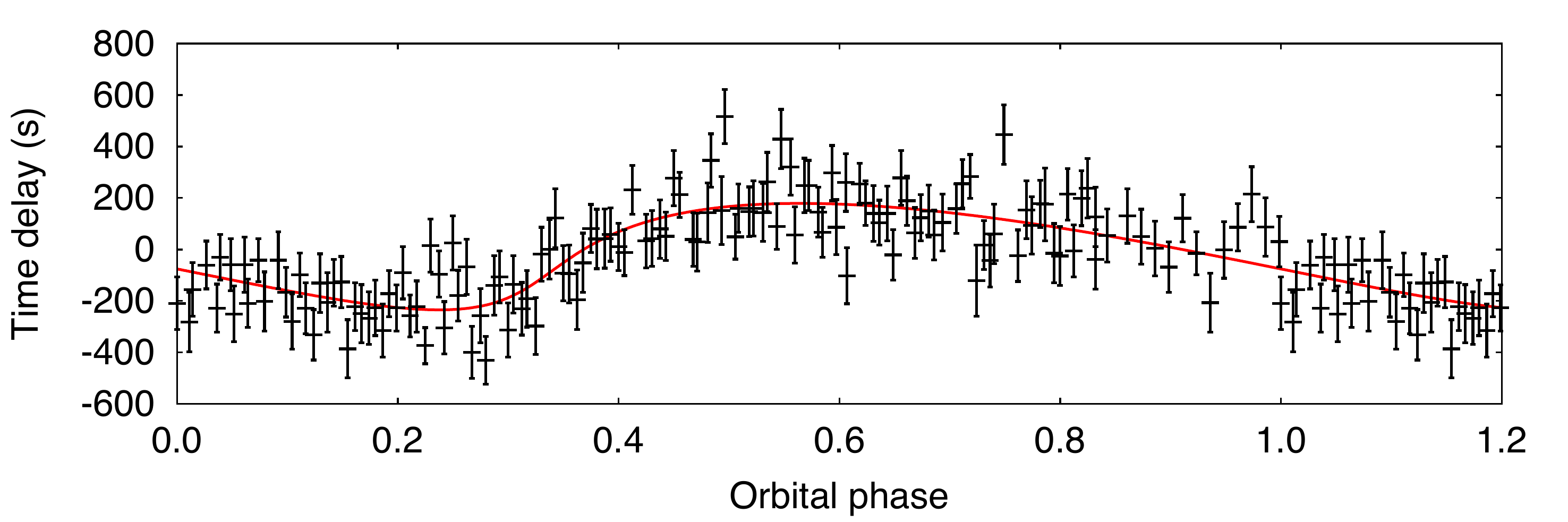}
\caption{PM analysis of the low-amplitude $\delta$\,Sct star KIC\,9172627. The strongest peak ($f_1$) in the Fourier transform of its light curve (top) has a close neighbour, unresolved in 10-d light-curve segments, so this peak is excluded. The remaining peaks above 0.01\,mmag, with frequencies of \mbox{$f_2=14.61$}, \mbox{$f_3=13.15$} and \mbox{$f_4=17.52$}\,d$^{-1}$, are used to extract time delays (middle panel). The bottom panel shows the weighted average time delay in each segment, folded on the orbital period, with the best fitting orbit ($P = 799$\,d, $a_1 \sin i / c = 250$\,s, $e=0.61$) shown as a red line.}
\label{fig:low-amp}
\end{center}
\end{figure}

\begin{figure}
\begin{center}
\includegraphics[width=0.48\textwidth]{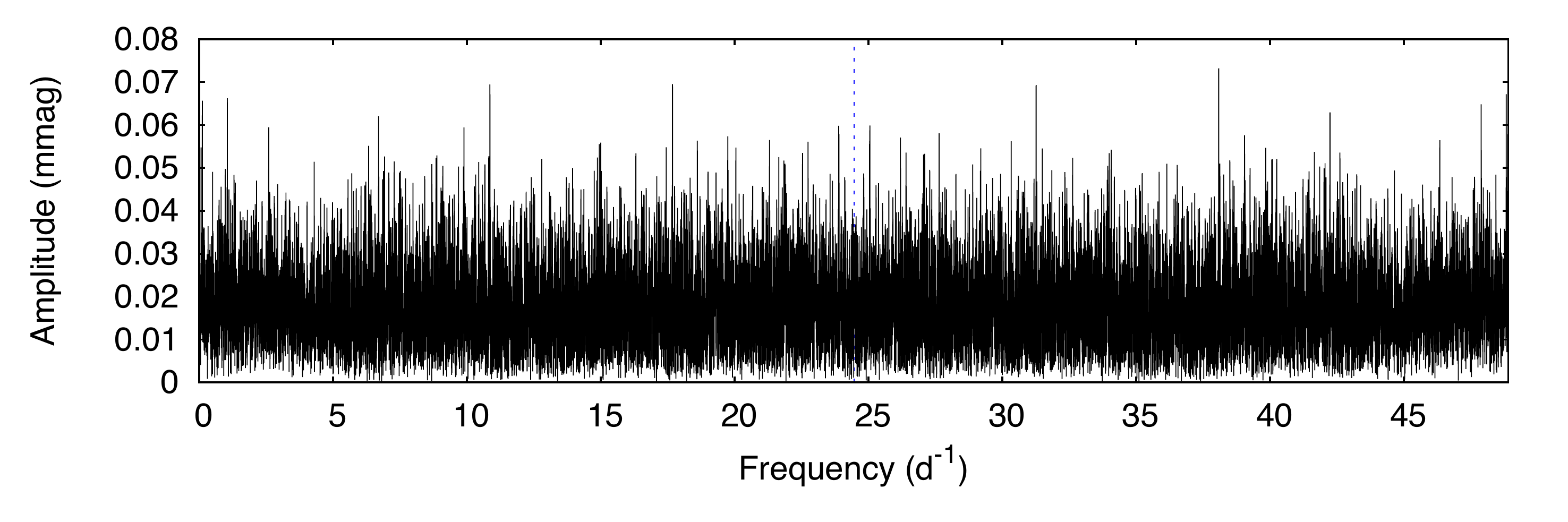}
\caption{Some non-pulsators have peak Fourier amplitudes above the 0.02\,mmag threshold, like KIC\,5264120 shown here. These were rejected manually.}
\label{fig:non-puls}
\end{center}
\end{figure}

Some pulsators had peak amplitudes above 0.02\,mmag, but with high noise levels such that the signal-to-noise ratio was deemed too low for any phase modulation to be detectable. We classified these as non-pulsators. No specific criterion was employed to distinguish between low-amplitude pulsators and non-pulsators, and one must also consider that the signal-to-noise ratio decreases markedly upon the segmentation of the light curve. Nonetheless, some very low amplitude pulsators that also had low noise levels did exhibit clear binarity. An example is KIC\,9172627 (Fig.\,\ref{fig:low-amp}), while a noisy non-pulsator exceeding the amplitude threshold is shown in Fig.\,\ref{fig:non-puls}. In our sample of 12\,649 stars, 2224 had usable p-mode pulsations. A breakdown is provided in Table\:\ref{tab:breakdown}.

\begin{table}
\centering
\begin{threeparttable}
\caption{Breakdown of the classification of 12\,649 targets analysed in this work. Eclipsing binaries, RR\,Lyr stars and stars without coherent p\:modes were not analysed. A further 84 stars had  `Insufficient data', usually where fewer than two consecutive Quarters of \textit{Kepler} data were available, and these were not analysed either. A total of 2224 stars had useful p\:modes (bottom three categories). Eclipsing binaries that were also non-pulsators were put in the `eclipsing' category. The `possible long period' systems look like binaries, but more data are needed to be certain.}
\label{tab:breakdown}
\begin{tabular}{l H H r c}
\toprule
Category & class & type & No. of stars & Percentage\tnote{a}\\
\midrule
Eclipsing / ellipsoidal variable & l & 0 & 439 & \phantom{1}3.5 \\
RR\,Lyrae stars & x & 1 & 33 & \phantom{1}0.3\\
No p\:modes (or weak p\:modes) & n, sn, not l & 2 & 9869 & 78.0 \\
\midrule
\multicolumn{4}{c}{\it p\:modes and...} \\
Insufficient data & f & 3 & 84 & \phantom{1}0.7 \\
Single (not PB1 or PB2) & s, s?, not sn / sw & 4 & 1756\tnote{b} & 13.9 \\
Possible long period & v,v? & 5 & 129 & \phantom{1}1.0 \\
PB1 or PB2 & w,sw & 6 & 339\tnote{c} &\phantom{1}2.7\\
\bottomrule
\end{tabular}
\begin{tablenotes}\footnotesize
\item [a] Rounding causes the percentage totals to differ from 100.0.
\item [b] This number includes 7 systems that were found to be single in the PM survey and hence in our statistics, but were found to be (mostly short-period) binaries by other methods and now have PM orbital solutions. They are described in Appendix\:\ref{sec:shortP} and Appendix\,\ref{sec:lampens} (C15).
\item [c] KIC\,5857714 and KIC\,8264588 have two PM orbits. Each orbit is counted in our statistics (Sect.\,\ref{sec:completeness} onwards) but each target is counted only once here. They are likely triple systems, detected as PB1s.
\end{tablenotes}
\end{threeparttable}
\end{table}

Our 10-d segmentation can only detect binaries with periods exceeding 20\,d. While it is possible to obtain PM solutions at shorter orbital periods \citep{schmidetal2015,murphyetal2016b}, this requires shorter segments, in turn raising the uncertainties on each time-delay measurement. Short-period orbits have correspondingly small values of $a_1 \sin i / c$, and therefore small time-delay variations. The signal-to-noise ratio in the time delays of these orbits is expected to be very small, and they will be detectable only if the binary has a mass ratio near unity. Such systems are likely to be detected from their orbital phase curves, i.e., via eclipses, ellipsoidal variation or reflection \citep{shporer2017}. Hence, a search with shorter segment sizes is not expected to return many new binaries. We discuss this further in Appendix\,\ref{sec:shortP}, where we provide time delay solutions for some \textit{Kepler} binaries discovered by other methods. Once binarity had been detected at 10-d sampling, we repeated the phase modulation analysis with shorter segments if the period was short and the orbit was undersampled, aiming for at least 10 time-delay measurements per orbit.

Visual inspection of the light curves and Fourier transforms led to the independent discovery of many eclipsing binaries and ellipsoidal variables.\footnote{The collection has good overlap with the Villanova eclipsing binary catalogue \citep[][\url{http://keplerebs.villanova.edu/}]{kirketal2016}.} Both have a long series of high-amplitude harmonics of the orbital frequency, continuing to frequencies above 5\,d$^{-1}$. These have been set aside for separate analysis in a future work. Light curve modelling of these systems can be used to provide an independent set of orbital constraints, and a PM analysis can provide the eccentricity and the orientation of the orbit. Since eclipses are geometrically more likely at short orbital periods and the PM method is more sensitive to longer orbits, these methods are complementary. Together they offer orbital detections over $\log P {\rm (d)} \approx 0$\,--\,3.3. Similarly, \citet{murphyetal2016b} showed that combining RV and time-delay data can provide orbital solutions for binaries with periods much longer than the 1500-d \textit{Kepler} data set, even with large gaps in observing coverage between the photometry and spectroscopy. Further, if the system is a hierarchical triple, a combination of PM with these other methods can lead to a very good orbital solution.

In addition to binary orbit statistics, such as the mass-ratio distributions presented from Sect.\,\ref{sec:MRD} onwards, a major output of this work is our classification of stars into the various aforementioned categories (eclipsing, single, non-pulsating, PM binary, etc.). We make these classifications available (online-only) with this article. A useful supplementary application of the classifications is the distinction of A and F stars into those with and without p-mode pulsation.


\section{Overview of the $\delta$\,Sct binaries}
\label{sec:overview}

\subsection{Effective temperature distribution}
\label{ssec:temp_range}

\begin{figure}
\begin{center}
\includegraphics[width=0.48\textwidth]{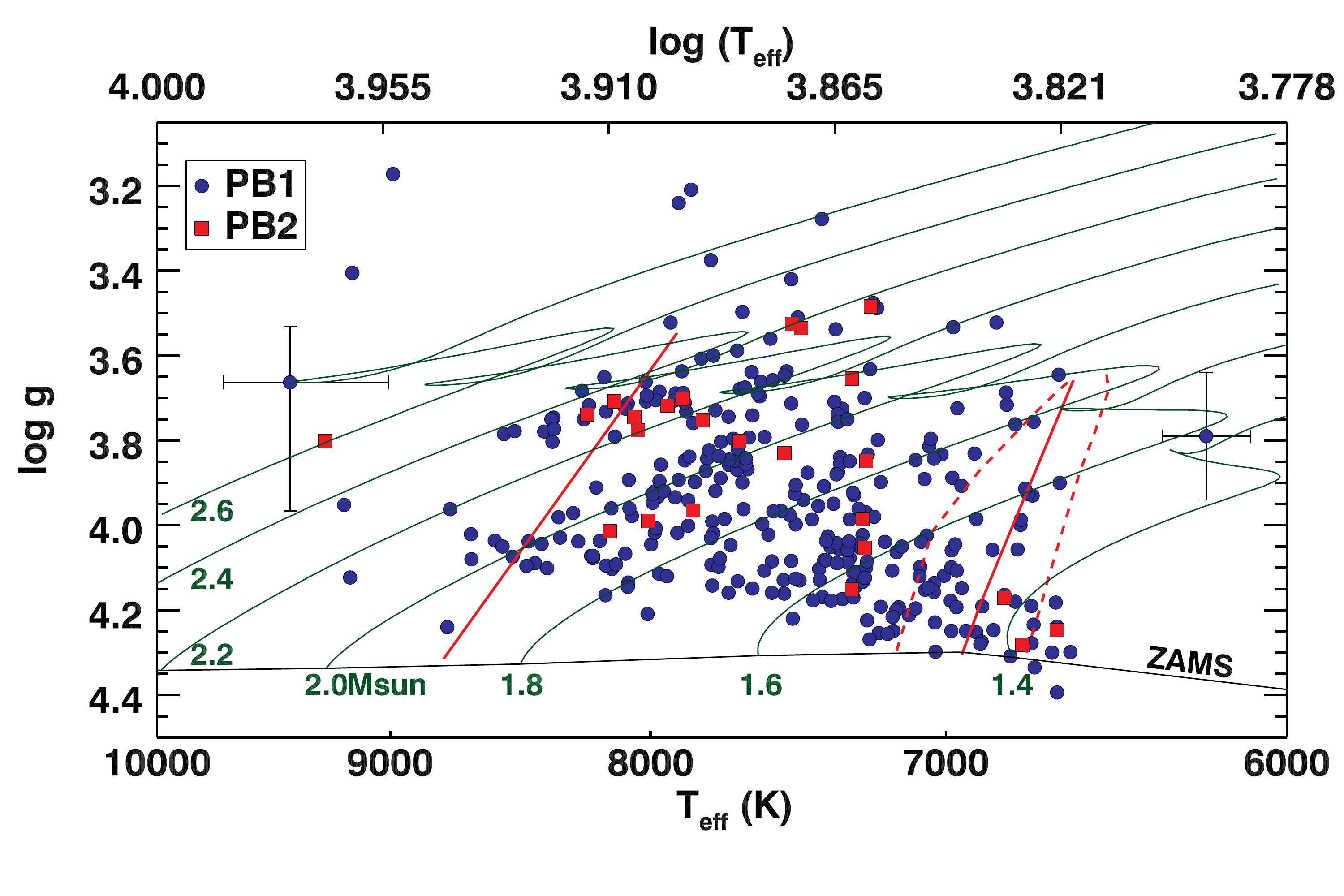}
\caption{The position of the PB1 (blue circles) and PB2 (red squares) systems on a $T_{\rm eff}$--$\log g$ diagram, according to revised KIC photometry \citep{huberetal2014}. The solid and dashed red lines delineate the $\delta$\,Sct and $\gamma$\,Dor instability strips, respectively, at solar metallicity. Evolutionary tracks of solar metallicity and mixing length $\alpha_{\rm MLT} = 1.8$ \citep{grigahceneetal2005} are also shown, with corresponding masses in M$_{\odot}$ written where the tracks meet the zero-age main-sequence (black line). Since these are binaries, the atmospheric parameters derived from photometry are unreliable, explaining the apparent location of pulsators outside of the instability strips. The hottest and coolest examples are shown with error bars, whose sizes increase with $T_{\rm eff}$.}
\label{fig:IS}
\end{center}
\end{figure}

The majority of our sample of $\delta$\,Sct stars in binaries appear to lie inside the $\delta$\,Sct instability strip according to broadband photometry, but there are some outliers. Fig.\,\ref{fig:IS} shows the position of the PB1 and PB2 stars on a $T_{\rm eff}$--$\log g$ diagram. The outliers are most likely the result of using photometry to determine $T_{\rm eff}$ for a star in a (blended) binary, although it should be noted that some well-studied $\delta$\,Sct stars lie far outside the instability strip (e.g.\ Vega; \citealt{butkovskaya2014}), that $\delta$\,Sct pulsation can be driven not only by the $\kappa$-mechanism but also by turbulent pressure \citep{antocietal2014}, and that the observed $T_{\rm eff}$--$\log g$ of intermediate-mass stars is a function of inclination angle, because their rapid rotation causes significant gravity darkening \citep{frematetal2005}. To be inclusive, we chose the lower $T_{\rm eff}$ cut-off of our sample such that stars were included if $T_{\rm eff} > 6600$ in either \citet{huberetal2014} {\em or} the original KIC \citep{brownetal2011}.

\subsection{Orbital parameters}
\label{ssec:orb_param}

\begin{table*}
\centering
\caption{Orbital parameters for the PB1 systems. The time of periastron, $t_{\rm p}$, is specified in Barycentric Julian Date. $K_1$ has been calculated from the other quantities (see Appendix\,\ref{app:RV_eq}). The full table is available online; here, the first 10 rows are shown for guidance on content and style.}
\label{tab:orbit}
\begin{tabular}{r r@{}l r@{}l r@{}l r@{}l r@{}l r@{}l r@{}l}
\toprule
\multicolumn{1}{c}{KIC number} & \multicolumn{2}{c}{$P$} & \multicolumn{2}{c}{$a_1 \sin i / c$} & \multicolumn{2}{c}{$e$} & \multicolumn{2}{c}{$\varpi$} & \multicolumn{2}{c}{$t_{\rm p}$} & \multicolumn{2}{c}{$f_{\rm M}$} & \multicolumn{2}{c}{$K_1$}\\
\multicolumn{1}{c}{} & \multicolumn{2}{c}{d} & \multicolumn{2}{c}{s} & \multicolumn{2}{c}{} & \multicolumn{2}{c}{rad} & \multicolumn{2}{c}{BJD} & \multicolumn{2}{c}{M$_{\odot}$} & \multicolumn{2}{c}{km\,s$^{-1}$}\\
\midrule
\vspace{1.5mm}
10001145 & $112.95$&$^{+0.42}_{-0.42}$ & $45.4$&$^{+4.1}_{-4.0}$ & $0.16$&$^{+0.15}_{-0.10}$ & $5.4$&$^{+1.3}_{-1.3}$ & $2\,455\,011$&$^{+26}_{-23}$ & $0.0079$&$^{+0.0021}_{-0.0021}$ & $8.65$&$^{+0.65}_{-0.49}$\\
\vspace{1.5mm}
10029999 & $185.95$&$^{+0.49}_{-0.49}$ & $106.5$&$^{+4.6}_{-4.4}$ & $0.271$&$^{+0.083}_{-0.081}$ & $3.97$&$^{+0.30}_{-0.24}$ & $2\,455\,059.1$&$^{+9.1}_{-7.6}$ & $0.0375$&$^{+0.0049}_{-0.0047}$ & $12.02$&$^{+0.50}_{-0.44}$\\
\vspace{1.5mm}
10031634 & $481.4$&$^{+3.3}_{-3.5}$ & $134$&$^{+71}_{-33}$ & $0.89$&$^{+0.071}_{-0.120}$ & $0.46$&$^{+0.20}_{-0.15}$ & $2\,455\,208.5$&$^{+8.9}_{-8.2}$ & $0.0111$&$^{+0.0180}_{-0.0081}$ & $2.76$&$^{+0.97}_{-1.50}$\\
\vspace{1.5mm}
10056297 & $619.9$&$^{+1.5}_{-1.6}$ & $163.2$&$^{+1.8}_{-1.8}$ & $0.033$&$^{+0.021}_{-0.018}$ & $5.33$&$^{+0.36}_{-0.25}$ & $2\,455\,427$&$^{+36}_{-25}$ & $0.01213$&$^{+0.00040}_{-0.00041}$ & $5.735$&$^{+0.043}_{-0.040}$\\
\vspace{1.5mm}
10056931 & $927.8$&$^{+3.3}_{-3.5}$ & $272.7$&$^{+2.3}_{-2.2}$ & $0.137$&$^{+0.019}_{-0.020}$ & $4.47$&$^{+0.12}_{-0.12}$ & $2\,455\,745$&$^{+17}_{-19}$ & $0.02529$&$^{+0.00067}_{-0.00064}$ & $6.348$&$^{+0.046}_{-0.046}$\\
\vspace{1.5mm}
10154094 & $893.1$&$^{+5.3}_{-5.5}$ & $254.8$&$^{+4.9}_{-4.9}$ & $0.142$&$^{+0.038}_{-0.038}$ & $4.02$&$^{+0.18}_{-0.23}$ & $2\,455\,506$&$^{+28}_{-33}$ & $0.0223$&$^{+0.0013}_{-0.0013}$ & $6.157$&$^{+0.099}_{-0.093}$\\
\vspace{1.5mm}
10206643 & $413.0$&$^{+2.6}_{-2.4}$ & $63.9$&$^{+3.6}_{-3.3}$ & $0.388$&$^{+0.096}_{-0.098}$ & $2.46$&$^{+0.20}_{-0.17}$ & $2\,455\,165$&$^{+16}_{-14}$ & $0.00164$&$^{+0.00028}_{-0.00026}$ & $3.11$&$^{+0.19}_{-0.17}$\\
\vspace{1.5mm}
10224920 & $1050$&$^{+14}_{-14}$ & $255.3$&$^{+14.0}_{-9.8}$ & $0.814$&$^{+0.050}_{-0.045}$ & $5.134$&$^{+0.056}_{-0.076}$ & $2\,455\,794$&$^{+19}_{-19}$ & $0.0162$&$^{+0.0026}_{-0.0019}$ & $3.08$&$^{+0.31}_{-0.29}$\\
\vspace{1.5mm}
10273384 & $2070$&$^{+220}_{-220}$ & $355$&$^{+85}_{-70}$ & $0.043$&$^{+0.053}_{-0.029}$ & $2.93$&$^{+0.18}_{-0.16}$ & $2\,455\,698$&$^{+230}_{-230}$ & $0.0112$&$^{+0.0083}_{-0.0071}$ & $3.73$&$^{+0.67}_{-0.54}$\\
\vspace{1.5mm}
10416779 & $1286$&$^{+71}_{-34}$ & $732$&$^{+41}_{-26}$ & $0.681$&$^{+0.027}_{-0.027}$ & $5.627$&$^{+0.042}_{-0.059}$ & $2\,455\,198$&$^{+53}_{-53}$ & $0.255$&$^{+0.045}_{-0.039}$ & $9.09$&$^{+0.55}_{-0.53}$\\
\bottomrule
\end{tabular}
\end{table*}

\begin{table*}
\centering
\caption{Orbital parameters for the PB2 systems, which have a measured $a_2 \sin i / c$ and thus have directly measured mass ratios, $q = (a_1 \sin i / c) / (a_2 \sin i / c) = a_1 / a_2$. The full table is available online; here, 10 rows from the middle of the table are shown for guidance on content and style.}
\label{tab:orbit2}
\begin{tabular}{r r@{}l r@{}l r@{}l r@{}l r@{}l r@{}l r@{}l r@{}l r@{}l}
\toprule
\multicolumn{1}{c}{KIC number} & \multicolumn{2}{c}{$P$} & \multicolumn{2}{c}{$a_1 \sin i / c$} & \multicolumn{2}{c}{$e$} & \multicolumn{2}{c}{$\varpi$} & \multicolumn{2}{c}{$t_{\rm p}$} & \multicolumn{2}{c}{$f_{\rm M}$} & \multicolumn{2}{c}{$K_1$} & \multicolumn{2}{c}{$a_2 \sin i / c$} & \multicolumn{2}{c}{$q$}\\
\multicolumn{1}{c}{} & \multicolumn{2}{c}{d} & \multicolumn{2}{c}{s} & \multicolumn{2}{c}{} & \multicolumn{2}{c}{rad} & \multicolumn{2}{c}{BJD} & \multicolumn{2}{c}{M$_{\odot}$} & \multicolumn{2}{c}{km\,s$^{-1}$} & \multicolumn{2}{c}{s} & \multicolumn{2}{c}{}\\
\midrule
\vspace{1.5mm}
2571868 & $158.1$&$^{+0.16}_{-0.17}$ & $155.3$&$^{+5.6}_{-5.7}$ & $0.0024$&$^{+0.0032}_{-0.0016}$ & $0.114$&$^{+0.130}_{-0.078}$ & $2\,455\,107.0$&$^{+2.9}_{-2.5}$ & $0.161$&$^{+0.017}_{-0.018}$ & $21.41$&$^{+0.77}_{-0.79}$ & $179.1$&$^{+2.8}_{-2.8}$ & $0.867$&$^{+0.024}_{-0.024}$ \\
\vspace{1.5mm}
2693450 & $634.1$&$^{+5.7}_{-5.7}$ & $384$&$^{+35}_{-37}$ & $0.205$&$^{+0.072}_{-0.076}$ & $3.07$&$^{+0.18}_{-0.38}$ & $2\,455\,116$&$^{+20}_{-37}$ & $0.151$&$^{+0.042}_{-0.043}$ & $13.5$&$^{+1.3}_{-1.3}$ & $395$&$^{+18}_{-18}$ & $0.973$&$^{+0.070}_{-0.072}$ \\
\vspace{1.5mm}
3661361 & $1031$&$^{+29}_{-22}$ & $663$&$^{+20}_{-20}$ & $0.482$&$^{+0.035}_{-0.033}$ & $1.752$&$^{+0.089}_{-0.100}$ & $2\,455\,821$&$^{+34}_{-41}$ & $0.294$&$^{+0.030}_{-0.032}$ & $15.99$&$^{+0.65}_{-0.60}$ & $692$&$^{+26}_{-24}$ & $0.957$&$^{+0.031}_{-0.033}$ \\
\vspace{1.5mm}
4471379 & $965.2$&$^{+2.5}_{-2.5}$ & $589$&$^{+17}_{-17}$ & $0.241$&$^{+0.015}_{-0.014}$ & $5.706$&$^{+0.050}_{-0.076}$ & $2\,454\,963.3$&$^{+7.7}_{-12.0}$ & $0.236$&$^{+0.0074}_{-0.0071}$ & $14.73$&$^{+0.18}_{-0.19}$ & $632.6$&$^{+6.6}_{-6.3}$ & $0.93$&$^{+0.02}_{-0.02}$ \\
\vspace{1.5mm}
4773851 & $67.129$&$^{+0.095}_{-0.095}$ & $74.0$&$^{+6.5}_{-6.5}$ & $0.29$&$^{+0.11}_{-0.10}$ & $4.0$&$^{+0.27}_{-0.26}$ & $2\,454\,996.5$&$^{+3.0}_{-3.1}$ & $0.096$&$^{+0.026}_{-0.026}$ & $25.2$&$^{+2.8}_{-2.5}$ & $74.5$&$^{+5.1}_{-4.9}$ & $0.993$&$^{+0.077}_{-0.079}$ \\
\vspace{1.5mm}
5310172 & $129.2$&$^{+0.20}_{-0.19}$ & $166.1$&$^{+4.6}_{-4.5}$ & $0.304$&$^{+0.054}_{-0.052}$ & $1.47$&$^{+0.11}_{-0.12}$ & $2\,455\,318.2$&$^{+2.3}_{-2.6}$ & $0.295$&$^{+0.063}_{-0.061}$ & $39.8$&$^{+3.1}_{-2.9}$ & $224$&$^{+16}_{-15}$ & $0.741$&$^{+0.039}_{-0.040}$ \\
\vspace{1.5mm}
5807415 & $1997$&$^{+190}_{-120}$ & $67.0$&$^{+8.4}_{-5.8}$ & $0.394$&$^{+0.064}_{-0.055}$ & $2.687$&$^{+0.120}_{-0.065}$ & $2\,455\,019$&$^{+170}_{-160}$ & $0.00008$&$^{+0.00005}_{-0.00006}$ & $0.91$&$^{+0.18}_{-0.20}$ & $76$&$^{+16}_{-18}$ & $0.88$&$^{+0.17}_{-0.14}$ \\
\vspace{1.5mm}
5904699 & $234.68$&$^{+0.26}_{-0.28}$ & $215.5$&$^{+6.2}_{-5.4}$ & $0.55$&$^{+0.03}_{-0.03}$ & $5.411$&$^{+0.041}_{-0.032}$ & $2\,455\,012.1$&$^{+1.7}_{-1.6}$ & $0.195$&$^{+0.016}_{-0.015}$ & $24.57$&$^{+1.20}_{-0.99}$ & $220.8$&$^{+6.0}_{-5.8}$ & $0.976$&$^{+0.027}_{-0.026}$ \\
\vspace{1.5mm}
6509175 & $222.01$&$^{+0.68}_{-0.66}$ & $175.4$&$^{+9.2}_{-8.7}$ & $0.389$&$^{+0.065}_{-0.067}$ & $6.21$&$^{+0.14}_{-0.12}$ & $2\,455\,156.6$&$^{+5.1}_{-5.7}$ & $0.118$&$^{+0.025}_{-0.024}$ & $28.0$&$^{+2.6}_{-2.3}$ & $263$&$^{+18}_{-18}$ & $0.67$&$^{+0.04}_{-0.04}$ \\
\vspace{1.5mm}
6784155 & $65.16$&$^{+0.10}_{-0.11}$ & $88$&$^{+11}_{-11}$ & $0.28$&$^{+0.15}_{-0.14}$ & $0.18$&$^{+0.52}_{-0.61}$ & $2\,454\,994.0$&$^{+5.4}_{-6.3}$ & $0.171$&$^{+0.050}_{-0.049}$ & $32.2$&$^{+4.5}_{-3.5}$ & $92.3$&$^{+9.0}_{-8.8}$ & $0.95$&$^{+0.11}_{-0.10}$ \\
\bottomrule
\end{tabular}
\end{table*}

\begin{figure}
\begin{center}
\includegraphics[width=0.48\textwidth]{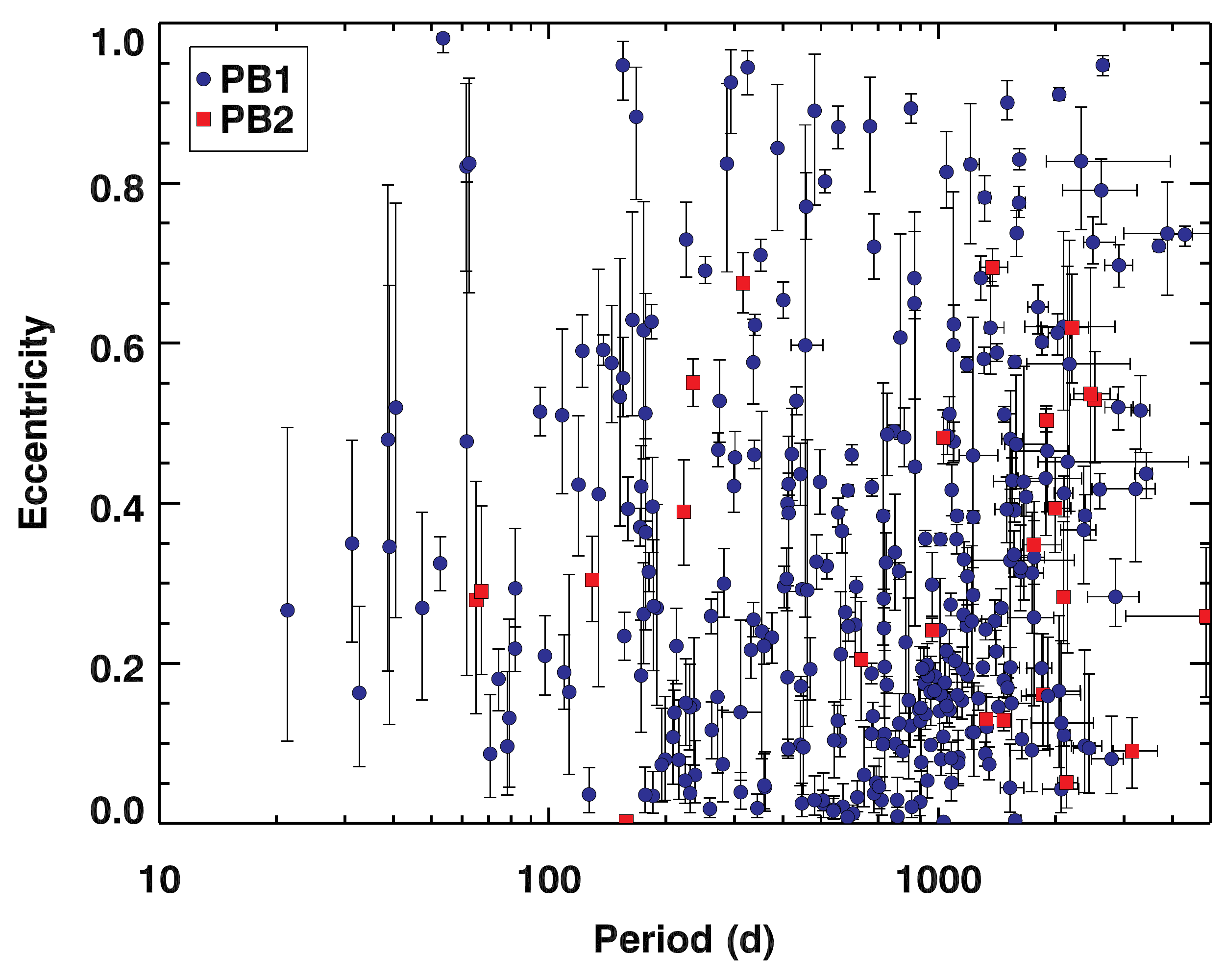}
\caption{Eccentricity vs. orbital period for the PB1 and PB2 systems (blue circles and red squares, respectively). Uncertainties are obtained via MCMC analysis, but are underestimated at long periods; see the text for details.}
\label{fig:pb_orbits}
\end{center}
\end{figure}

\begin{figure}
\begin{center}
\includegraphics[width=0.48\textwidth]{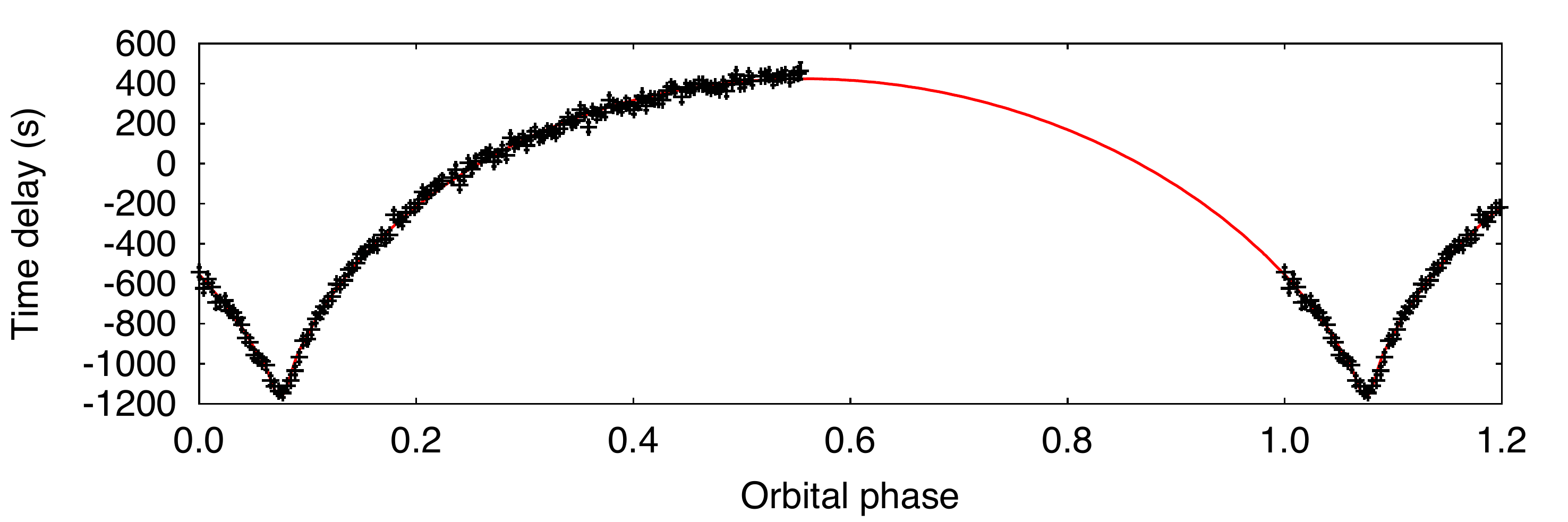}
\includegraphics[width=0.48\textwidth]{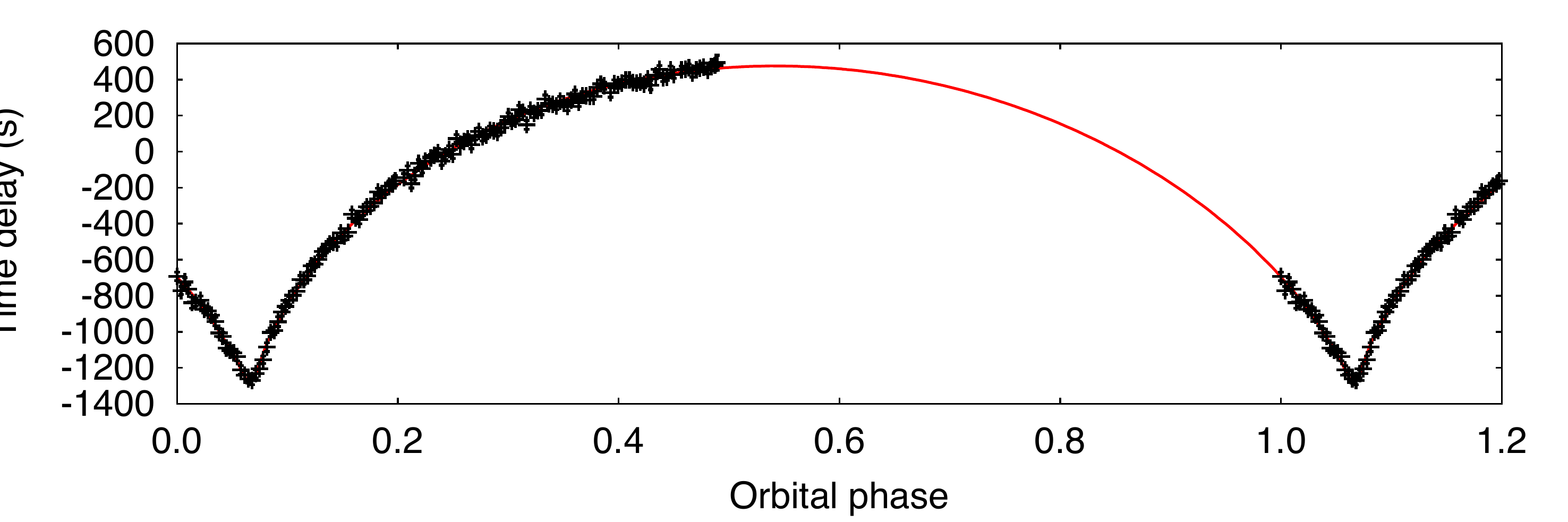}
\caption{Multiple orbital solutions can be found for binaries with periods unbounded by the four-year \textit{Kepler} data set, as shown here for KIC\,9108615. The top panel shows a solution at $P = 2644$\,d, $e = 0.95$ and $a_1 \sin i / c = 818$\,s, while the bottom panel has $P = 2994$\,d, $e = 0.92$ and $a_1 \sin i / c = 897$\,s. The differences are larger than the 1$\sigma$ uncertainties, although the correlated period and semi-major axis lead to a similar binary mass function.}
\label{fig:longp}
\end{center}
\end{figure}

The orbital parameters of the PB1s and PB2s are given in Tables\:\ref{tab:orbit} and \ref{tab:orbit2} (the full table is available online only). 
The orbital periods of our binaries range from just over 20\,d to a few thousand days, while the eccentricities span the full range from 0 to almost 1 (Fig.\,\ref{fig:pb_orbits}). The uncertainties for all parameters were obtained via Markov-chain Monte Carlo (MCMC) analysis of the time delays \citep{murphyetal2016b}. When the orbital periods are longer than the \textit{Kepler} data set, those periods are unbounded and the orbital solutions are multi-modal: degeneracies occur between some of the orbital parameters (especially $P$ and $a_1 \sin i / c$) and it is no longer possible to obtain unique solutions. Good solutions (small $\chi^2$) can still be obtained, but they are not the only solutions, as Fig.\,\ref{fig:longp} shows. Systematic errors are not adequately represented by the displayed error bars in these cases, and we advise strong caution in using them. We perform our analysis of binary statistics (Sect.\,\ref{sec:MRD} onwards) for periods $P<1500$\,d, only.

Uncertainties on eccentricities display a wide range, from 0.002 to 0.3. These are governed by the quality of the time delay observations; high-amplitude oscillations that are well separated in frequency give the lowest noise \citep{murphyetal2016b}. There is also some dependence on the orbital period, with long-period binaries tending to have better-determined eccentricities, up to the $P=1500$\,d limit. This is largely a result of increased scatter in the time delays at short periods, caused by poorer resolution and by having fewer pulsation cycles per orbit. Conversely, uncertainties on orbital periods increase towards longer orbital periods.

The use of 10-d sampling prevents binaries with $P<20$\,d from being found by the PM method. With short-period binaries also having smaller orbits, the binaries with periods in the range 20--100\,d are difficult to detect and the sample suffers considerable incompleteness (further described in Sect.\,\ref{sec:completeness}).

\subsubsection{Spectroscopic binary sample for comparison}
\label{sssec:sb9}

We have collected a sample of spectroscopic binary systems with known orbital parameters for later comparison with our pulsating binaries.
We selected all spectroscopic binaries from the ninth catalogue of spectroscopic binary orbits (``SB9 Catalogue'', \citealt{pourbaixetal2004}, accessed at VizieR\footnote{\url{http://cdsarc.u-strasbg.fr/viz-bin/Cat?B/sb9}} on 2016~May~09), with primary stars similar to our PBs. We filtered by spectral type, selecting systems classified between B5 and F5, which is somewhat broader than the temperature range of our \textit{Kepler} sample in order to enlarge the spectroscopic sample. We made no restriction on luminosity class, but the majority are class IV or V. We removed systems where the eccentricity was not known, or where it was `0.0000', implying that no eccentricity has been measured. We also removed multiple systems for a more direct comparison to our sample. This is necessary because we excluded ellipsoidal variables and eclipsing binaries from our PM analysis. These are usually found at short periods, and are therefore much more likely to be in hierarchical triples \citep{tokovininetal2006,moe&distefano2017}. Finally, we rejected any system for which no uncertainty on the orbital period was provided. The final SB9 sample considered here contained 164 binaries, from 100 different literature sources.

\subsubsection{Circularisation at short periods}
\label{sssec:circularisation}

\begin{figure}
\begin{center}
\includegraphics[width=0.48\textwidth]{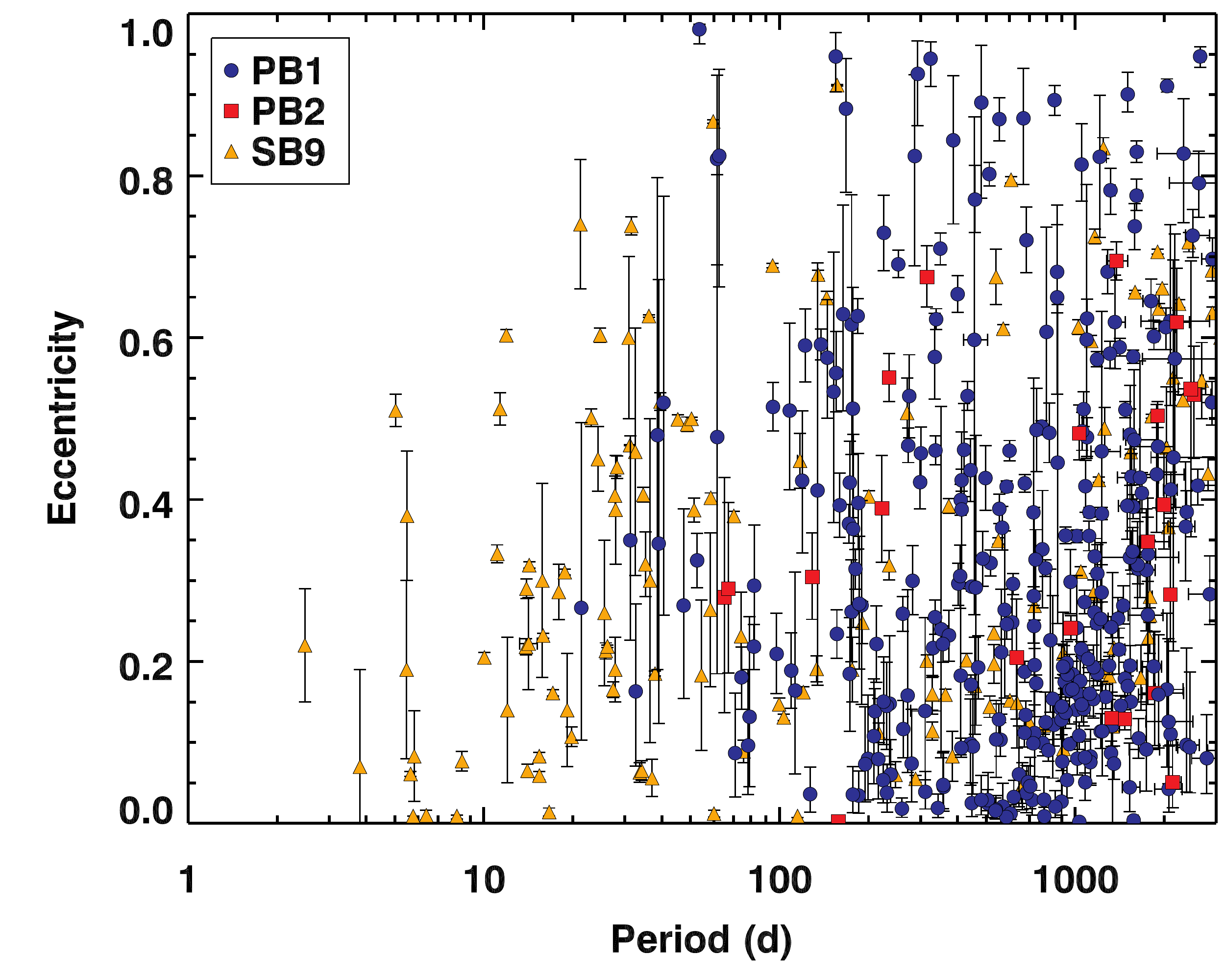}
\includegraphics[width=0.48\textwidth]{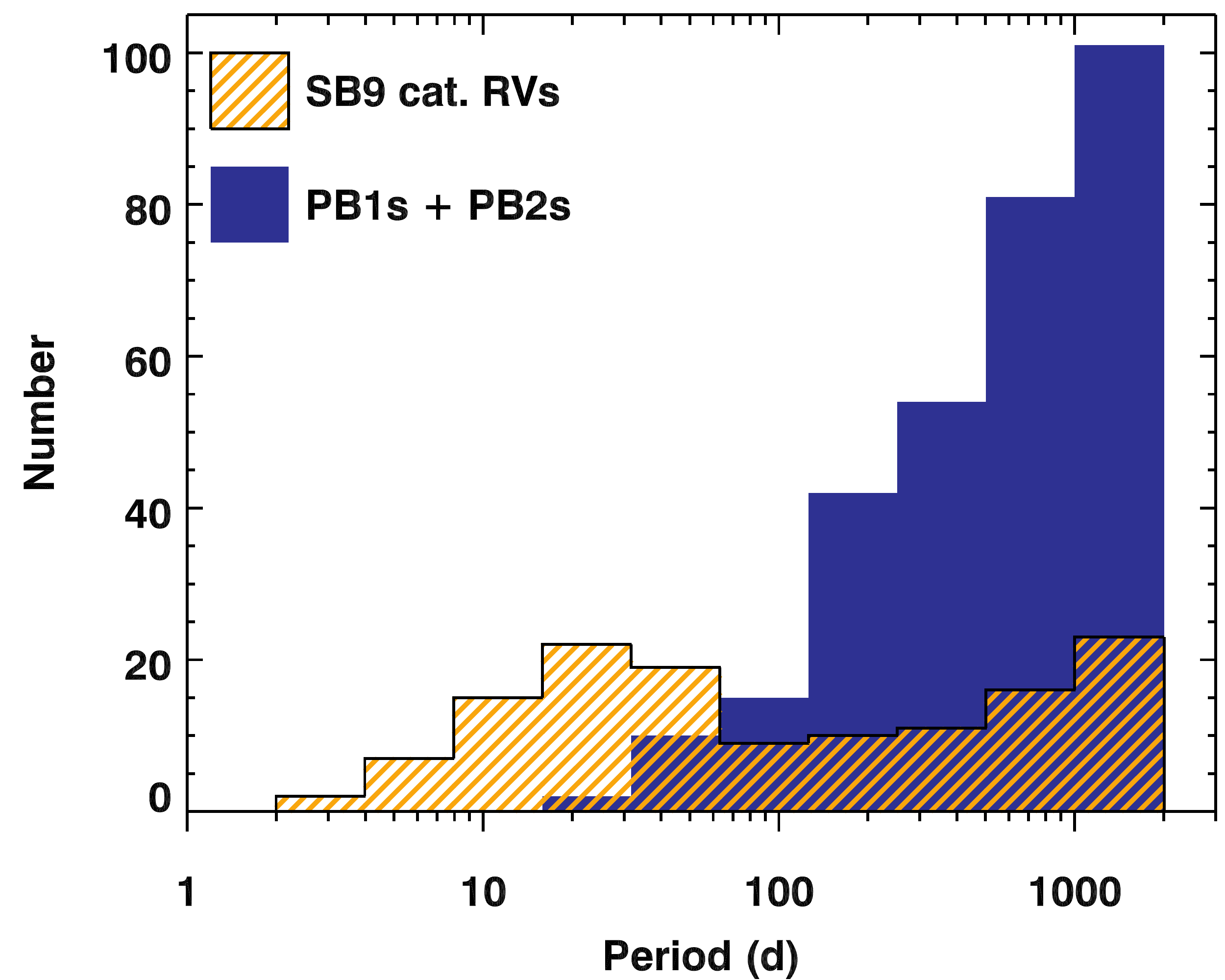}
\caption{Upper panel: The orbital periods and eccentricities of the PB1 systems (blue circles), PB2 systems (red squares) and the SB1 systems (orange triangles). The one-dimensional distribution in orbital period (for periods shorter than 2000\,d) is collapsed into an overlapping histogram in the lower panel.}
\label{fig:pe}
\end{center}
\end{figure}

The period--eccentricity diagram is shown in Fig.\,\ref{fig:pe} with the addition of the 164 spectroscopic binaries. \citet{pourbaixetal2004} noted circularisation of the spectroscopic binaries at short periods; the observed distribution is more heavily circularised than simple theory predicts \citep{hut1981}. We see some evidence of circularisation in our PB sample. At periods below $\sim$150\,d, Fig.\:\ref{fig:pb_orbits} shows few systems have high eccentricities ($e>0.7$), while there are many more systems with low eccentricity ($e < 0.3$). The statistical significance of this observation is enhanced by adding the SB1 systems, which contain many binaries with $P < 100$\,d (Fig.\,\ref{fig:pe}).

The details of tidal effects in binaries -- which are responsible for the circularisation -- are only partly understood. For solar-type stars with radiative cores and convective envelopes, the weak-friction equilibrium tide model adequately describes the eccentricity evolution of a population of binaries \citep{zahn1977,hut1981}, although there are slight differences in the tidal efficiencies between theory and those inferred from observations \citep{meibom&mathieu2005,belczynskietal2008,moe&kratter2017}. For more massive stars with convective cores and radiative envelopes, including our $\delta$\,Sct systems, tidal energy dissipation is less well understood and the observed efficiency of tidal interactions \citep{abt&boonyarak2004} greatly exceeds classical predictions \citep{zahn1975,zahn1984,tassoul&tassoul1992a}. Redistribution of angular momentum likely proceeds via dynamical oscillations \citep{zahn1975,witte&savonije2001,fuller2017}. These modes can be excited to large amplitudes in a variety of systems \citep{fuller&lai2011,fulleretal2013,hambletonetal2015}, and have been observed in abundance in \textit{Kepler} data of eccentric ellipsoidal variables (`heartbeat stars'; \citealt{thompsonetal2012}). Our sample contains the longest-period heartbeat stars in the \textit{Kepler} data set, which lie at the upper-left envelope of the period--eccentricity diagram (Fig.\,\ref{fig:pe}, see also \citealt{shporeretal2016}), some of which were previously unknown. Further examination of our pulsating binaries, including the heartbeat stars, for the presence of tidally excited modes is therefore highly worthwhile, and we will be performing spectroscopic follow-up to constrain the atmospheric parameters of the components.

\subsubsection{Circularisation at long periods}

There is also evidence for circularisation at long periods in Figs~\ref{fig:pb_orbits}~and~\ref{fig:pe}. We argue in Sect.\,\ref{sec:MRD} that the excess of orbits at long period and low eccentricity is caused by post-mass-transfer binaries. These orbits bias the observed eccentricity distribution towards small values. We discuss the eccentricity distribution of A/F stars at intermediate periods in Sect.\,\ref{sec:eccentricity}.


\section{Detection efficiency (Completeness)}
\label{sec:completeness}

The completeness of our search for binary companions is a function of the orbital period and the mass ratio of the stars. Generally, it does not depend on the eccentricity, except for a small bias against detecting highly eccentric orbits at very low values of $a_1 \sin i/c$ \citep{murphyetal2016b}. The PM method can only be applied to pulsating stars, but we assume that the binary properties of pulsators and non-pulsators are alike. This assumption does not affect our detection efficiency calculations but is relevant to the binary statistics of A/F stars, so we discuss it below, along with the interplay between chemical peculiarity, binarity and pulsation in metallic-lined (Am) stars. Then we estimate the detection efficiency (completeness) for stars that are found to pulsate.

\subsection{Applicability to all main-sequence A/F binaries}
\label{ssec:proxies}

We would like to extend our results to all main-sequence A/F stars, beyond the pulsators. For this, we need to confirm that there is no interplay between pulsations and binarity at the periods over which our statistics are derived in Sect.\,\ref{sec:MRD} (100\,--\,1500\,d).

It is known, for instance, that metallic-lined (Am) stars, which comprise up to 50\% of late-A stars \citep{wolff1983}, are preferentially found in short-period binaries ($P\sim10$\,d, \citealt{abt1961,vauclair1976,debernardi2000}), and are less likely to pulsate than normal stars \citep{breger1970,kurtzetal1976}. Tidal braking and atomic diffusion are the cause \citep{baglinetal1973}. However, recent theoretical developments and new observations of Am stars, including by \textit{Kepler} and \textit{K2}, have narrowed the discrepancy in the pulsation incidence between Am and normal stars \citep{smalleyetal2011,antocietal2014,smalleyetal2017}. Moreover, the typical orbital periods of Am stars are much shorter than those we consider.

Pulsation and binarity are independent at $P>100$\,d. \citet{liakos&niarchos2017} recently compiled a list of 199 eclipsing binaries with $\delta$\,Sct components. They found that orbital and pulsation periods were correlated below $P=13$\,d (\citealt{kahramanetal2017} found the correlation to extend a little higher, to $\sim$25\,d) but not above that.

Binaries can also excite pulsation, such as in heartbeat stars, but the tidally excited modes have frequencies lower than the range considered in our analysis (\citealt{fuller2017}; our Sect.\,\ref{sec:method}) and so do not bias our sample. At $100<P<200$\,d, only the most eccentric systems ($e\gtrsim0.9$) may be heartbeat stars. The fact that we detect the time delays of $\delta$\,Sct pulsations in these long-period heartbeat systems suggests that we do not have any selection effects caused by tides.

The implications of excluding eclipsing binaries and ellipsoidal variables from our sample are small, since we are focusing only on systems with $P > 100$\,d and the geometric probability of eclipses decreases rapidly with orbital period. For $P = 100$\,d and $M_1 = M_2 = 1.8$\,M$_{\odot}$ ($R_{1,2}\sim1.5$\,R$_{\odot}$), grazing eclipses are seen when $i > 88.8^{\circ}$ (the probability of eclipse is $\sim$2.1\%). At $P = 1000$\,d, this changes to $i>89.7^{\circ}$ and a probability < 0.5\%. The greatest implication is a bias against triple systems. This is because a tight pair of stars, hypothetically orbiting a $\delta$\,Sct tertiary, have a higher probability of eclipse due to their short period. These eclipses would dominate the light curve and cause a rejection from our sample. For solar-type systems, only $\sim$5\% of companions across $P = 100$\,--\,1000\,d are the outer tertiaries to close, inner binaries, the majority of which do not eclipse \citep{tokovinin2008,moe&distefano2017}. For A stars, the effect cannot be much larger.

In summary, at $P>100$\,d where we derive our statistics, binarity does not affect whether or not a main-sequence A/F star pulsates as a $\delta$\,Sct star, hence our statistics are applicable to all main-sequence A/F stars.


\subsection{Completeness assessment}
\label{ssec:detection_efficiency}

The completeness was assessed as follows. We used an algorithm to measure the noise in the Fourier transforms of the time delays. After a satisfactory manual inspection of those results for $\sim$100 stars, we applied the algorithm to the pulsating `single' stars with no binary detection. This formed our benchmark distribution of time-delay noise amplitudes.

We used the same algorithm to find the noise in the time delays of detected binaries, and divided their $a_1 \sin i/c$ values by that noise to determine the signal-to-noise ratio (SNR) distribution of the binary detections. Only 8 binaries were detected at a SNR below 4. We decided that binaries would be detected in most cases where the SNR exceeded 4, and used that as our SNR criterion.

\begin{figure}
\begin{center}
\includegraphics[width=0.48\textwidth]{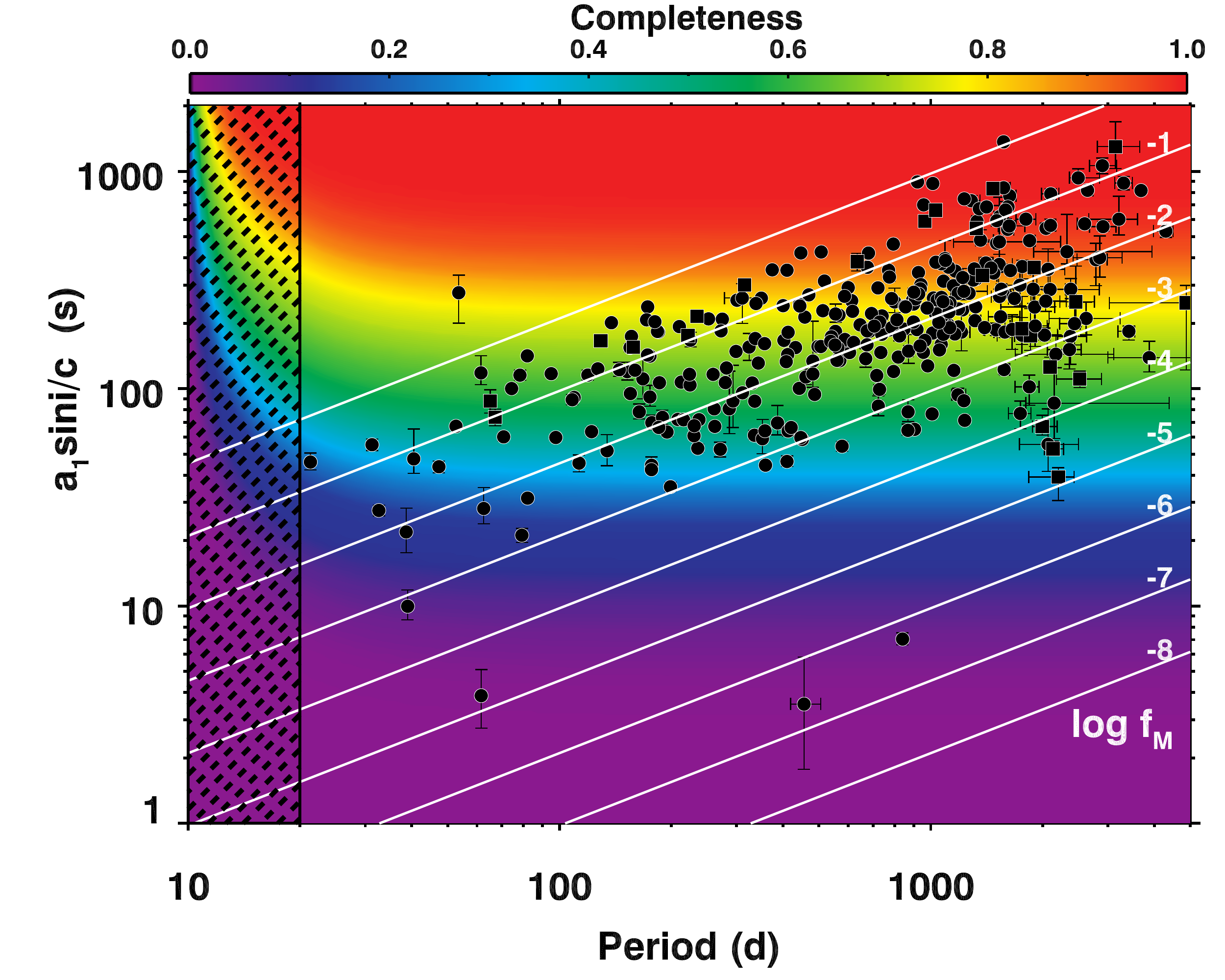}
\caption{Completeness across the $P$ -- $a_1 \sin i / c$ parameter space, with the detected binaries overlaid as circles (PB1s) and squares (PB2s). White lines follow constant binary mass function \mbox{log$_{10}$~$f_{\rm M}$}.\protect\footnotemark\ The hashed region below 20\,d is the unexplored region beyond the time-resolution limit of our survey. Exceptionally low-noise data are required to detect planetary-mass companions ($\log f_{\rm M} < -6$), one of which was characterised by \citet{murphyetal2016c}.}
\label{fig:detection_rate}
\end{center}
\end{figure}

The completeness was calculated as the fraction of stars in which a particular signal would have been detected, had it been present. The signal was computed over a logarithmically-distributed grid of orbital period and orbit size, spanning 10 $<$ $P$ (d) $<$ 5000, and 1 $<$ $a_1 \sin i/c$ (s) $<$  2000 (Fig.\,\ref{fig:detection_rate}). This parameter space is particularly informative because lines of constant $\log f_{\rm M}$ are readily overlaid, visualising the completeness as a function of the binary mass function. The weak dependence on eccentricity was not accounted for, but we did account for attenuation of the orbital signal due to `smearing' of the orbit in the 10-d segments \citep{murphyetal2016b}.

\footnotetext{For PB2s, $f_{\rm M}$ is calculated using the primary, as if it were a single star.}

To forward model the binary population (Sect.\,\ref{ssec:ms_pop}), the mean systematic uncertainty on the completeness is required. We estimated this by varying the slightly arbitrary SNR criterion from 4 to 3 and 5, and re-evaluating the completeness at the position in the grid, finding a mean systematic uncertainty of $\pm$5\%. We also need to know the detection efficiency, $D$, of each binary, which is simply the completeness at its specific $P$ and $a_1 \sin i / c$ values. 

In some binaries, both stars are pulsating (PB2 systems). In PB2s, the additional mode density can make the binary more difficult to detect, introducing a slight bias against them. Further, since their mass ratios are close to unity and their time delays are in anti-phase, their contributions to the weighted average time delay tend to cancel, further hindering their detection. While the observational signature of two time-delay series in anti-phase is easy to spot at longer periods (e.g. the \mbox{$\sim$960-d} PB2 in \citealt{murphyetal2014}), this is not the case below $\sim$100\,d. In this section, we have not explicitly calculated the bias against PB2s, but we suspect it is at least partly responsible for the overestimated completeness at periods $P<100$\,d in Fig.\,\ref{fig:detection_rate}. That overestimation leads us to examine the binary statistics of our sample from $P>100$\,d in Sect.\,\ref{sec:MRD}.

It is clear from Fig.\,\ref{fig:detection_rate} that our sensitivity to low-mass companions increases towards longer orbital periods. This is a direct result of having more pulsation cycles per orbit. It therefore has a similar dependence on orbital period to astrometry, but the inverse of the dependence in RV surveys. 
Using our grid of detection efficiencies, we calculate that at $P \sim 1000$\,d and for $M_1 = 1.8$\,M$_{\odot}$, we are sensitive to $q = 0.02$ ($q=0.10$) binaries in edge-on orbits in 13\% (62\%) of our pulsators.


\section{Determination of the mass-ratio distributions}
\label{sec:MRD}

Not all of our companions to $\delta$\,Sct stars are on the main-sequence. The original \textit{Kepler} field lies out of the galactic plane, and does not sample stars at the ZAMS. The most massive stars in the field have already evolved off the main sequence, beyond the red giant phase and become compact objects. We expect that a significant fraction of the long-period, low-eccentricity binaries are Sirius-like systems \citep{holbergetal2013}, featuring an evolved compact object orbiting an A/F star. The presence of these systems causes the mass ratio and eccentricity distributions to peak sharply at $q=0.3$ and $e\approx0$. To access the primordial distributions, we had to separate these systems from the main-sequence pairs.

\subsection{Separation of the populations}
\label{ssec:sep_pop}

Specific types of post-mass-transfer binaries, including post-AGB stars \citep{vanwinckel2003}, blue stragglers \citep{geller&mathieu2011}, and barium stars \citep{boffin&jorissen1988,jorissenetal1998,vanderswaelmenetal2017}, cluster near small to moderate eccentricities and intermediate periods $P$~=~200\,--\,5000\,d. These binaries all previously experienced Roche-lobe overflow or efficient wind accretion involving AGB donors and main-sequence F/G accretors \citep{karakasetal2000}.  During this process, the main-sequence F/G companions accreted sufficient material to become slightly more massive main-sequence A/F field blue stragglers.  These will later evolve into cooler GK giants and appear as barium stars due to the significant amounts of s-process-rich material they accreted from the AGB donors.  Radial velocity measurements of these types of post-mass-transfer binaries reveal secondaries with dynamical masses $M_2$~$\approx$~0.5\,M$_{\odot}$ \citep{jorissenetal1998,vanwinckel2003,geller&mathieu2011}, consistent with those expected for white dwarfs, i.e., the cores of the AGB donors.  A significant fraction of our main-sequence A/F $\delta$\,Sct stars with secondaries across $P$~$\approx$~200\,--\,2000\,d may therefore be field blue stragglers containing white-dwarf companions.  

\begin{figure}
\begin{center}
\includegraphics[width=0.48\textwidth]{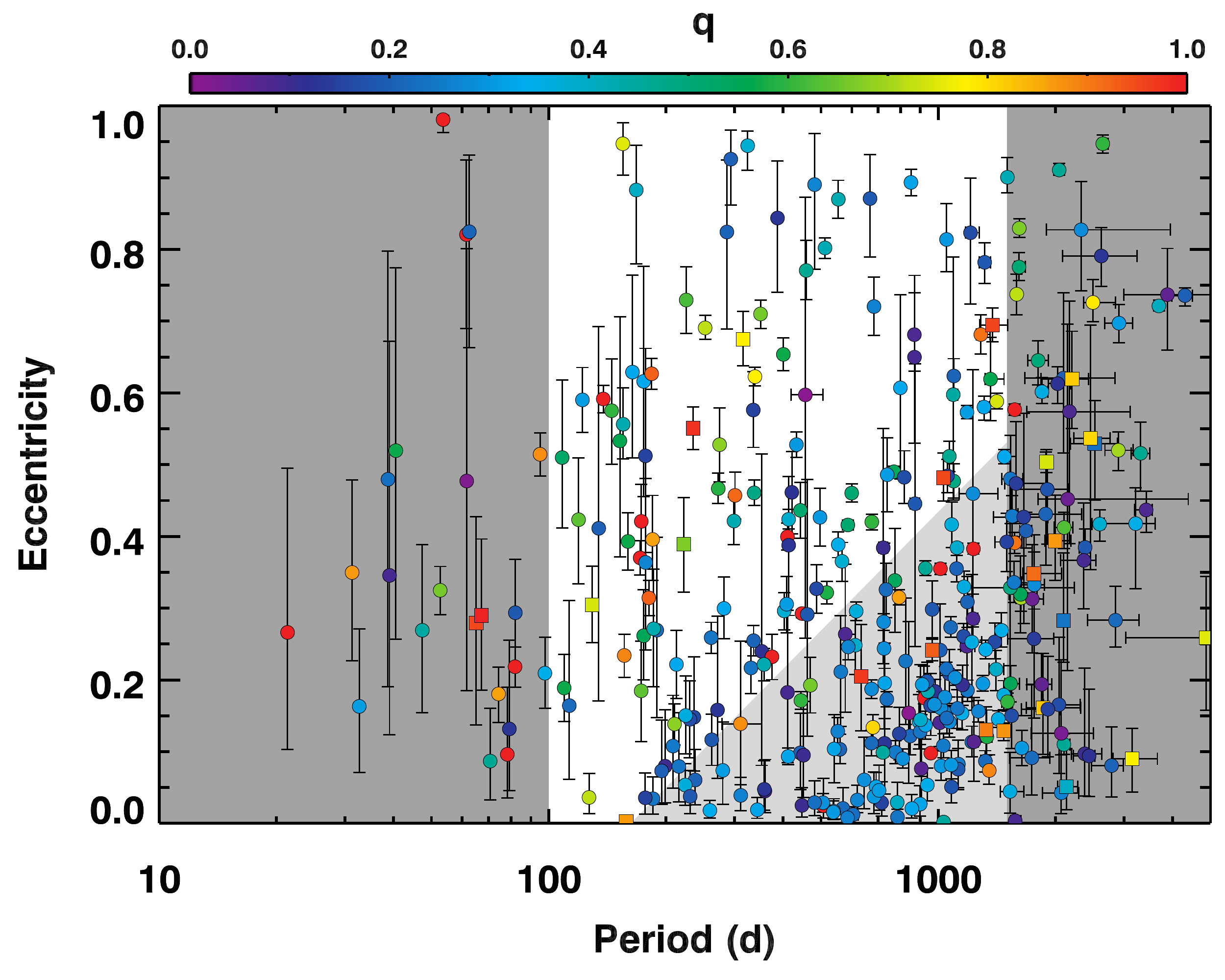}
\caption{The PB1 systems (circles) and PB2 systems (squares), separated into a `clean' population of main-sequence companions to $\delta$\,Sct stars (short $P$, high $e$, white background) and a `mixed' population that consists of both main-sequence pairs and post-mass-transfer systems (long $P$, low $e$, light-grey background). Orbital periods below $100$\,d have overestimated completeness rates, and those beyond 1500\,d cannot be determined reliably; these systems were not included in either subsample (dark-grey background). Mass ratios are encoded with colour; for PB2s these are directly measured, but for PB1s we approximated using $i=60^{\circ}$ and taking $M_1$ from \citet{huberetal2014} for each PB1. The existence of white-dwarf companions in the `mixed' subsample is evident from the clustering of systems with small mass ratios ($q \approx 0.3$; $M_2 \approx 0.5$\,M$_{\odot}$).}
\label{fig:separate}
\end{center}
\end{figure}

During the mass transfer process, binaries tidally circularise towards smaller eccentricities.  The observed populations of post-AGB binaries, blue stragglers, and barium stars all lie below a well-defined line in the $e$ vs. log\,$P$ parameter space \citep{jorissenetal1998,vanwinckel2003,geller&mathieu2011}.  This line extends from $e$~=~0.05 at log\,$P$\,(d)~=~2.3 to $e$~=~0.6 at log\,$P$\,(d)~=~3.3, as shown in Fig.\,\ref{fig:separate}.  In general, while binaries that lie below this $e$ vs. log\,$P$ relation may contain white-dwarf companions, we surmise that short-period, highly eccentric binaries above this line almost exclusively contain unevolved main-sequence companions. This follows from a straightforward comparison of periastron separations, which depend on eccentricity, with the Roche lobe geometries of post-main-sequence stars. For our observed companions to $\delta$\,Sct stars, we used the relation $e = 0.55 \log P ({\rm d}) - 1.21$ to separate a `clean' subsample in which nearly all the binaries have main-sequence companions from a `mixed' subsample that contains a large number of white-dwarf companions (see Fig.\,\ref{fig:separate}). 

Our statistical analysis covers the orbital period range \mbox{$P$~=~100\,--\,1500\,d.} Binaries with $P$~$<$~100\,d have small detection efficiencies (Sect.\,\ref{ssec:detection_efficiency}), and binaries with $P$~$>$~1500\,d have unreliable orbital elements, given the 4-year duration of the main {\it Kepler} mission (Sect.\,\ref{ssec:orb_param}). We removed the two outlier systems with very small detection efficiencies $D$~=~0.01\,--\,0.04; they are not likely stellar companions \citep{murphyetal2016c}, and it avoids division by small numbers when applying our inversion technique (see below). The remaining 245 binaries all have $D$~$>$~0.27.  Our short-period, large-eccentricity `clean' main-sequence subsample contains 115 systems (109 PB1s and 6 PB2s) with periods $P$~=~100\,--\,1500\,d, eccentricities above the adopted $e$ vs. log\,$P$ relation, and detection efficiencies $D$~$>$~0.27.  Meanwhile, our long-period, small-eccentricity `mixed' subsample includes white-dwarf companions and contains 130 binaries (126 PB1s and 4 PB2s) with periods $P$~=~200\,--\,1500\,d, eccentricities below the adopted $e$ vs. log\,$P$ relation, and detection efficiencies $D$~$>$~0.36.

\subsection{The mass-ratio distribution for main-sequence companions}
\label{ssec:ms_pop}

We investigated the mass-ratio distribution of main-sequence binaries based on our `clean' subsample of 115 observed systems (109 PB1s and 6 PB2s) with \mbox{$P$ = 100\,--\,1500\,d} and eccentricities large enough to ensure they have unevolved main-sequence companions.  Our detection methods become measurably incomplete toward smaller mass ratios ($q$~$<$~0.4), so observational selection biases must be accounted for. To assess the systematic uncertainties that derive from accounting for incompleteness, we used a variety of techniques to reconstruct the intrinsic mass-ratio distribution from the observations, consistent with parametrizations used in the literature.  In the following, we compare the mass ratios inferred from: (1)~a simple inversion technique that accounts for incompleteness, (2)~an MCMC Bayesian forward modelling method assuming a multi-step prior mass-ratio distribution, and (3)~a similar MCMC Bayesian technique assuming a segmented power-law prior mass-ratio distribution.

\begin{figure}
\begin{center}
\includegraphics[width=0.48\textwidth]{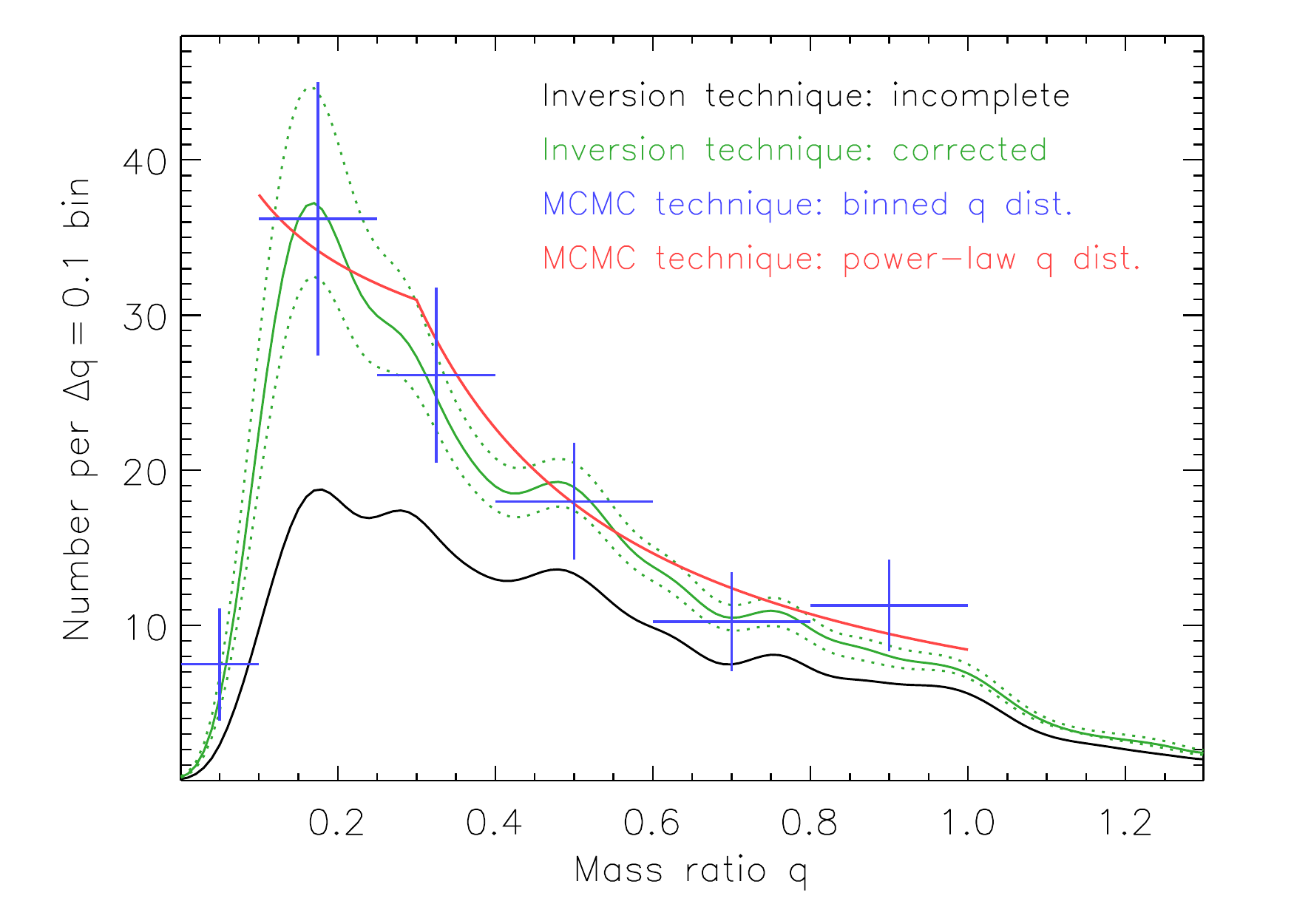}
\caption{The mass-ratio distribution based on the observed subsample of 115 binaries (109 PB1s and 6 PB2s) with $P = 100$\,--\,1500\,d and sufficiently large eccentricities that guarantee they have main-sequence companions. Our results from the population inversion technique are shown with completeness corrections (green) and without (black). Our MCMC Bayesian forward modelling method assuming a binned mass-ratio distribution (blue), and the MCMC Bayesian forward modelling technique assuming a segmented power-law mass-ratio distribution (red) agree well with the completeness-corrected inversion technique.  They yielded a total corrected number of 179\,$\pm$\,28 binaries and a mass-ratio distribution that is skewed significantly toward small values $q=0.1$\,--\,0.3 with a rapid turnover below $q \lesssim 0.10$\,--\,0.15.  This represents the first robust measurement of the mass-ratio distribution of binaries with intermediate orbital periods.}
\label{fig:qdist_clean}
\end{center}
\end{figure}

\subsubsection{Inversion Technique}
\label{ssec:inversion}

Population inversion techniques are commonly used to recover the mass-ratio distribution from observed binary mass functions (\citealt{mazeh&goldberg1992}; references therein). Here we describe our specific approach.

For each PB1, we have measured the binary mass function $f_{\rm M}$ from the pulsation timing method and its primary mass $M_1$ is taken from \citet{huberetal2014}, who estimated stellar properties from broadband photometry.  Given these parameters and assuming random orientations, i.e., $p(i) = \sin i$ across $i = 0$\,--\,$90^{\circ}$, we measured the mass-ratio probability distribution $p_j(q)$ for each $j$th PB1. For each of the six PB2s, we adopted a Gaussian mass-ratio probability distribution $p_j(q)$ with mean and dispersion that matched the measured value and uncertainty, respectively. By summing the mass-ratio probability distribution $p_j(q)$ of each of the 115 systems, we obtained the total mass-ratio distribution function $f(q)$ without completeness corrections, shown as the solid black line in Fig.~\ref{fig:qdist_clean}.

For each of the 115 systems, we also calculated the estimated detection efficiency $D_j$ in Sect.\,\ref{sec:completeness}. The total, bias-corrected mass-ratio distribution is simply $f(q)=\sum_j p_j(q)/D_j$. The detection efficiencies all satisfied $D>0.27$ in our `clean' subsample of 115 systems, so that we never divided by a small number.  The corrected mass-ratio distribution $f(q)$ based on this inversion technique is the solid green curve in Fig.~\ref{fig:qdist_clean}.  One of the dominant sources of error arose from the systematic uncertainties in the completeness rates.  For example, for $\log a_1 \sin i/c$\,(s)~=~1.6 across our orbital periods of interest ($P = 100$\,--\,1500\,d), we  measured a detection efficiency of $D$~=~0.32$_{-0.09}^{+0.11}$.  So for each detected system with $\log a_1 \sin i/c$\,(s)~=~1.6, there may be 1/0.23~=~4.3 (1\,$\sigma$ upper) or 1/0.43~=~2.3 (1\,$\sigma$ lower) total binaries with similar properties in the true population.  To account for this, we increased (decreased) each detection efficiency $D_j$ by its estimated 1\,$\sigma$ upper (lower) systematic uncertainty, and then repeated our inversion technique. The resulting upper and lower $f(q)$ distributions are shown as the dotted green lines in Fig.~\ref{fig:qdist_clean}.  By integration of $f(q)$, the total number of binaries was 180 $\pm$\,17\,(stat.) $\pm$\,21\,(sys.). We interpret the mass-ratio distribution in Sect.\,\ref{sssec:interp}, after describing the remaining techniques.

\subsubsection{MCMC method with step-function mass-ratio distribution}
\label{sssec:MCMC_step}

We next employed an MCMC Bayesian forward modelling technique by generating a population of binaries with various combinations of $M_1$, $P$, $i$, and $q$. To synthesize a binary, we selected $M_1$ and $P$ from the observed probability density functions of the 115 systems in our subsample.  In the `clean' subsample, the observed log\,$P$ distribution was roughly uniform across our selected interval 2.00 $<$ log\,$P$\,(d) $<$ 3.18, and the observed distribution of $M_1$ was approximately Gaussian with a mean $\langle M_1 \rangle = 1.8$\,M$_{\odot}$ and standard deviation $\sigma$ = 0.3\,M$_{\odot}$.  We generated binaries with random orientations, and so selected inclinations from the prior distribution $p(i) = \sin i$ across \mbox{$i=0^{\circ}$\,--\,$90^{\circ}$}. We adopted a six-parameter step-function model to describe the mass-ratio distribution: one bin of width $\Delta q = 0.1$ across \mbox{$q = 0.0$\,--\,0.1}, two bins of width $\Delta q = 0.15$ across \mbox{$q = 0.1$\,--\,0.4}, and three bins of width $\Delta q = 0.2$ across \mbox{$q = 0.4$\,--\,1.0}.  The number of binaries within each of the six mass-ratio bins represented a free parameter. For each mass-ratio bin, we selected $q$ assuming the mass-ratio distribution was uniform across the respective interval. For each synthesized binary, we calculated its binary mass function $f_{\rm M}$ and detection efficiency $D$ according to its physical properties $M_1$, $P$, $i$, and $q$.  By weighting each synthesized system by its detection efficiency, we simulated the posterior probability distribution of binary mass functions.

\begin{figure}
\begin{center}
\includegraphics[width=0.48\textwidth]{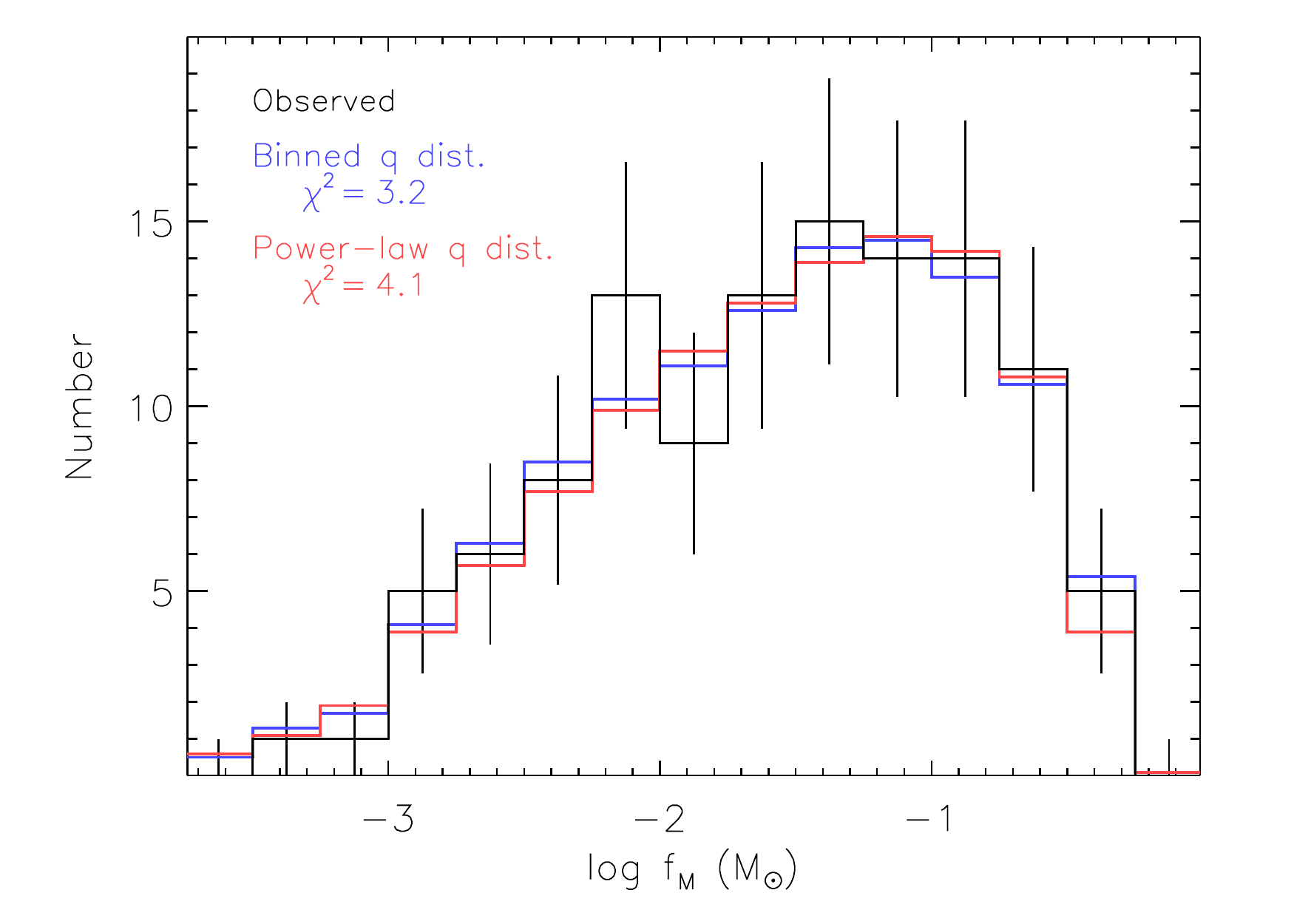}
\caption{Distribution of binary mass functions $f_{\rm M}$ for the 109 PB1 and 6 PB2 systems that compose our main-sequence subsample. The observations (black) were compared to the best-fit Bayesian models assuming either a binned mass-ratio distribution (blue) or a segmented power-law mass-ratio distribution (red).}
\label{fig:fm_clean}
\end{center}
\end{figure}

To fit the six free parameters that describe the mass-ratio distribution, we minimized the $\chi^2$ statistic between the observed and simulated posterior distributions of binary mass functions $f_{\rm M}$. Since we used the observed distribution of binary mass functions $f_{\rm M}$ to constrain the mass-ratio distribution, we essentially treated all 115 observed systems in our subsample as PB1s. This was necessary because one cannot assess {\em a priori} whether a binary with a certain combination of physical and orbital parameters will manifest itself as a PB1 or a PB2, in part because not all stars in the $\delta$\,Sct instability strip pulsate. Figure~\ref{fig:fm_clean} shows the observed distribution of $\log f_{\rm M}$ for our 115 systems divided into 15 intervals of width $\Delta$\,log\,$f_{\rm M}$\,(M$_{\odot}$)~=~0.25 across $-$3.75~$<$~log\,$f_{\rm M}$\,(M$_{\odot}$)~$<$~0.00. We adopted Poisson errors, but for the two bins with no observed members, we set the uncertainties to unity. 

We utilized a random walk Metropolis--Hastings MCMC algorithm and the probabilities $p \propto$ exp($-\chi^2$/2) of the models to explore the parameter space of our six-bin mass-ratio distribution. Figure~\ref{fig:fm_clean} compares our best-fitting model ($\chi^2$ = 3.2) to the observed distribution of $\log f_{\rm M}$. Given 15 bins of $\log f_{\rm M}$ and 6 fitted parameters, there are $\nu$ = 15\,$-$\,6 = 9 degrees of freedom, which results in a rather small $\chi^2$/$\nu$ = 0.4. However, many of the bins in the observed distribution of $\log f_{\rm M}$ have only one or even zero members, and so the {\it effective} number of degrees of freedom is considerably smaller.  Counting only the bins with $>$5 elements, there are $\nu_{\rm eff}$ = 3 effective degrees of freedom, which gives a more believable $\chi^2$/$\nu_{\rm eff}$ = 1.1.

In Fig.~\ref{fig:qdist_clean}, we show the average values of our six fitted parameters based on this MCMC technique (blue data points). For each mass-ratio bin, the standard deviation of the values obtained in our Markov chain provides the 1$\sigma$ measurement uncertainty. For the three bins that span $0.0 < q < 0.4$, however, the dominant source of error came from the systematic uncertainties in the detection efficiencies. We therefore repeated our MCMC routine, shifting the detection efficiency of each synthesized binary by its systematic uncertainty. In this manner, we measured the systematic uncertainties for each of the six parameters. For the six blue data points displayed in Fig.~\ref{fig:qdist_clean}, the error bars represent the quadrature sum of the random and systematic uncertainties. The total number of binaries was 178 $\pm$\,19\,(stat.) $\pm$\,23\,(sys.).

\subsubsection{MCMC technique with segmented power-law mass-ratio distribution} 
\label{sssec:MCMC_power}

It is common to fit the mass-ratio distribution as a segmented power-law \citep[e.g.][]{moe&distefano2017}, so we conducted a final MCMC forward-modelling routine incorporating a different Bayesian prior. We adopted a segmented power-law probability distribution p $\propto q^{\gamma}$ with three parameters: the power-law slope $\gamma_{\rm smallq}$ across small mass ratios $q = 0.1$\,--\,0.3, the power-law slope $\gamma_{\rm largeq}$ across large mass ratios $q = 0.3$\,--\,1.0, and the total number of binaries with $q>0.1$. Figure\,\ref{fig:fm_clean} compares our best-fitting model ($\chi^2 = 4.1$) of the posterior distribution of binary mass functions to the observed distribution. While the $\chi^2 = 4.1$ statistic for the segmented power-law mass-ratio distribution is slightly larger than the $\chi^2 = 3.2$ statistic for the binned mass-ratio distribution, we were fitting only three parameters instead of six. Hence, here there are $\nu_{\rm eff}$~=~$9-3$~=~6 effective degrees of freedom, and so the reduced  $\chi^2$/$\nu_{\rm eff}$~=~0.7 statistic is actually smaller for our segmented power-law mass-ratio distribution. 

Accounting for both measurement and systematic uncertainties in our MCMC algorithm, we obtained \mbox{$\gamma_{\rm smallq} = -0.2\pm0.4$}, \mbox{$\gamma_{\rm largeq}=-1.1 \pm 0.3$}, and 174 $\pm$\,16\,(stat.) $\pm$\,18\,(sys.) total binaries with $q>0.1$.  Our best-fitting segmented power-law distribution is shown as the red line in Fig.\,\ref{fig:qdist_clean}. Clearly, all three methods that account for completeness are in good agreement.

\subsubsection{Interpretation of the mass-ratio distribution}
\label{sssec:interp}

The intrinsic mass-ratio distribution is weighted significantly toward $q=0.3$, peaking at $q=0.2$ and turning over rapidly below $q=0.1$. All three methods yield a consistent total number of binaries. By taking an average, we adopt 179\,$\pm$\,28 total binaries, 173\,$\pm$\,24 of which have $q$ $>$ 0.1.

Using our step-function MCMC method (Sect.\,\ref{sssec:MCMC_step}), we can investigate the statistical significance of this turnover.  There are $7.2\pm2.0$ times more binaries with $q=0.10$--0.25 than binaries with $q<0.1$, i.e., we are confident in the turnover at the 3.6$\sigma$ significance level. We emphasize that the observational methods are sensitive to companions with $q<0.1$, and so the turnover at extreme mass ratios is intrinsic to the real population. To illustrate, in our `clean' sample of 115 systems, 30 binaries have detection efficiencies with $D=0.4$--0.6. These 30 binaries span binary mass functions $-$3.5~$<$~log\,$f_{\rm M}$\,(M$_{\odot}$)~$<$~$-1.1$. Meanwhile, we detect only 4 binaries with $D$~=~0.25\,--\,0.40, which have $-$3.0~$<$~log\,$f_{\rm M}$\,(M$_{\odot}$)~$<$~$-2.1$. Even after accounting for the slightly smaller detection efficiencies toward smaller mass ratios, we would expect several times more systems with $D$~=~0.25\,--\,0.40 and log\,$f_{\rm M}$\,(M$_{\odot}$)~$<$~$-$2.5 if binaries with $q$~$<$~0.1 were as plentiful as systems with $q$~=~0.1\,--\,0.2. We conclude that for intermediate-mass primaries with $M_1$~$\approx$~1.8\,M$_{\odot}$, there is a real deficit of extreme-mass-ratio companions at $q<0.1$ across intermediate orbital periods log\,$P$\,(d)~$\approx$~2\,--\,3.

It has been known for more than a decade that solar-type primaries with $M_1 \approx 1.0$\,M$_{\odot}$ exhibit a near-complete absence of extreme-mass-ratio companions with $q\approx0.02$\,--\,0.09 at short and intermediate orbital periods $P < 2000$\,d (\citealt{grether&lineweaver2006}; references therein). This dearth of closely orbiting extreme-mass-ratio companions is commonly known as the brown dwarf desert. It is believed that, if such brown dwarf companions existed in the primordial disc of T~Tauri stars, they would have either accreted additional mass to form a stellar companion or migrated inward and merged with the primary \citep{armitage&bonnell2002}. In contrast, for more massive early-B main-sequence primaries with $M_1 > 8$\,M$_{\odot}$, closely orbiting companions with extreme mass ratios $q=0.06$\,--\,0.10 have been detected on the pre-main-sequence via eclipses \citep{moe&distefano2015a}. There is no indication of a turnover across extreme mass ratios, at least down to $q \approx 0.06$.  \citet{moe&distefano2015a} argued that extreme-mass-ratio companions to more massive $M_1$~$>$~8\,M$_{\odot}$\ primaries can more readily survive at close separations without merging due to their larger orbital angular momenta and more rapid disc photoevaporation timescales. 

Based on our MCMC model, the true number of binaries with $q<0.1$ in our `clean' subsample is $7.5\pm3.6$, which differs from zero at the 2.1$\sigma$ level. Either we underestimated the systematic uncertainties, or $\delta$\,Sct stars have at least some companions at extreme mass ratios ($q<0.1$). Main-sequence A primaries with $M_1$ $\approx$ 1.8\,M$_{\odot}$\ may therefore represent the transition mass where extreme mass ratio binaries with $q<0.1$ can just begin to survive at intermediate separations during the binary star formation process. There is certainly a significant deficit of extreme-mass-ratio (brown dwarf) companions to main-sequence A stars at intermediate separations, but it may not necessarily be a complete absence as is observed for solar-type systems.

\subsubsection{Comparison with non-A-type primaries and other surveys} 
\label{ssec:lit_comp}

Based on a meta-analysis of dozens of binary star surveys, \citet{moe&distefano2017} compared the power-law slopes $\gamma_{\rm smallq}$ and $\gamma_{\rm largeq}$ of the mass-ratio distribution as a function of $M_1$ and $P$. Across intermediate periods $\log P$~(d)~$\approx$~2\,--\,4, they measured \mbox{$\gamma_{\rm largeq} = -0.5 \pm 0.4$} for solar-type main-sequence primaries with \mbox{$M_1 \approx 1$\,M$_{\odot}$}, and \mbox{$-2.3 < \gamma_{\rm largeq} < -1.4$} for mid-B through O main-sequence primaries with $M_1 > 5$\,M$_{\odot}$, depending on the survey. Our measurement of \mbox{$\gamma_{\rm largeq} = -1.3 \pm 0.4$} for $M_1 = 1.8$\,M$_{\odot}$ main-sequence A primaries is between the solar-type and OB values.  This demonstrates that the mass-ratio distribution of binaries with intermediate orbital periods gradually becomes weighted toward smaller mass ratios $q \approx 0.3$ with increasing primary mass $M_1$.

While the measurements of $\gamma_{\rm largeq}$ reported by \citet{moe&distefano2017} are robust, their estimates of the power-law slope $\gamma_{\rm smallq}$ across small mass ratios $q = 0.1$\,--\,0.3 and intermediate periods $\log P$\,(d) = 2\,--\,4 suffer from large selection biases. Specifically, the SB1 samples investigated by \citet{moe&distefano2017} are contaminated by binaries with white-dwarf companions. Although they accounted for this in their analysis, the uncertainty in the frequency of compact remnant companions leads to large systematic uncertainties in the intrinsic power-law slope $\gamma_{\rm smallq}$ for stellar companions. For example, by incorporating the \citet{raghavanetal2010} survey of solar-type binaries and estimating the fraction of solar-type stars with white-dwarf companions, \citet{moe&distefano2017} measured \mbox{$\gamma_{\rm smallq} = -0.1\pm0.7$} across intermediate periods \mbox{$\log P$\,(d) = 2\,--\,4}.  Similarly, they reported \mbox{$\gamma_{\rm smallq} = 0.2 \pm 0.9$} across \mbox{$\log P$\,(d) = 2.6\,--\,3.6} based on a sample of SB1 companions to Cepheids that evolved from mid-B $M_1 \approx 5$\,--\,9\,M$_{\odot}$ primaries \citep{evansetal2015}. These SB1 samples indicate that the power-law slope \mbox{$\gamma_{\rm smallq} \approx 0.0$} is close to flat, but the uncertainties \mbox{$\delta \gamma_{\rm smallq} \approx 0.7$\,--\,0.9} are quite large due to contamination by white-dwarf companions.  

\citet{gulliksonetal2016} obtained extremely high SNR spectra of A/B primaries and directly detected the spectroscopic signatures of unresolved extreme-mass-ratio companions down to $q \approx 0.1$.  They recovered an intrinsic mass-ratio distribution that is narrowly peaked near $q \approx 0.3$.  Our mass-ratio distribution is also skewed toward $q = 0.3$, but it rises towards $q = 0.2$ and doesn't turnover until \mbox{$q \lesssim 0.10$\,--\,0.15} (see Fig.\,\ref{fig:qdist_clean}).  The inferred completeness rates based on the direct spectral detection technique developed by \citet{gulliksonetal2016} depended critically on the assumed degree of correlation in the rotation rates of the binary components. Accounting for this uncertainty, \citet{moe&distefano2017} measured \mbox{$\gamma_{\rm smallq} = 0.7\pm0.8$} across \mbox{$\log P$\,(d) = 1.3\,--\,4.9} based on the \citet{gulliksonetal2016} survey. Once again, the uncertainties are rather large.

In summary, for binaries with intermediate periods log\,$P$\,(d)~$\approx$~2\,--\,4, it is currently impossible to measure $\gamma_{\rm smallq}$ for massive stars (see also figure\:1 of \citealt{moe&distefano2017} and our Sect.\,\ref{sec:intro}), and all previous attempts to measure $\gamma_{\rm smallq}$ for solar-type and intermediate-mass stars have been plagued by large systematic uncertainties. Our analysis of companions to $\delta$\,Sct stars provides the first direct and reliable measurement of the intrinsic binary mass-ratio distribution across intermediate periods. By focusing on binaries with large eccentricities, we are confident that our subsample of 115 systems contains only binaries with stellar main-sequence companions. The uncertainty in our measurement of \mbox{$\gamma_{\rm smallq} = -0.2\pm0.4$} is a factor of two smaller than all of the previous measurements. For A-star primaries at intermediate periods, we find the intrinsic mass-ratio distribution is significantly skewed toward small mass ratios $q \approx 0.3$ \mbox{($\gamma_{\rm largeq} = -1.1 \pm 0.3$)}, peaks near $q \approx 0.2$ \mbox{($\gamma_{\rm smallq} = -0.2 \pm 0.4$)}, and then rapidly turns over below \mbox{$q \lesssim 0.10$\,--\,0.15}.


\subsection{The mass-ratio distribution for the mixed sample}

Here we describe the mass-ratio distribution of the 130 observed systems (126 PB1s and 4 PB2s) that have orbital periods $P$~=~200\,--\,1500\,d, detection efficiencies $D$~$>$~0.36, and eccentricities below our adopted $e$ vs. log\,$P$ relation.  Unlike the previous large-eccentricity `clean' subsample, some of the companions in this `mixed' subsample are expected to be white dwarfs that tidally decayed towards smaller eccentricities during the previous mass-transfer phase.  We utilized similar methods and procedures as applied in Sect.\,\ref{ssec:ms_pop}.

\subsubsection{Application of the three methods}

\begin{figure}
\centering
\includegraphics[width=0.5\textwidth]{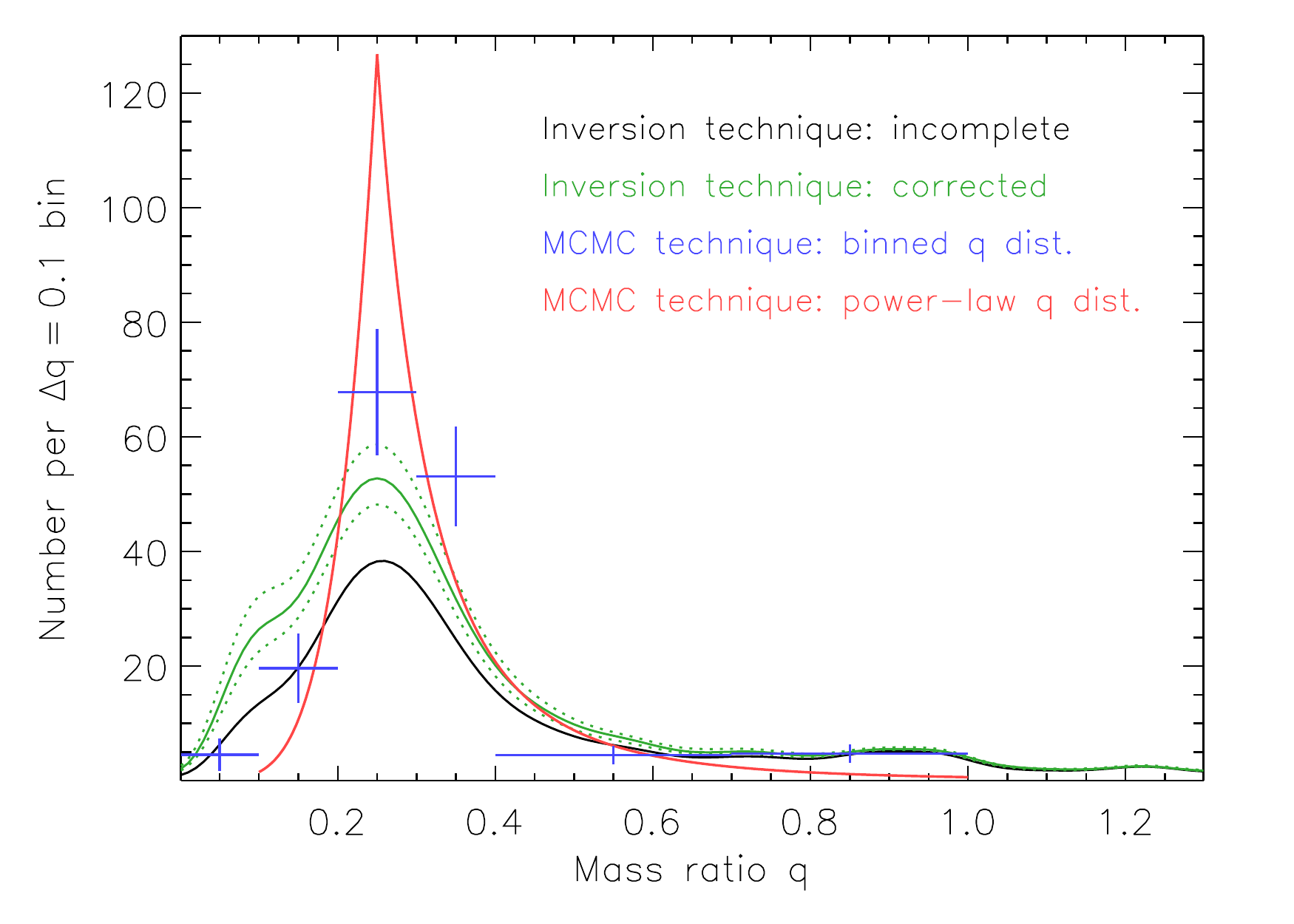}
\caption{Similar to Fig.\,\ref{fig:qdist_clean}, but for the long-period, small-eccentricity subsample of 130 observed binaries (126 PB1s and 4 PB2s).  Unlike the previous subsample that contains only main-sequence companions, this subsample includes white-dwarf companions that tidally decayed towards smaller eccentricities during the previous mass transfer phase.  After accounting for incompleteness and the distributions of binary inclinations, we reconstructed a mass-ratio distribution that narrowly peaks across $q$~=~0.2\,--\,0.4, i.e., $M_2$~=~0.3\,--\,0.7\,M$_{\odot}$ given $\langle M_1 \rangle$~=~1.7\,M$_{\odot}$.  Approximately 40\% of this subsample are white-dwarf companions with $M_2$~$\approx$~0.5\,M$_{\odot}$.}
\label{fig:qdist_contaminated}
\end{figure}

We reconstructed the mass-ratio distribution using our inversion technique and found it to peak narrowly at $q$~=~0.25 (Fig.~\ref{fig:qdist_contaminated}). It yielded a total of 177 $\pm$~16~(stat.) $\pm$~16~(sys.) binaries after accounting for incompleteness, which is only moderately larger than the observed number of 130 systems. This is because the long-period, small-eccentricity subsample contains more PB1s with $P$~=~500\,--\,1500\,d that have systematically larger detection efficiencies.

\begin{figure}
\centering
\includegraphics[width=0.5\textwidth]{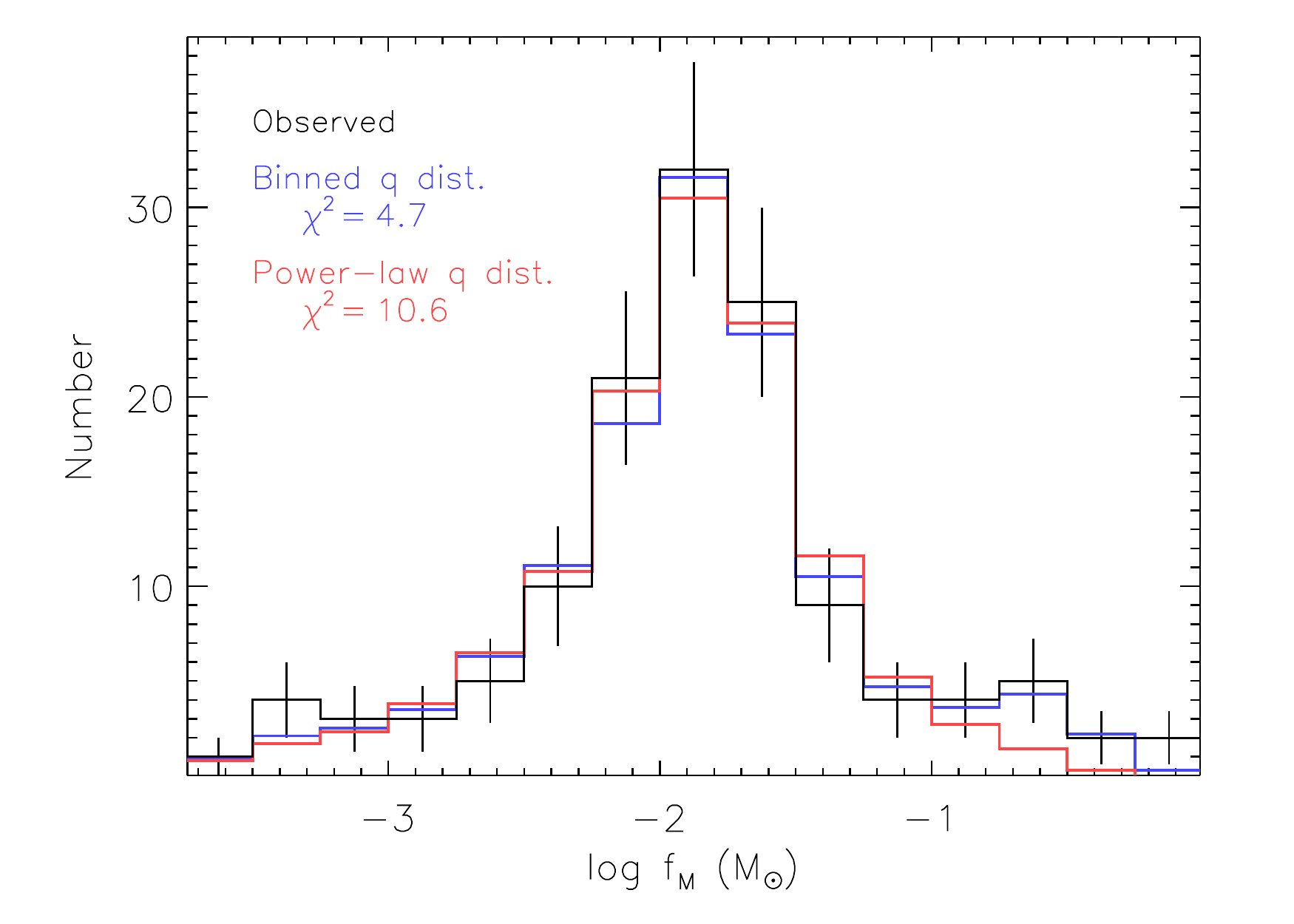}
\caption{Similar to Fig.\,\ref{fig:fm_clean}, but for the long-period, small-eccentricity subsample of 130 observed binaries.  While the previous `clean' main-sequence subsample exhibits a broad binary mass-function distribution, approximately 40\% of this subsample are white-dwarf companions that contribute a narrow peak at log\,$f_{\rm M}$\,(M$_{\odot}$)~=~$-$1.9.}
\label{fig:fm_contaminated}
\end{figure}

For our MCMC Bayesian models, we selected primary masses $M_1$ and periods $P$ from the observed distributions.  For the long-period, small-eccentricity subsample, the distribution of $M_1$ is still approximately Gaussian with a similar dispersion $\sigma$ = 0.3\,M$_{\odot}$, now with $\langle M_1 \rangle = 1.7$\,M$_{\odot}$.  The distribution of logarithmic orbital periods, however, is skewed toward larger values, linearly rising from log\,$P$\,(d) = 2.35 to log\,$P$\,(d) = 3.18. We show in Fig.~\ref{fig:fm_contaminated} the observed distribution of binary mass functions, which narrowly peaks at log\,$f_{\rm M}$\,(M$_{\odot}$)~=~$-$1.9.  By minimizing the $\chi^2$ statistic between the simulated and observed distributions of binary mass function $f_{\rm M}$, we measured the intrinsic mass-ratio distribution $f_q$ of our small-eccentricity subsample.

Using our Bayesian model with the six-step mass-ratio distribution as defined in Sect.\,\ref{sssec:MCMC_step}, we could not satisfactorily fit the observed distribution of binary mass functions.  Specifically, the two steps across $q$~=~0.1\,--\,0.4 were too wide to reproduce the narrow peak in the observed $f_{\rm M}$ distribution.  For this subsample, we therefore adopted a different mass-ratio prior distribution, still with six total steps, but with finer $\delta q$~=~0.1 spacings across $q$~=~0.0\,--\,0.4 (four steps), and coarser $\delta q$~=~0.3 spacings across $q$~=~0.4\,--\,1.0 (two steps).  With this prior, we satisfactorily fitted the binary mass function distribution ($\chi^2$ = 4.7; see Fig.~\ref{fig:fm_contaminated}).  This fit was dominated by the 12 bins across $-$3.5~$<$~log\,$f_{\rm M}$\,(M$_{\odot}$)~$<$~$-$0.5, providing $\nu_{\rm eff}$~=~12~bins $-$ 6~parameters = 6~effective degrees of freedom ($\chi^2$/$\nu_{\rm eff}$~=~0.8).  The resulting multi-step mass-ratio distribution is shown in Fig.~\ref{fig:qdist_contaminated}. After accounting for incompleteness, we estimated a total of 173 $\pm$~15~(stat.) $\pm$~19~(sys.) binaries, of which $\sim$169 have $q$~$>$~0.1.

Our Bayesian model with the segmented power-law mass-ratio distribution was again described by three parameters but with a different boundary: the power-law slope $\gamma_{\rm smallq}$ across small mass ratios 0.1~$<$~$q$~$<$~$q_{\rm trans}$, the power-law slope $\gamma_{\rm largeq}$ across large mass ratios $q_{\rm trans}$~$<$~$q$~$<$~1.0, and the total number of binaries with $q$~$>$~0.1.  By setting the transition mass ratio to $q_{\rm trans}$~=~0.3 as we did in Sect.\,\ref{sssec:MCMC_power}, we could not fit the observed $f_{\rm M}$ distribution of the `mixed' subsample.  We instead used $q_{\rm trans}$~=~0.25, which resulted in a good fit ($\chi^2$~=~10.6; $\chi^2$/$\nu_{\rm eff}$~=~1.2 given $\nu_{\rm eff}$ = 12~bins $-$ 3~parameters = 9~effective degrees of freedom; see Fig.~\ref{fig:fm_contaminated}), with best-fit parameters of $\gamma_{\rm smallq}$~=~4.9\,$\pm$\,0.8, $\gamma_{\rm largeq}$~=~$-$4.0\,$\pm$\,0.5, and 166 $\pm$~14~(stat.) $\pm$~20~(sys.) total binaries with $q$~$>$~0.1.  This power-law mass-ratio distribution is also shown in Fig.~\ref{fig:qdist_contaminated}.

\subsubsection{Interpretation of the mass-ratio distribution}

All three techniques yielded a mass-ratio distribution that narrowly peaks across $q$~=~0.2\,--\,0.4, but there are some minor differences.  With the step-function model, we found slightly fewer systems across $q$~=~0.1\,--\,0.2 and more systems across $q$~=~0.2\,--\,0.4 when compared with the inversion method.  Nevertheless, across all mass ratios, the differences between these two methods were smaller than the 2$\sigma$ uncertainties of the step-function model. The total number of binaries was consistent across all three methods, having an average 174\,$\pm$\,23 binaries, of which 169\,$\pm$\,21 have $q>0.1$.

White dwarfs have minimum masses $M_{\rm WD} > 0.17$\,Msun \citep{kilicetal2007a}, and so extreme-mass-ratio companions with $q < 0.1$ in our `mixed' subsample must be unevolved stars. We found $4.5\pm2.9$ such systems, which we added to the $7.5\pm3.6$ systems from the `clean' subsample to give a total of $12.0\pm4.6$ systems with $q<0.1$. This strengthens our conclusion that A/F primaries have a dearth, but not a complete absence (non-zero at 2.6$\sigma$), of extreme-mass-ratio companions (see \ref{sssec:interp}).


\subsection{Comparison of the two subsamples}
\label{sec:comp}

The `mixed' subsample contains 169\,$\pm$\,21 binaries with $q$~$>$~0.1, compared with 173\,$\pm$\,24 binaries with $q$~$>$~0.1 in the `clean' subsample. Hence, among our full sample of 2224 $\delta$\,Sct stars, 342\,$\pm$\,32 stars (15.4\%\,$\pm$\,1.4\%) have companions with $q$~$>$~0.1 and periods $P$~=~100\,--\,1500\,d, after accounting for incompleteness.

\begin{figure}
\centering
\includegraphics[width=0.5\textwidth]{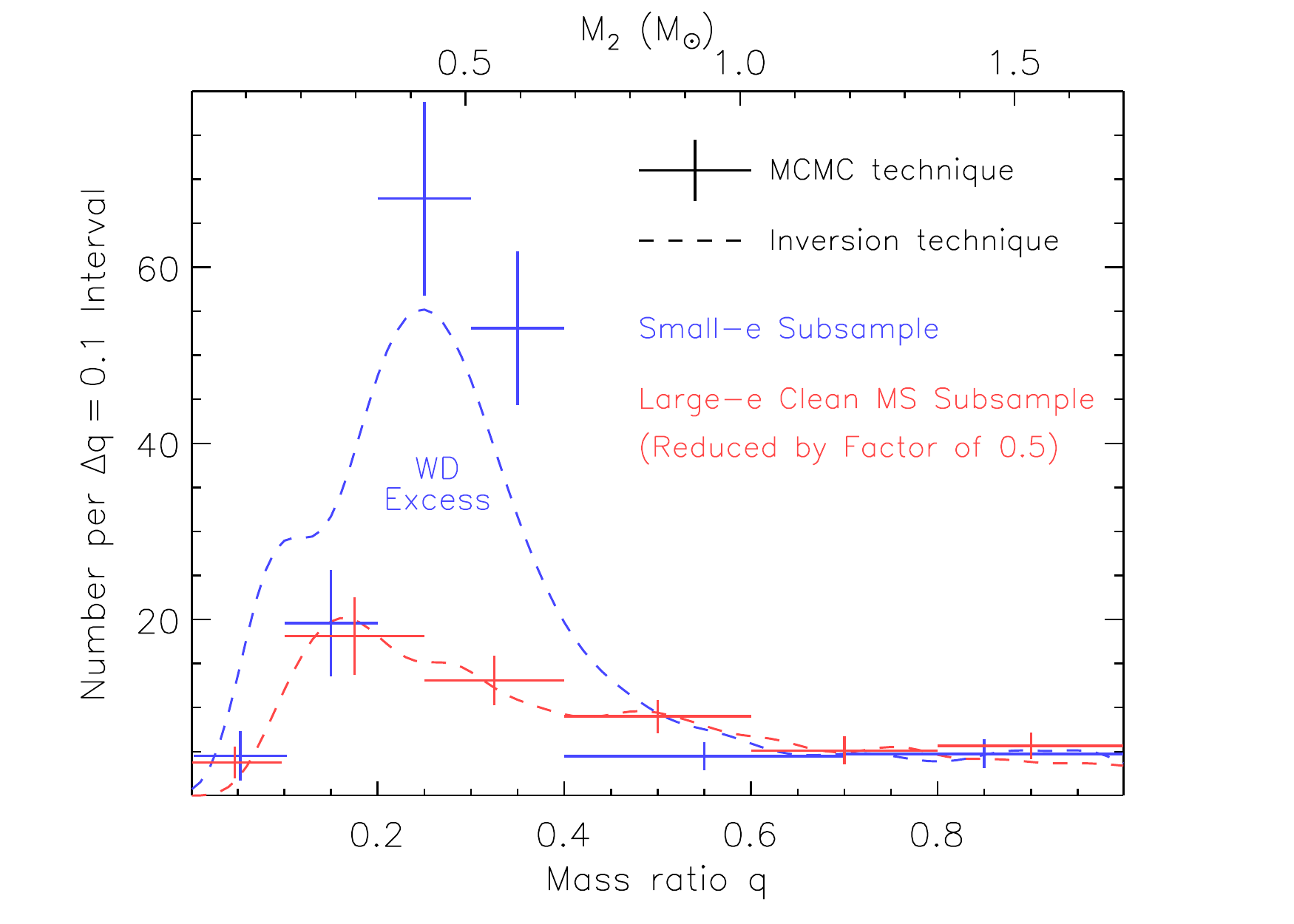}
\caption{Corrected mass-ratio distributions of our long-period, small-eccentricity subsample (blue) and our short-period, large-eccentricity `clean' main-sequence subsample (red) determined by our MCMC Bayesian forward-modelling technique (data points) and population inversion technique (dashed lines).  By scaling the corrected `clean' main-sequence mass-ratio distribution down by a factor of 0.5, both tails ($q$~$<$~0.2 and $q$~$>$~0.4) of the two distributions are consistent with each other.  In our small-eccentricity subsample, we measure an excess of 73\,$\pm$\,18 white-dwarf companions with periods $P$~=~200\,--\,1500\,d and mass ratios $q$~$\approx$~0.2\,--\,0.4 ($M_2$~$\approx$~0.3\,--\,0.7\,M$_{\odot}$ given $\langle M_1 \rangle$~=~1.7\,M$_{\odot}$). }
\label{qcomp}
\end{figure}

However, not all of these binaries contain main-sequence companions. The number of white dwarfs can be estimated by comparing the corrected mass-ratio distributions of our two subsamples. 
Although the existence of a binary companion generally truncates the growth of the white dwarf slightly \citep{pyrzasetal2009}, the majority of white dwarfs have masses between 0.3 and 0.7\,M$_{\odot}$, whether they are in close binaries \citep{rebassa-mansergasetal2012} or not \citep{tremblayetal2016}.
No significant white-dwarf contribution should therefore occur at binary mass ratios $q$~$\lesssim$~0.2 ($M_2$~$\lesssim$~0.3\,M$_{\odot}$) or $q$~$\gtrsim$~0.4 ($M_2$~$\gtrsim$~0.7\,M$_{\odot}$).  In Fig.~\ref{qcomp}, we have scaled the mass-ratio distribution of the clean main-sequence sample by a factor of 0.5, so that the small-$q$ and high-$q$ tails of the two distributions are mutually consistent.  By subtracting the two mass-ratio distributions inferred from our population inversion technique, we found an excess of $\sim$\,60 white-dwarf companions across $q$~$\approx$~0.2\,--\,0.4 in the `mixed' subsample.  Similarly, using the multi-step mass-ratio distributions we found an excess of $\sim$\,85 white-dwarf companions. Considering both statistical and systematic uncertainties, our full sample therefore contains 73\,$\pm$\,18 white-dwarf companions with periods $P$~=~200\,--\,1500\,d and masses $M_2$~$\approx$~0.3\,--\,0.7\,M$_{\odot}$.

Based on various lines of observational evidence, \citet{moe&distefano2017} estimated that 30\%\,$\pm$\,10\% of single-lined spectroscopic binaries (SB1s) in the field with B-type or solar-type primaries contain white-dwarf companions.  About 30\,--\,80\% of spectroscopic binaries appear as SB1s, depending on the sensitivity of the observations, implying 10\,--\,25\% of spectroscopic binaries in the field have white-dwarf companions. These estimates are significantly model dependent.  For example, \citet{moe&distefano2017} incorporated the observed frequency of hot white-dwarf companions that exhibit a UV excess \citep{holbergetal2013} and models for white-dwarf cooling times to determine the total frequency of close white-dwarf companions to main-sequence AFGK stars.  Similarly, they utilized the observed frequency of barium stars \citep[1\%,][]{macconnelletal1972,jorissenetal1998} and a population synthesis method to estimate the fraction of all \mbox{$M_1$~$\approx$~1\,--\,2\,M$_{\odot}$} main-sequence stars harbouring white-dwarf companions (see also below). We {\it measure} (73\,$\pm$\,18)/(342\,$\pm$\,32) = 21\%\,$\pm$\,6\% of binaries with periods $P$~=~200\,--\,1500\,d and main-sequence A/F primaries actually host white-dwarf secondaries.  Our measurement is consistent with the estimates by \citet{moe&distefano2017}, but {\em is not model dependent} and has substantially smaller uncertainties. 

We also measure (73\,$\pm$\,18)/2224 = 3.3\%\,$\pm$\,0.8\% of main-sequence A/F $\delta$\,Sct stars have white-dwarf companions across $P$~=~200\,--\,1500\,d.  Meanwhile, only $\sim$\,1.0\% of GK giants appear as barium stars with white-dwarf companions across a slightly broader range of orbital periods $P$~=~200\,--\,5000\,d \citep{macconnelletal1972,boffin&jorissen1988,jorissenetal1998,karakasetal2000}.  According to the observed period distribution of barium stars, we estimate that $\sim$\,0.7\% of GK giants are barium stars with white-dwarf companions across $P$~=~200\,--\,1500\,d.  Hence, roughly a fifth (0.7\%\,/\,3.3\%\,=\,21\%) of main-sequence A/F stars with white-dwarf companions across $P$~=~200\,--\,1500\,d will eventually evolve into barium GK giants. The measured difference is because not all main-sequence A/F stars with white-dwarf companions at $P$~=~200\,--\,1500\,d experienced an episode of significant mass transfer involving thermally pulsing, chemically enriched AGB donors.  Instead, some of them will have experienced mass transfer when the donor was less evolved and had only negligible amounts of barium in their atmospheres.  The donors could have been early-AGB, RGB, or possibly even Hertzsprung Gap stars if the binary orbits were initially eccentric enough or could sufficiently widen to $P$~$>$~200\,d during the mass transfer process.  In other cases, mass transfer involving AGB donors may have been relatively inefficient and non-conservative (especially via wind accretion), and so the main-sequence accretors may not have gained enough mass to pollute their atmospheres (see mass transfer models by \citealt{karakasetal2000}).  In any case, only a fifth of main-sequence A/F stars with white-dwarf companions across $P$~=~200\,--\,1500\,d become chemically enriched with enough barium to eventually appear as barium GK giants. This conclusion is in agreement with the study by \citet{vanderswaelmenetal2017}, who directly observed that 22\% (i.e. a fifth) of binaries with giant primaries and intermediate periods have WD companions.  This measurement provides powerful insight and diagnostics into the efficiency and nature of binary mass transfer involving thermally-pulsing AGB donors.

Our determination that 3.3\%\,$\pm$\,0.8\% of main-sequence A/F stars have white-dwarf companions across $P$~=~200\,--\,1500\,d also provides a very stringent constraint for binary population synthesis studies of Type Ia supernovae (SN~Ia).  In both the symbiotic single-degenerate scenario \citep{patatetal2011,chenetal2011} and the double-degenerate scenario \citep{iben&tutukov1984,webbink1984}, the progenitors of SN~Ia were main-sequence plus white-dwarf binaries with periods $P$~$\approx$~100\,--\,1000\,d at some point in their evolution.  Granted, the majority of our observed binaries with $\langle M_1 \rangle = 1.7$\,M$_{\odot}$ and $M_{\rm WD}$~=~0.3\,--\,0.7\,M$_{\odot}$ have masses too small to become SN~Ia.  Nevertheless, several channels of SN~Ia derive from immediately neighbouring and partially overlapping regions in the parameter space.  For instance, in the symbiotic SN~Ia channel, $M_1$~$\approx$~1\,--\,2\,M$_{\odot}$ stars evolve into giants that transfer material via winds and/or stable Roche-lobe overflow to $M_{\rm WD}$ = 0.7\,--\,1.1\,M$_{\odot}$ carbon-oxygen white dwarfs with periods $P$~$\approx$~100\,--\,1000\,d \citep{chenetal2011}.  Similarly, in the double-degenerate scenario, slightly more massive giant donors $M_1$~$\approx$~2\,-\,4\,M$_{\odot}$ overfill their Roche lobes with white-dwarf companions across $P$~=~100\,--\,1000\,d, resulting in unstable common envelope evolution that leaves pairs of white dwarfs with very short periods $P$~$\lesssim$~1\,d \citep{ruiteretal2009,mennekensetal2010,claeysetal2014}.  The cited binary population synthesis models implement prescriptions for binary evolution that are not well constrained, and so the predicted SN~Ia rates are highly uncertain.  By anchoring binary population synthesis models to our measurement for the frequency of white-dwarf companions to intermediate-mass stars across intermediate periods, the uncertainties in the predicted rates of both single-degenerate and double-degenerate SN~Ia can be significantly reduced.  Related phenomena, such as blue stragglers, symbiotics, R~CrB stars and barium stars will benefit similarly.


\subsection{The binary fraction of A/F stars at intermediate periods, compared to other spectral types}

We now calculate the fraction of original A/F primaries that have main-sequence companions across $P$~=~100\,--\,1500\,d. We must remove the systems with white-dwarf companions, i.e. those where the A/F star was not the original primary but in many cases was an F/G-type secondary that accreted mass from a donor. In Sect.\,\ref{sec:comp} we calculated the fraction of current A stars that have any companions across $P$~=~100\,--\,1500\,d as $F_{\rm total} = (i+j)/(X+Y) = 15.4$\%, where $i+j = 342$ is the corrected total number of companions across $P$~=~100\,--\,1500\,d and $X + Y = 2224$ is the total number of A/F stars in our sample. To find the fraction of original A/F primaries, \mbox{$F_{\rm orig.} = i/X$,} we must remove the $j$ detected white-dwarf companions across $P$~=~100\,--\,1500\,d, and the $Y$ targets in our sample that have white-dwarf companions at any period, including those with $P<100$\,d or $P>1500$\,d that are undetected by our method.

We first remove the measured number of $j = 73$\,$\pm$\,18 white dwarfs across $P$~=~200\,--\,1500\,d, leaving $i$ = (342\,$\pm$\,32) - (73\,$\pm$\,18) = 269\,$\pm$\,37 systems with A/F main-sequence $\delta$\,Sct primaries and main-sequence companions with $q > 0.1$ and $P$~=~100\,--\,1500\,d. To estimate the number of white-dwarf companions, $Y$, in the total sample, we turn to other surveys. For solar-type main-sequence primaries in the field, \citet{moe&distefano2017} estimated that 11\%\,$\pm$\,4\% have white-dwarf companions.  Similarly, they determined that 20\,\%\,$\pm$\,10\% of $M_1$~$\approx$~10\,M$_{\odot}$ main-sequence B ``primaries'' in a volume-limited sample are actually the secondaries in which the true primaries have already evolved into compact remnants. For A/F stars we interpolated between these two estimates, and inferred that 13\%\,$\pm$\,5\% of our 2224 $\delta$\,Sct field stars with $M_1$~=~1.8\,M$_{\odot}$ have white-dwarf companions.  This corresponds to $Y = 290$\,$\pm$\,110 total systems with white-dwarf companions, which is 4.0\,$\pm$\,1.8 times more than that directly measured across $P$~=~200\,--\,1500\,d, leaving $X$ = 2224\,(0.87\,$\pm$\,0.05) = 1934\,$\pm$\,110 main-sequence $\delta$\,Sct stars without white-dwarf companions. The binary fraction of original A/F primaries with main-sequence companions is therefore $F_{\rm orig.}$ = $i/X$ = (269\,$\pm$\,37)\,/\,(1934\,$\pm$\,110) = 13.9\%\,$\pm$\,2.1\% at $q$~$>$~0.1 and $P$~=~100\,--\,1500\,d. 

After accounting for selection biases as we did above, the binary fraction of solar-type ($M_1$~=~0.8\,--\,1.2\,M$_{\odot}$) main-sequence primaries is 6.3\%\,$\pm$\,1.6\% across the same interval of mass ratios $q$~$=$~0.1\,--\,1.0 and periods $P$~=~100\,--\,1500\,d (average of the \citealt{raghavanetal2010} and \citealt{moe&distefano2017} results). Our measurement of 13.9\%\,$\pm$\,2.1\% for main-sequence A/F primaries is significantly higher (2.9$\sigma$).  Extending toward smaller primary masses, the frequency of companions to early M-dwarfs ($M_1$~$\approx$~0.3\,--\,0.5\,M$_{\odot}$) with $P$~=~100\,--\,1500\,d is 0.05\,$\pm$\,0.02 \citep[][see their figure 2b]{fischer&marcy1992}. For lower-mass M-dwarfs ($M_1$~$\approx$~0.1\,--\,0.3\,M$_{\odot}$), the corrected cumulative binary fraction is $\sim$\,4\% for $a$~$<$~0.4~au, $\sim$\,9\% for $a$~$<$~1\,au, and $\sim$\,11\% for $a$~$<$~6\,au \citep{clarketal2012,guenther&wuchterl2003,joergens2006,basri&reiners2006}, giving a binary fraction of 6\,$\pm$\,3\% across $P$~=~100\,--1500\,d. \citet{joergens2008} showed that 10$^{+18}_{-8}$\% of very low-mass stars and brown dwarfs ($M_1$~$\approx$~0.06\,--\,0.10\,M$_{\odot}$) have companions with $a$~$<$~3~au, $\sim$70\% of which have separations $a$~$<$~0.3~au.  Based on these observations, we adopt a binary fraction of 3\%\,$\pm$\,2\% across $P$ = 100\,--\,1500\,d for low-mass stars and brown dwarfs.  The uncertainties in the binary fractions of M-dwarfs and brown dwarfs are dominated by the large statistical errors due to the small sample sizes, and so systematic uncertainties from correcting for incompleteness and the presence of white-dwarf companions are negligible.

\begin{figure}
\centering
\includegraphics[width=0.5\textwidth]{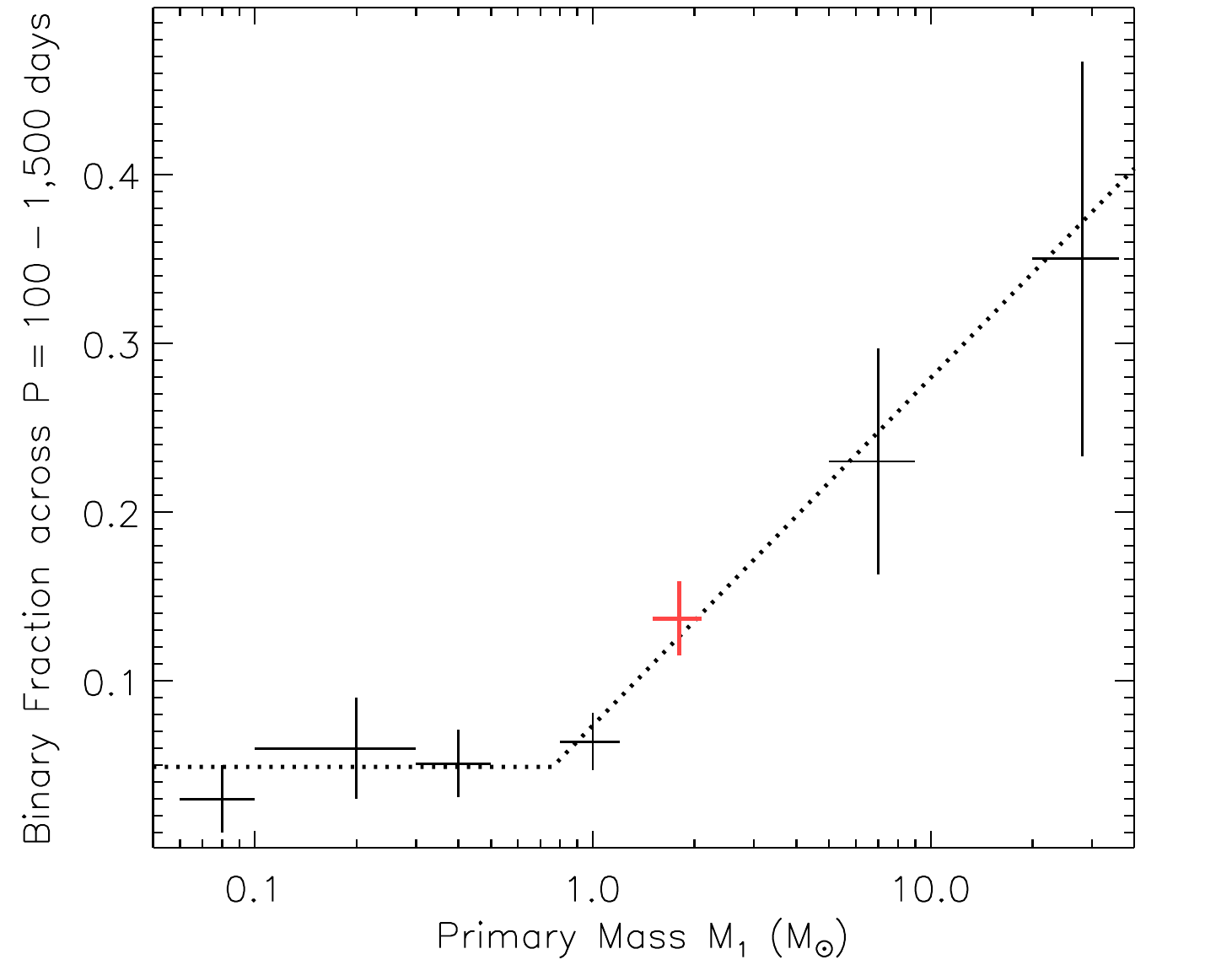}
\caption{Corrected frequency of companions with $q$~$>$~0.1 and $P$~=~100\,--\,1500\,d as a function of primary mass $M_1$.  The binary fraction across intermediate periods is relatively constant at $\sim$5\% for $M_1$~$<$~0.8\,M$_{\odot}$ and then increases linearly with respect to $\log M_1$ above $M_1$~$>$~0.8\,M$_{\odot}$ (dotted line). This is consistent with analytic and hydrodynamical models of the formation of binary stars with intermediate periods (see the main text).}
\label{binfrac}
\end{figure}

By combining long baseline interferometric observations of main-sequence OB stars \citep{rizzutoetal2013,sanaetal2014} and spectroscopic radial velocity observations of Cepheids \citep{evansetal2015}, \citet{moe&distefano2017} found the binary fraction across intermediate periods of more massive primaries $M_1$~$>$~5~M$_{\odot}$ is considerably larger (see their figure~37).  As mentioned in Sect.\,\ref{ssec:lit_comp}, observations of massive binaries across intermediate orbital periods are sensitive down to only moderate mass ratios $q$~$\gtrsim$~0.3.  Nevertheless, the frequency of companions with $q$~$>$~0.3 at intermediate periods, where the observations are relatively complete, is still measurably higher for main-sequence OB stars than for solar-type main-sequence stars.  After accounting for incompleteness and selection biases, the binary fraction across $q$~=~0.1\,--\,1.0 and $P$~=~100\,--\,1500\,d is 23\%\,$\pm$\,7\% for mid-B main-sequence primaries with $M_1$~=~7\,$\pm$\,2\,M$_{\odot}$, and is 35\%\,$\pm$\,12\% for main-sequence O primaries with $M_1$~=~28\,$\pm$\,8\,M$_{\odot}$ \citep[][see their table~13]{moe&distefano2017}.  Figure\,\ref{binfrac} shows the measured binary fractions across intermediate periods $P$~=~100\,--\,1500\,d as a function of primary mass.  Our measurement of 13.9\%\,$\pm$\,2.1\% for main-sequence A/F stars is between the GKM and OB main-sequence values.

\subsection{Implication for binary star formation mechanisms}

While fragmentation of molecular cores produces wide binaries with $a$~$\gtrsim$~200~au, disc fragmentation results in stellar companions across intermediate periods \citep{tohline2002,kratter2011,tobinetal2016,moe&distefano2017,guszejnovetal2017}.  According to both analytic \citep{kratter&matzner2006} and hydrodynamical \citep{kratteretal2010a} models, the primordial discs around more massive protostars, especially those with $M_1$~$\gtrsim$~1\,M$_{\odot}$, are more prone to gravitational instabilities and subsequent fragmentation.  While it was previously known that the binary fraction across intermediate periods was considerably larger for main-sequence OB stars than for solar-type main-sequence stars \citep{abtetal1990,sanaetal2013,moe&distefano2017}, the transition between these two regimes was not well constrained.  By combining our measurement with those reported in the literature, we see an interesting trend.  Specifically, while the binary fraction across intermediate periods is relatively constant for $M_1$~$\lesssim$~0.8\,M$_{\odot}$, the fraction increases linearly with respect to log\,$M_1$ for $M_1$~$>$~0.8\,M$_{\odot}$ (see Fig.~\ref{binfrac}).  This trend is consistent with the models \citep{kratter&matzner2006,kratteretal2010a}, which show that disc fragmentation is relatively inefficient for $M_1$~$\lesssim$~1\,M$_{\odot}$ but becomes progressively more likely with increasing mass above $M_1$~$\gtrsim$~1M$_{\odot}$.  In addition, while the mass-ratio distributions of binaries with intermediate periods are either uniform or weighted towards twin components for GKM dwarf primaries \citep{duquennoy&mayor1991,fischer&marcy1992,raghavanetal2010}, we found the companions to main-sequence A/F stars across intermediate periods are weighted towards smaller mass ratios $q$~=~0.1\,--\,0.3 (see Sect.\,\ref{ssec:ms_pop} and Fig.\,\ref{fig:qdist_clean}).  The mass-ratio distribution of more massive OB main-sequence binaries with intermediate periods is even further skewed toward smaller mass ratios, although the frequency of $q$~=~0.1\,--\,0.3 is uncertain at these separations \citep{abtetal1990,moe&distefano2017}.  We conclude that $M_1$~$\approx$~0.8\,M$_{\odot}$ represents a transition mass in the binary star formation process.  Above $M_1$~$\gtrsim$~0.8\,M$_{\odot}$, disc fragmentation is increasingly more likely,  the binary fraction across intermediate periods increases linearly with respect to $\log M_1$, and the binary mass ratios become progressively weighted toward smaller values $q$~=~0.1\,--\,0.3.


\section{Eccentricity distribution of A/F binaries at intermediate periods}
\label{sec:eccentricity}

The existence of tidally circularised orbits at short periods \mbox{($P\lesssim100$\,d)} among main-sequence pairs, and at long periods due to mass transfer from post-main-sequence stars, demands a careful selection of binaries to investigate the eccentricity distribution of A/F binaries. We chose two subsamples of our pulsating binaries accordingly. In the `narrow-period' subsample, we attempted to minimise the contribution from post-mass-transfer systems by selecting all PBs with periods in the narrow interval $P = 100$\,--\,400\,d. Some white-dwarf companions inevitably remain, but low-mass main-sequence stars dominate. In the `high-$q$' subsample we selected all systems in the 100\,--\,1500\,d range for which $q>0.5$, assuming $i=60^{\circ}$. This cuts out He-core and CO-core white-dwarf companions effectively, but also (undesirably) removes some low-mass main-sequence companions. We did not otherwise discriminate by companion mass, since that has been shown to have no effect on circularisation at periods $P \gtrsim 10$\,d \citep{vaneylenetal2016}.

\begin{figure}
\begin{center}
\includegraphics[width=0.48\textwidth]{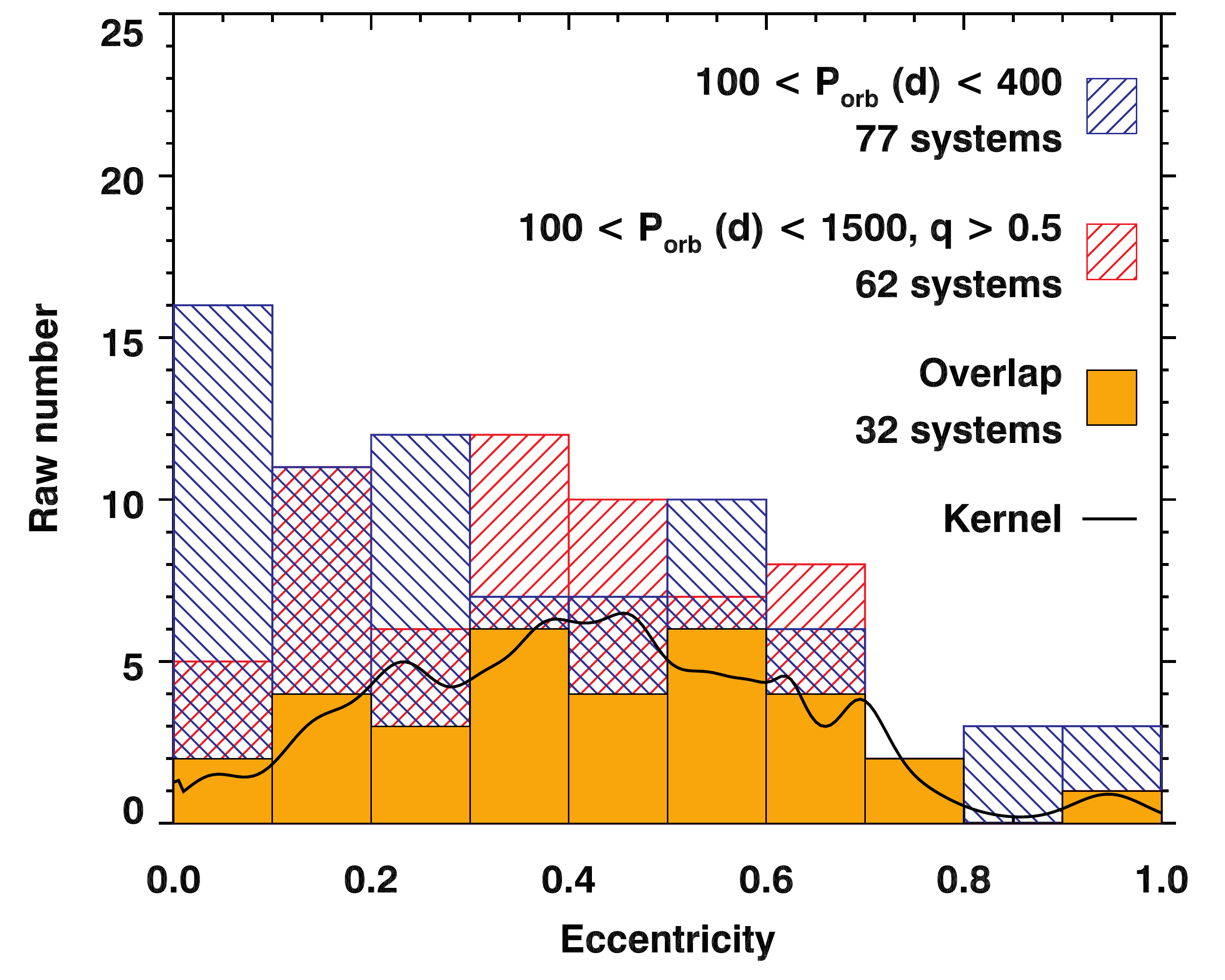}
\caption{Histograms of the eccentricities of our two subsamples chosen to select main-sequence pairs with minimal tidal effects (see the main text). These two subsamples have 32 systems in common, shown as the orange histogram, and highlight a non-uniform eccentricity distribution. The solid black curve is a Kernel density estimate that accounts for the measured eccentricity uncertainties of each of the 32 common systems.}
\label{fig:e_hist}
\end{center}
\end{figure}

Figure\,\ref{fig:e_hist} shows the eccentricity histograms of the two subsamples. The peak at low eccentricity for the narrow-period subsample is mostly caused by white-dwarf companions. Neither histogram shows a `flat' eccentricity distribution [$f(e) =$ const.], and the distribution is certainly not `thermal' [$f(e) \propto 2e$] \citep{ambartsumian1937,kroupa2008}. This is even more apparent when one considers only the systems in common between the two subsamples. Since histograms often obscure underlying trends upon binning, and because they do not adequately incorporate measurement uncertainties, we also show in Fig.\,\ref{fig:e_hist} the Kernel density estimate for the overlapping $N=32$ systems. This takes the functional form
\begin{eqnarray}
f(e) \propto \frac{1}{N}\sum_{i=1}^N \sigma_i K(e,e_i,\sigma_i),
\end{eqnarray}
where
\begin{eqnarray}
K(e,e_i,\sigma_i) = \frac{1}{\sigma_i \sqrt{2 \uppi}}~{\rm exp} \left[ -\frac{1}{2} \left(\frac{e - e_i}{\sigma_i}\right)^2 \right]
\end{eqnarray}
and $e_i$ and $\sigma_i$ are the $i$th measured eccentricity and its uncertainty. The similarity between the kernel density estimate and the histogram suggests that the bin widths are well-matched to the eccentricity uncertainties.

\begin{figure}
\begin{center}
\includegraphics[width=0.48\textwidth]{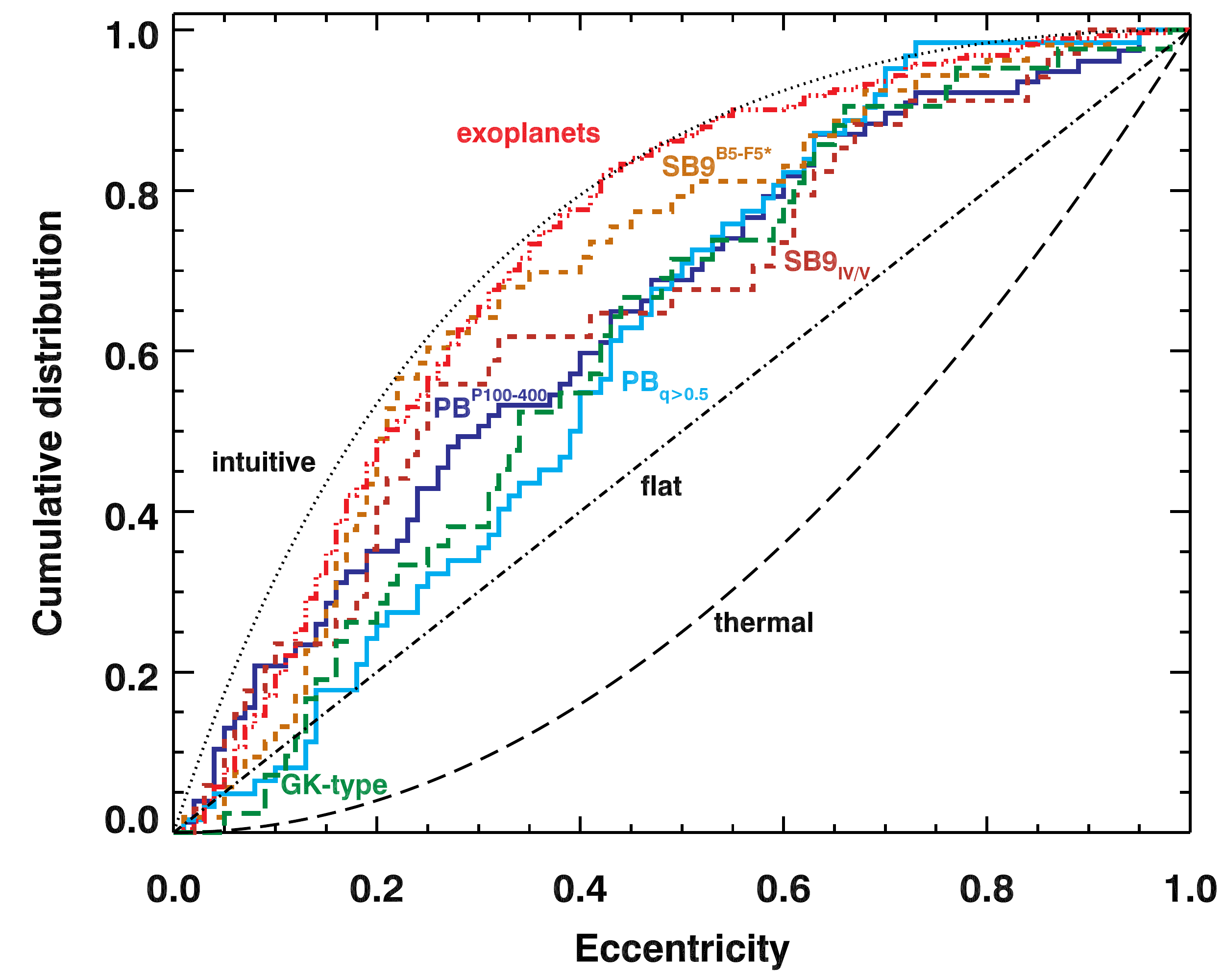}
\caption{Cumulative distribution of eccentricities, for orbits with $100 < P < 1500$\,d, except for the `narrow-period' PB subsample spanning 100\,--\,400\,d (solid dark-blue curve). The solid cyan curve is the `high-$q$' PB subsample. B5--F5 primaries from the SB9 catalogue with luminosity classes of IV or V are shown as a dashed brown curve, and the same temperature classes without the restriction on luminosity class is the dashed yellow-orange curve. GK primaries of luminosity class V or IV from the SB9 catalogue are shown as a long-dashed green curve, and exoplanet orbits from the Exoplanet Orbits Database \citep{hanetal2014} are in dash-dotted red. The thermal distribution [$f(e) = 2e$; \citealt{ambartsumian1937,kroupa2008}], flat distribution [$f(e) = e$], and so-called intuitive\protect\footnotemark\ distribution [$f(e) = (1+e)^{-4} - e/2^4$; \citet{shen&turner2008}] are shown in black, as dashed, dash-dotted and dotted lines, respectively. Styled on figure\:4 of \citet{duchene&kraus2013}.}
\label{fig:cumulative_e}
\end{center}
\end{figure}

We compare these results against binaries of other spectral types, and against the `SB9' sample outlined in Sect.\,\ref{sssec:sb9}, for orbits with $P=100$\,--\,1500\,d in a cumulative distribution in Fig.\,\ref{fig:cumulative_e}. Here we added 34 GK primaries, extracted from the SB9 catalogue \citep{pourbaixetal2004}, restricting the luminosity class to IV, V or similar (e.g. IV-V). This is necessary to exclude red giants. We included two subsamples of the SB9 catalogue for B5--F5 stars. The smaller of those two subsamples contains 34 systems and makes the same restriction on luminosity class as the GK stars. The full SB9 sample from Sect.\,\ref{sssec:sb9} does not make that luminosity class restriction because it contains many chemically peculiar stars, including Am stars which are often in binaries, whose spectral types are sometimes given without luminosity classes. Since there is no large population of A giants akin to GK giants, this subsample retains validity. Finally, we included exoplanet orbits from the Exoplanet Orbits Database \citep{hanetal2014}, accessed 2017\,July\,09, with no filter of host spectral type.

\footnotetext{The name `intuitive' was not given by \citeauthor{shen&turner2008}, but by \citet{hoggetal2010b}. The distribution is `intuitive' in that it is a model chosen to match the observation that there are many planetary orbits with $e\sim0$ but naturally none at $e=1$.}

The aforementioned presence of white-dwarf contaminants in the narrow-period subsample is mostly responsible for its excess of low-eccentricity systems. Fig.\,\ref{fig:separate} contains nine objects below our $P$\,--\,$e$ relation in the narrow-period subsample, and 8\,--\,10 more with low mass ratios lie just above that relation. These objects bias the narrow-period subsample towards lower eccentricity, explaining its difference from the high-$q$ subsample.

The high-$q$ subsample steepens at moderate eccentricity, particularly compared to the SB9 subsamples of A/F stars. It is interesting to note that, like the pulsating binaries, the GK stars also have few low-eccentricity systems at intermediate periods, despite the measured eccentricities in the SB9 catalogue being biased towards low eccentricity \citep{tokovinin&kiyaeva2016}. Our observations of PBs are unbiased with respect to eccentricity across $0.0 < e < 0.8$ (see below).  The eccentricity distribution for the high-$q$ subsample therefore suggests low-eccentricity binaries ($e < 0.1$) with intermediate periods are difficult to form, at least for A/F primaries.

\begin{figure}
\begin{center}
\includegraphics[width=0.48\textwidth]{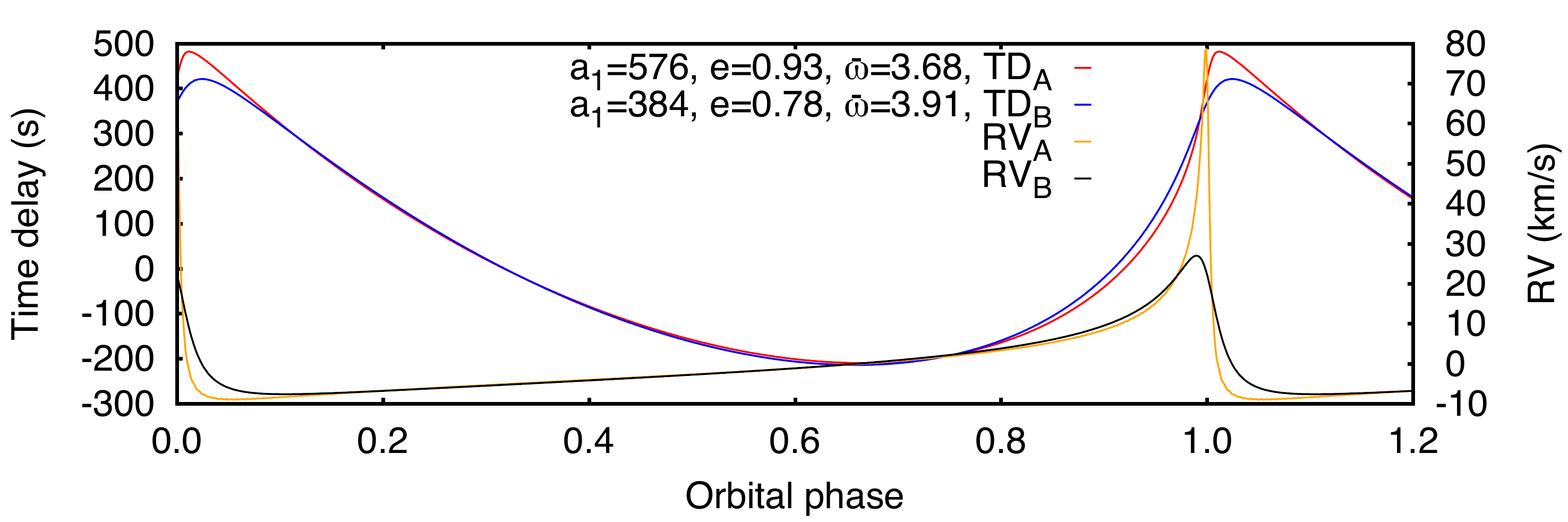}
\caption{Highly eccentric orbits with modestly different parameters can sometimes appear similar, shown here with two hypothetical orbits, `A' and `B'. A combination of RVs and time delays can help to discriminate between them.}
\label{fig:high-e_orbits}
\end{center}
\end{figure}

The large selection biases with regards to eccentricity in spectroscopic and visual binaries led \citet{moe&distefano2017} to analyse the eccentricity distribution only up to $e=0.8$ in their meta-analysis. The PM method is very efficient at finding high-eccentricity binaries, but known computational selection effects in conventional binary detection methods (e.g.\ \citealt{finsen1936,shen&turner2008,hoggetal2010b,tokovinin&kiyaeva2016}) prompted us to consider whether PM suffers similarly. Expanding on a previous hare-and-hounds exercise \citep{murphyetal2016b}, we have found that different highly eccentric orbits are sometimes poorly distinguished. For example, an orbit with $e=0.93$ and $\varpi=3.68$\,rad can appear similar to an orbit with $e=0.78$ and $\varpi=3.91$\,rad (Fig.\,\ref{fig:high-e_orbits}). They only differ substantially at periastron, which is not well sampled by time-averaged data. These binaries are best studied in combination with RVs, for which the sampling and the discriminatory power at periastron is superior. We will therefore use a combination of time delays, RVs and light-curve modelling in future detailed analyses of the heartbeat stars in our sample. Meanwhile, we restrict our analysis of the eccentricity distribution in this work to $e < 0.8$, where our sample of PBs is relatively unbiased.

\begin{figure}
\begin{center}
\includegraphics[width=0.48\textwidth]{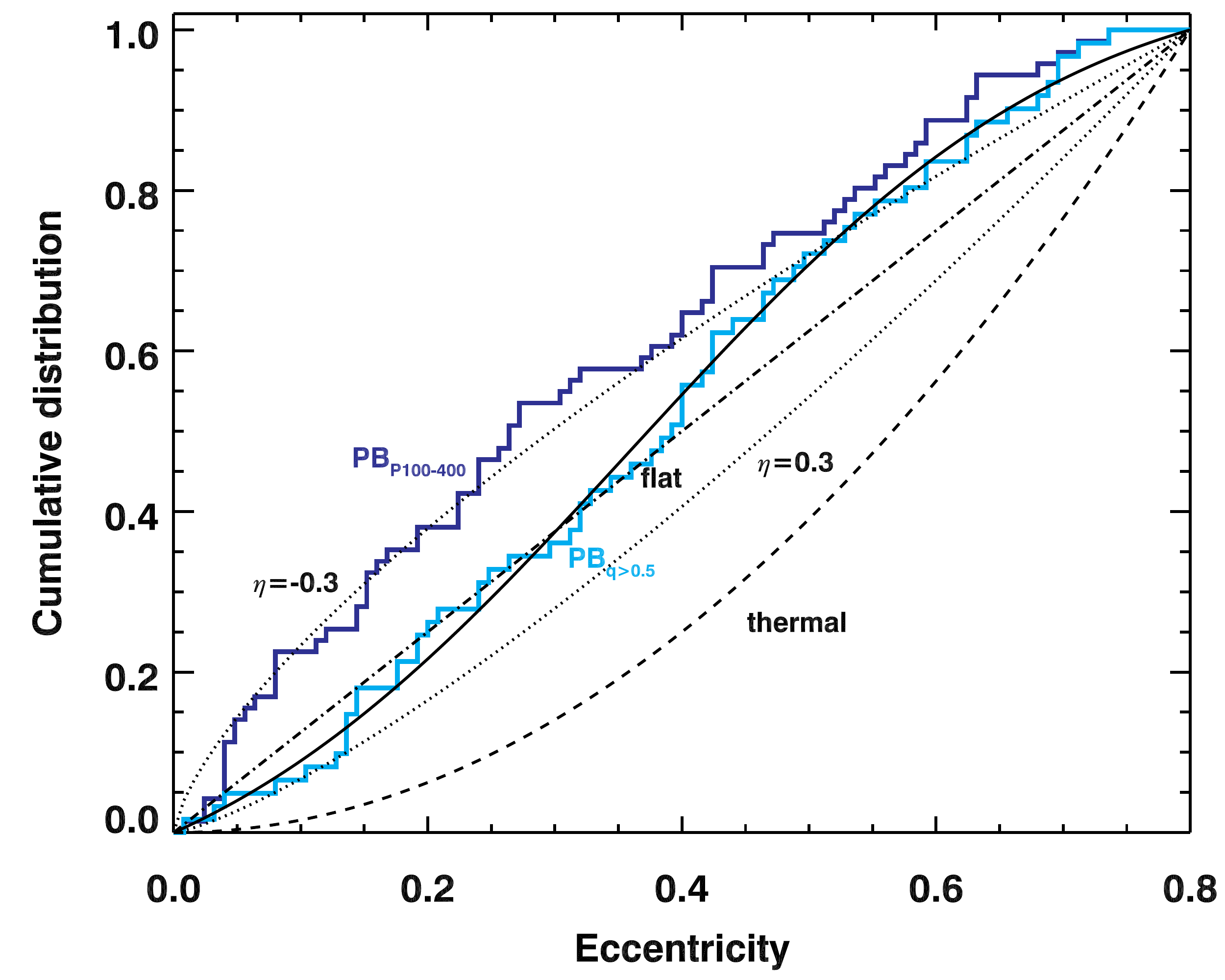}
\caption{Cumulative distribution of eccentricities, for the `narrow-period' and `high-$q$' subsamples, up to $e=0.8$. Examples of power-law distributions $p_e \sim e^{\eta}$ with $\eta = \pm0.3$ are shown as dotted lines, in addition to the thermal distribution and flat distributions as dashed and dash-dotted lines, respectively. We measured $\eta = -0.1\pm0.3$ for the high-$q$ (PB$_{\rm q>0.5}$) subsample. An example Gaussian fit to that subsample with $\mu=0.36$ and $\sigma=0.27$ is shown as a solid black line.}
\label{fig:cumulative_e2}
\end{center}
\end{figure}

Using a maximum-likelihood method, we fitted a single-parameter power-law distribution to the high-$q$ subsample across the interval $0.0 < e < 0.8$, and measured $\eta = -0.1\pm0.3$.
We found that a two-parameter Gaussian distribution (with $\mu \approx 0.36$ and $\sigma \approx 0.27$) gave a slightly better fit (Fig.\,\ref{fig:cumulative_e2}), as has been noted elsewhere \citep[e.g.][]{moe&distefano2017}, but the results are clear regardless of the specific parametrization: our observed eccentricity distribution is nearly flat across $0.0 < e < 0.8$ and has either an intrinsic deficit at or bias against very large eccentricities $e > 0.8$. In either case, the eccentricity distribution of binaries with intermediate periods and A/F primaries is inconsistent with a thermal distribution ($\eta =1$) at the 3.7$\sigma$ significance level. Previous studies of binaries with similar periods and masses have suggested non-thermal distributions that are possibly weighted toward larger $e$ compared to a flat distribution (\citealt{abt2005}, $\eta = 0.0$; \citealt{moe&distefano2017}, $\eta = +0.3 \pm 0.3$), as is the case for solar-type primaries \citep{tokovinin&kiyaeva2016,moe&distefano2017}. The results suggest that A/F binaries with intermediate periods do not form strictly through dynamical processing or tidal capture, but instead the natal discs significantly moderate the eccentricities early in their evolution. Only the most massive OB binaries, which have extremely short disc lifetimes, may have an initial eccentricity distribution across intermediate separations that is consistent with thermal (\citealt{moe&distefano2015b,moe&distefano2017}, $\eta = 0.8 \pm 0.3$).

\section{Conclusions}
\label{sec:conclusions}

We monitored the pulsation phases of 2224 $\delta$\,Sct stars, from a sample of 12\,649 targets in the original \textit{Kepler} field with effective temperatures between 6600 and 10\,000\,K (covering spectral types from early F to early A), and looked for periodic phase shifts attributable to binary motion.
We classified all 12\,649 stars according to their pulsational and binary properties. Among the 2224 $\delta$\,Sct pulsators we found a total of 317 single-pulsator binaries, 24 double-pulsator binaries, and a further 129 pulsating binaries with periods too long to attempt an orbital solution. We also found over 400 eclipsing binaries or ellipsoidal variables that we set aside for future analysis.

Among the 341 measured orbits, we found a clear excess of low-eccentricity, long-period systems, which we identified as post-mass-transfer binaries. We separated these from main-sequence pairs via an empirically determined period--eccentricity relation based on post-AGB binaries: $e = 0.55 \log P ({\rm d}) - 1.21$. We then calculated orbital parameter distributions of our two subsamples in the period range $P = 100$\,--\,1500\,d.

After completeness correction, the main-sequence subsample contains $173\pm24$ systems. It shows that the primordial mass-ratio distribution of intermediate-mass (1.5--2.5\,M$_{\odot}$) stars rises gradually from $q=1.0$ towards a peak at $q=0.2$, then exhibits a rapid turn-over between $q=0.15$ and $q=0.10$. This contrasts sharply with the almost-flat distribution for GK-dwarfs. Our results represent the first robust measurement of the mass-ratio distribution of binaries with A/F primaries at intermediate orbital periods, with uncertainties two times smaller than previous measurements at low $q$.

The low-eccentricity subsample consists of $174\pm23$ binaries, after completeness, and its mass-ratio distribution peaks sharply at $q=0.25$. For primary masses of $1.7\pm0.3$\,M$_{\odot}$, the companion masses therefore span $M_2 \approx 0.3$\,--\,0.7\,M$_{\odot}$, typical for white dwarfs. We calculated an excess of $73\pm18$ white dwarfs in this subsample, and made a model-independent estimate that 21\%\,$\pm$\,6\% of A/F primaries have white-dwarf companions at periods $P=200$\,--\,1500\,d. Across all A/F main-sequence stars, 3.2\%\,$\pm$\,0.8\% therefore have white-dwarf companions in the same period range, compared to the $\sim$0.7\% of GK giants that appear as barium stars with white-dwarf companions at those periods. This result provides a very stringent constraint for binary population synthesis studies of Type Ia supernovae, as well as barium stars, symbiotics and a plethora of related phenomena.

Across periods $P=100$\,--\,1500\,d and at $q>0.1$, we measured a binary fraction of 15.4\%\,$\pm$\,1.4\% for current A/F primaries in the \textit{Kepler} field, but this number is age (and field) dependent because the most massive stars quickly become white dwarfs and often transfer mass in the process. By subtracting those systems with white-dwarf companions, we found that \mbox{13.9\%\,$\pm$\,2.1\%} of {\it original} A/F primaries have main-sequence companions with $q>0.1$ and $P=100$\,--\,1500\,d. This is much higher than the 6.3\%\,$\pm$\,1.6\% for solar-type primaries, at the 2.9$\sigma$ significance level. From the literature we compiled the binary fraction at intermediate periods for M-dwarf up to OB primaries and found a transition mass at $\sim$1\,M$_{\odot}$, above which this binary fraction becomes a linear function of $\log M_1$, providing firm evidence for disc fragmentation becoming more dominant with increasing primary mass.

The eccentricity distribution of main-sequence pairs with A/F primaries differs very significantly from a thermal distribution. The same is true for solar-type primaries with intermediate periods, but not OB primaries. This suggests that A/F binaries with intermediate periods do not form entirely through dynamical processing or tidal capture, but instead the natal discs significantly moderate the eccentricities early in their evolution. Only the most massive (OB) binaries with extremely short disc lifetimes are the exception to this.

Finally, a scarcity of extreme-mass-ratio companions is apparent around main-sequence A stars, but is not as significant as the ``brown-dwarf desert'' of solar-type stars. We measured the true number of extreme-mass-ratio companions ($q<0.1$) in our sample to be $12.0\pm4.6$, inconsistent with zero at 2.6\,$\sigma$. Main-sequence A primaries with $M_1 \approx 1.8$\,M$_{\odot}$ may therefore represent the transition mass where extreme-mass-ratio binaries can just begin to survive at intermediate separations during the binary star formation process.


\section*{Acknowledgements}
This research was supported by the Australian Research Council. Funding for the Stellar Astrophysics Centre is provided by The Danish National Research Foundation (grant agreement no.: DNRF106). The research is supported by the ASTERISK project (ASTERoseismic Investigations with SONG and Kepler) funded by the European Research Council (grant agreement no.: 267864). M.M. acknowledges financial support from NASA's Einstein Postdoctoral Fellowship program PF5-160139. H.S. thanks the financial support from JSPS Grant-in-Aid for Scientific Research (16K05288). We thank the referee, John Telting, for his careful reading of the manuscript. S.J.M. thanks Sanjib Sharma for discussion. We are grateful to the entire \textit{Kepler} team for such exquisite data. This research has made use of the Exoplanet Orbit Database and the Exoplanet Data Explorer at \url{exoplanets.org}.


\bibliographystyle{mnras}
\bibliography{sjm_bibliography}


\appendix


\section{PM analysis of binaries found by other methods}
\label{sec:shortP}

The binaries detailed in the appendices were found via other methods. Appendix~\ref{sec:shortP} contains mostly short period binaries that have been found via the FM method \citep{shibahashi&kurtz2012,shibahashietal2015}, with two additional systems that have published PM orbits but whose orbital periods are too short to have been picked up in our survey at 10-d sampling. Some of the FM binaries simply have pulsation properties unsuited to a PM orbital solution. Their binary status with respect to our survey is discussed in each of the subsections here. Appendices~\ref{sec:lampens} and \ref{app:single} discuss the dedicated RV search for binarity among 50 hybrid $\delta$\,Sct--$\gamma$\,Dor pulsators that was recently conducted by \citet{lampensetal2017}. We also manually searched the heartbeat stars observed by \citet{shporeretal2016} for $\delta$\,Sct p\:modes on which to apply the PM method, but none appeared to be a $\delta$\,Sct star.

\begin{table*}
\centering
\caption{Orbital parameters for the PM binaries in Appendix\,\ref{sec:shortP}. The time of periastron, $t_{\rm p}$, is specified in Barycentric Julian Date. $K_1$ is calculated from the other quantities and is convolved with $\sin i$.}
\label{tab:appendix_a_orbits}
\begin{tabular}{r r@{}l r@{}l r@{}l r@{}l r@{}l r@{}l r@{}l}
\toprule
\multicolumn{1}{c}{KIC number} & \multicolumn{2}{c}{$P$} & \multicolumn{2}{c}{$a_1 \sin i / c$} & \multicolumn{2}{c}{$e$} & \multicolumn{2}{c}{$\varpi$} & \multicolumn{2}{c}{$t_{\rm p}$} & \multicolumn{2}{c}{$f_{\rm M}$} & \multicolumn{2}{c}{$K_1$}\\
\multicolumn{1}{c}{} & \multicolumn{2}{c}{d} & \multicolumn{2}{c}{s} & \multicolumn{2}{c}{} & \multicolumn{2}{c}{rad} & \multicolumn{2}{c}{BJD} & \multicolumn{2}{c}{M$_{\odot}$} & \multicolumn{2}{c}{km\,s$^{-1}$}\\
\midrule
\vspace{1.5mm}
3952623 & $19.512$&$^{+0.019}_{-0.021}$ & $25.7$&$^{+4.6}_{-4.2}$ & $0.40$&$^{+0.26}_{-0.22}$ & $5.21$&$^{+0.92}_{-1.20}$ & $2\,454\,960.8$&$^{+3.2}_{-3.3}$ & $0.048$&$^{+0.026}_{-0.023}$ & $31.3$&$^{+7.3}_{-4.6}$\\
\vspace{1.5mm}
5641711 & $38.730$&$^{+0.074}_{-0.075}$ & $27.1$&$^{+4.1}_{-3.8}$ & $0.29$&$^{+0.27}_{-0.18}$ & $4.8$&$^{+1.6}_{-1.5}$ & $2\,454\,964.1$&$^{+10.0}_{-9.0}$ & $0.0143$&$^{+0.0076}_{-0.0053}$ & $16.2$&$^{+4.0}_{-2.6}$\\
\vspace{1.5mm}
9229342 & $21.8957$&$^{+0.0045}_{-0.0046}$ & $35.4$&$^{+1.4}_{-1.3}$ & $0.23$&$^{+0.074}_{-0.074}$ & $3.05$&$^{+0.29}_{-0.53}$ & $2\,454\,967.6$&$^{+1.0}_{-1.8}$ & $0.099$&$^{+0.012}_{-0.011}$ & $36.2$&$^{+1.3}_{-1.1}$\\
\bottomrule
\end{tabular}
\end{table*}

\subsection{KIC\,3952623} 

This binary system has a period of only 19.5\,d, so it was missed in our PM survey with 10-d sampling, but found by FM. The short period and minimum mass of $\sim$0.6\,M$_{\odot}$ make the orbit quite uncertain. We adopted 3-d sampling and used the frequencies $f_{1,2,3,5,6} =$ [19.74, 18.73, 18.58, 17.97, 19.14]~d$^{-1}$. Two spurious signals were prewhitened from the time delays. The orbit is given in Table\:\ref{tab:appendix_a_orbits}.

This object suggests there is some value in a systematic PM survey with shorter segment sizes, but also highlights the difficulty in spotting and extracting binary orbits from such a survey.

\subsection{KIC\,5641711} 

This binary was found by FM. The pulsation spectrum is quite crowded, resulting in poor resolution and noisy time delays that hide the binary signal when sampling at 10-d intervals. However, the short orbital period (39\,d), demands short segments. Best results were achieved using the two strongest p\:modes only, at 21.11 and 30.48\,d$^{-1}$, at 3-d sampling, but the orbital parameters have large uncertainties. The mass function indicates a late-type companion with a mass, $m_{\rm 2,min} = 0.4$\,M$_{\odot}$. The orbit is given in Table\:\ref{tab:appendix_a_orbits}. A better solution may be achievable with the FM method.

\subsection{KIC\,6780873} 

This 9.1-d binary was found by the RV method \citep{nemecetal2017}, but has a good PM solution using both time delays and RVs \citep{murphyetal2016b}. This is one of the targets classified as single in our survey (Table\:\ref{tab:breakdown}), but which now has a PM orbital solution.

\subsection{KIC\,7900367} 

An FM analysis found a 9.23-d orbit. The triplet for the one high-amplitude p\:mode peak is equally split with a phase difference of $-1.45$\,rad, hence shows pure frequency modulation with the 0.1083-d$^{-1}$ splitting. At 3-d sampling the time delays of that same p\:mode are multiperiodic, and these periodicities are obscured by the other p\:modes, hence no PM orbital solution was derived. It is classed as single in Table\:\ref{tab:breakdown}. The lower frequency peaks and the series of peaks near 8\,d$^{-1}$ in the Fourier transform of the light curve make this target worthy of further investigation.

\subsection{KIC\,8332664} 

This is a $\gamma$\,Dor star with one isolated p\:mode at $f_1=20.1$\,d$^{-1}$ that is clearly split into an FM triplet. There are also a few Fourier peaks between 5 and 10\,d$^{-1}$ at low signal-to-noise. When all of the peaks are considered, the periodicity in the time delays of $f_1$ is not replicated across the (very noisy) time delays of other peaks, hence we classified it as single in Table\:\ref{tab:breakdown}. PM could produce a good quality orbital solution from $f_1$ alone, but we do not base solutions on just one peak. The period appears to be about 81\,d, but it is not necessarily of binary origin.

\subsection{KIC\,9229342} 

We found this binary by using FM on its strongest Fourier peak, $f_1$. It was missed in the PM survey because the 21.9-d orbit is not obvious under 10\,d sampling and because there are numerous close frequency pairs that obfuscate the binary signal. The usable frequencies at 4-d sampling are $f_{1,3,4,6,7}$ = [25.52, 25.03, 17.01, 27.53, and 17.31]~d$^{-1}$. The frequency of $f_2 = 24.473$\,d$^{-1}$ is so close to the \textit{Kepler} LC Nyquist frequency, $f_{\rm Ny} = 24.470$\,d$^{-1}$, that it is unresolved from its own Nyquist alias in a 10-d light-curve segment. The peak at 19.67\,d$^{-1}$ ($=f_5$) is also unresolved from another with similar amplitude at 19.63\,d$^{-1}$, and so on. Two further spurious signals must then be prewhitened from the time delays due to other close frequency pairs before the binarity becomes the strongest modulating peak in the weighted-average time delays, after which a good solution can be obtained (Table\:\ref{tab:appendix_a_orbits}).

The star also has a 5-ppt peak at the orbital frequency, $f_{\rm orb} = 0.04570$\,d$^{-1}$, from which we conclude it is an ellipsoidal variable.

\subsection{KIC\,10080943} 

The g\:modes of this target clearly belong to two different stars \citep{keenetal2015}, and binarity is also evident in RVs \citep{schmidetal2015}. A joint time-delay and RV analysis of this PB2/SB2 system shows a period of 15.3\,d \citep{murphyetal2016b}. This is one of the targets classified as single in our survey (Table\:\ref{tab:breakdown}), but that now has a PM orbital solution.


\section{Equations for radial velocities and their uncertainties}
\label{app:RV_eq}

Equation (\ref{eq:K1}) is the RV equation, which was typeset incorrectly in equation (16) of the PM4 paper \citep{murphyetal2016b}. There should have been a `/' before the term $\sqrt{1-e^2}$. We have reformulated it here for clarity:
\begin{eqnarray}
	K_1 = \frac{(2\uppi G)^{1/3}}{\sqrt{1-e^2}} \left\{ {{ f(m_1,m_2,\sin i) }\over{P_{\rm orb}}} \right\}^{1/3},
\label{eq:K1}
\end{eqnarray}
where the symbols have their usual meaning, and
\begin{eqnarray}
	f(m_1,m_2,\sin i) 	&\equiv& \frac{(m_2 \sin i)^3}{(m_1 + m_2)^2}\label{eq:fm}\\
					&\equiv& \frac{4\uppi^2}{G}\frac{(a_1 \sin i)^3}{P_{\rm orb}^2}.\label{eq:fmorb}
\end{eqnarray}
Note that $K_1$ is the {\it projected} (i.e. convolved with $\sin i$) {\it semi}-amplitude of the RV variation of the pulsating star, $m_1$, whose companion is $m_2$. Unlike $a_1 \sin i$, the $\sin i$ factor is omitted from $K_1$ by definition \citep{aitken1918}.

We use the full Markov chain of orbital parameters to calculate our uncertainties on $K_1$.


\section{Analysis of stars with radial velocities from Lampens et al.}
\label{sec:lampens}

\citet{lampensetal2017} recently analysed 50 hybrid $\delta$\,Sct--$\gamma$\,Dor pulsators for binarity using RVs from the HERMES spectrograph at the Mercator telescope (La Palma, Spain) and the ACE spectrograph at the RCC telescope (Konkoly, Hungary). They also discovered time-delay variations in nine of their targets. Their 50 targets overlap almost entirely with our sample, prompting us to analyse them further. We were also motivated by the benefits offered by analysing a combined RV and time-delay data set, namely the longer observing span and the complementarity of the mathematical functions governing the orbits \citep{murphyetal2016b}. The RVs particularly help in constraining the orbital periods of binaries with periods longer than the \textit{Kepler} data set. For this reason, we enhanced the weights of the RV data by a factor of 10 (equivalent to shrinking the error bars by $\sqrt{10}$). This appears to be necessary anyway, because the uncertainties on these RV data seem to be overestimated: almost all of the data are much closer to the orbital fit than the range of the 1$\sigma$ uncertainties (see Figs \ref{fig:4480321}, \ref{fig:9775454}, \ref{fig:4044353} and \ref{fig:7668791}).

Combining the PM and RV methods allowed us to reclassify their targets with greater accuracy, and has increased the number of confirmed triple systems among those objects from two to four. However, we have not used these refined orbital solutions in the statistics presented in the main body of this paper, in order to retain homogeneity and robust completeness estimates. The RV and PM methods have very different sensitivities with respect to orbital period, and here we have prior information on whether these targets are binaries based on their status as spectroscopic binaries. Hence, we had to keep the samples separate.

The following subsections are ordered identically to those in section 6 of \citet{lampensetal2017}. Our new classifications are given in the subsection headings. \citeauthor{lampensetal2017} also classified 29 objects as single, three of which we found to be PBs. We discuss these in Appendix\,\ref{app:single}.

\begin{table*}
\centering
\caption{Orbital parameters for the PM binaries. The time of periastron, $t_{\rm p}$, is specified in Barycentric Julian Date. $K_1$ is calculated from the other quantities and is convolved with $\sin i$. For KIC\,4480321 the orbit is the outer orbit of a known triple.}
\label{tab:appendix_c_orbits}
\begin{tabular}{r r@{}l r@{}l r@{}l r@{}l r@{}l r@{}l r@{}l}
\toprule
\multicolumn{1}{c}{KIC number} & \multicolumn{2}{c}{$P$} & \multicolumn{2}{c}{$a_1 \sin i / c$} & \multicolumn{2}{c}{$e$} & \multicolumn{2}{c}{$\varpi$} & \multicolumn{2}{c}{$t_{\rm p}$} & \multicolumn{2}{c}{$f_{\rm M}$} & \multicolumn{2}{c}{$K_1$}\\
\multicolumn{1}{c}{} & \multicolumn{2}{c}{d} & \multicolumn{2}{c}{s} & \multicolumn{2}{c}{} & \multicolumn{2}{c}{rad} & \multicolumn{2}{c}{BJD} & \multicolumn{2}{c}{M$_{\odot}$} & \multicolumn{2}{c}{km\,s$^{-1}$}\\
\midrule
\vspace{1.5mm}
4480321 & $2271$&$^{+63}_{-52}$ & $1288$&$^{+75}_{-60}$ & $0.0087$&$^{+0.0130}_{-0.0061}$ & $6.105$&$^{+0.075}_{-0.074}$ & $2\,456\,460$&$^{+69}_{-69}$ & $0.445$&$^{+0.060}_{-0.045}$ & $12.37$&$^{+0.46}_{-0.38}$\\
\vspace{1.5mm}
6756386 & $136.64$&$^{+0.39}_{-0.40}$ & $56.6$&$^{+9.4}_{-6.8}$ & $0.48$&$^{+0.18}_{-0.16}$ & $2.37$&$^{+0.24}_{-0.28}$ & $2\,455\,087$&$^{+7}_{-7}$ & $0.0104$&$^{+0.0052}_{-0.0037}$ & $10.3$&$^{+3.3}_{-1.7}$\\
\vspace{1.5mm}
6951642 & $1867$&$^{+12}_{-13}$ & $458.2$&$^{+8.6}_{-8.5}$ & $0.473$&$^{+0.046}_{-0.047}$ & $4.74$&$^{+0.059}_{-0.059}$ & $2\,455\,608$&$^{+20}_{-19}$ & $0.0296$&$^{+0.0017}_{-0.0017}$ & $6.07$&$^{+0.21}_{-0.19}$\\
\vspace{1.5mm}
9775454 & $1686$&$^{+13}_{-12}$ & $463.0$&$^{+11.0}_{-9.9}$ & $0.233$&$^{+0.011}_{-0.011}$ & $3.029$&$^{+0.031}_{-0.023}$ & $2\,454\,984$&$^{+18}_{-16}$ & $0.0375$&$^{+0.0028}_{-0.0025}$ & $6.16$&$^{+0.14}_{-0.13}$\\
\vspace{1.5mm}
9790479 & $230.34$&$^{+0.71}_{-0.74}$ & $57.3$&$^{+3.2}_{-2.1}$ & $0.254$&$^{+0.061}_{-0.059}$ & $1.38$&$^{+0.12}_{-0.13}$ & $2\,455\,041$&$^{+7}_{-7}$ & $0.0038$&$^{+0.00065}_{-0.00042}$ & $5.61$&$^{+0.40}_{-0.27}$\\
\bottomrule
\end{tabular}
\end{table*}

\subsection{KIC\,3429637 -- single}
This object was studied by \citet{murphyetal2012}, who used FM and found no evidence for binarity. We have used PM on the full 4-yr \textit{Kepler} data set and arrived at the same conclusion, because the time delays for each mode, while variable, show no consistent behaviour. \citeauthor{lampensetal2017} found a `long-term RV variability', but also not consistent with binarity.

\subsection{KIC\,3453494 -- long-period binary}
The PM analysis of this system shows a period much longer than the \textit{Kepler} data set. We classed it as a long period binary in Table\:\ref{tab:breakdown}. The possibility exists (from one mode only) that it is a PB2. \citeauthor{lampensetal2017} classified it as single. Given the long period, the presumed low $K_1$ value and the rapidly rotating primary ($v\sin i \sim 220$\,km\,s$^{-1}$), a second star with moderate or rapid rotation could hide easily in the spectrum. The high $v\sin i$ causes the derived RVs to have large error bars. They are therefore unable to contribute to a refined orbital solution.

\subsection{KIC\,4480321 -- triple}
\begin{figure}
\begin{center}
\includegraphics[width=0.4750\textwidth]{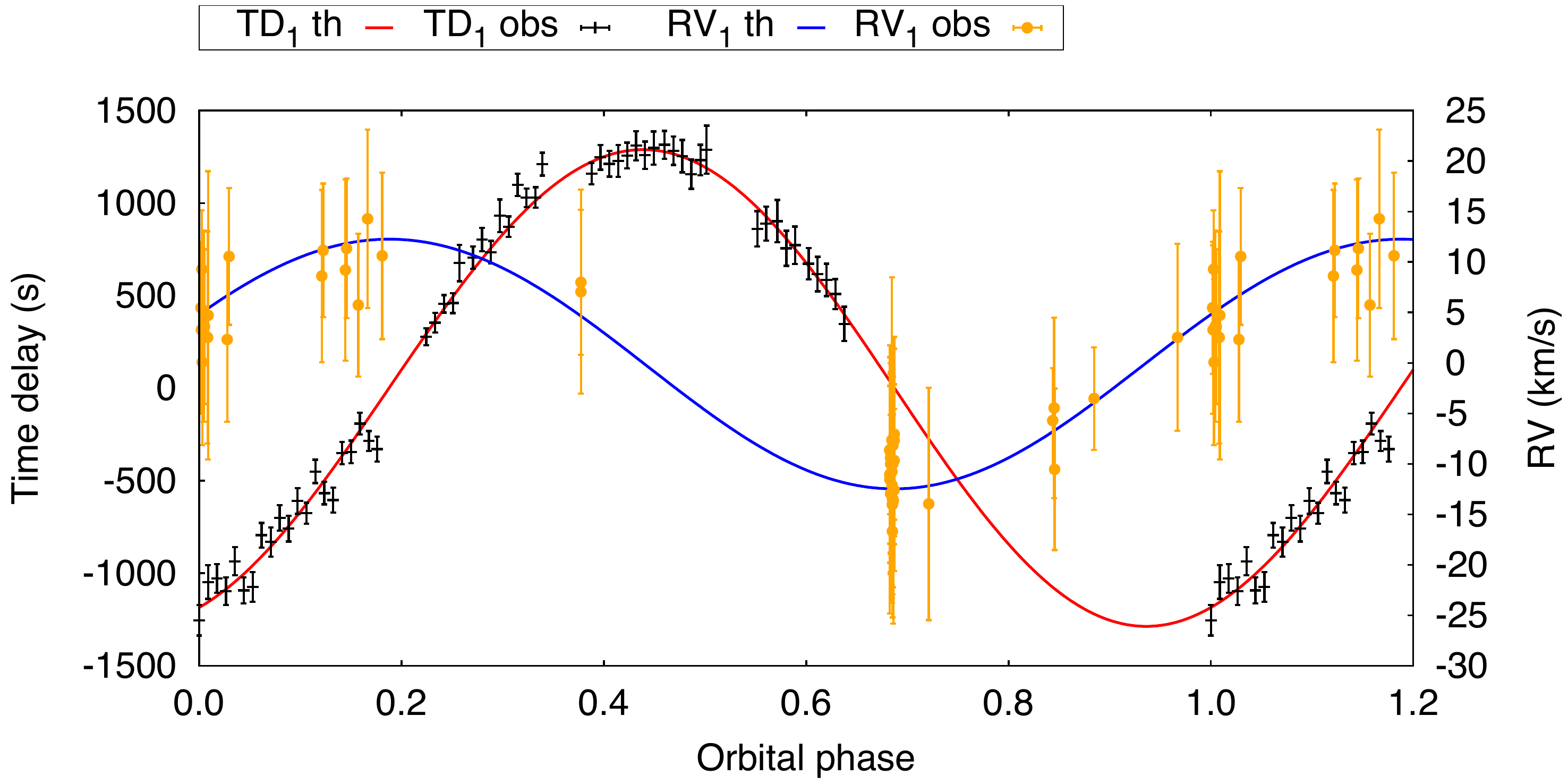}
\caption{The combined RV and time delay data set for the outer component of KIC\,4480321, and the best fitting orbit.}
\label{fig:4480321}
\end{center}
\end{figure}

\citeauthor{lampensetal2017} reported a triple system (SB3) with a mid-A star in a long orbit around a twin-like inner pair of early F stars. They found the F stars to have a period of 9.17\,d. We did not observe the system as a PB2 or PB3 and so we only saw the long-period orbit of the mid-A $\delta$\,Sct star. That period was too long for us to attempt an orbital solution from the time delays alone, and we classified it as a long-period binary in Table\:\ref{tab:breakdown}. \citeauthor{lampensetal2017} inferred a preliminary period of $2280$\,d for the tertiary component. Our combined data set incorporating their RV measurements and our time delays spans 2682.9\,d, from which we obtained $P=2270\,\pm\,60$\,d for the outer orbit (Table\:\ref{tab:appendix_c_orbits}). We calculated $K_1 = 12.4$\,km\,s$^{-1}$, which is comparable to the size of the unweighted uncertainty per RV measurement (Fig.\,\ref{fig:4480321}). There remains some possibility that the orbital period is 100--200\,d longer than the observed data set instead.

\subsection{KIC\,5219533 -- triple}
Our PM analysis showed this system to have a very high mass function, so began our own RV monitoring programme to determine the components of this system (Murphy et al. in prep.). 
\citeauthor{lampensetal2017} obtained 21 RV measurements for this system and found it to be an SB3, although only two components were parametrized. The inner pair have a 31.9-d period according to their analysis, and are probably mid-A stars ($T_{\rm eff} \sim 8300$\,K). \citeauthor{lampensetal2017} inferred the presence of a third body because the spectrum was not well-fit with two components; via PM we found the third body is a $\delta$\,Sct star with an orbital period $>$1500\,d. The mass function we measured for the pulsator (tertiary) is high because it has two companions of 1.5--2.0\,M$_{\odot}$ each. We did not detect time delays from either of the 31.9-d period pair when using 5-d sampling, but have determined that they would have been detectable if either star were a $\delta$\,Sct star. It is possible that they pulsate in p\:modes with very low amplitude (below 50\,ppm), whose intrinsic amplitudes are higher but have been diluted in the blended \textit{Kepler} photometry. Alternatively, the inner components may be hotter than the blue edge of the $\delta$\,Sct instability strip, and the broad-lined tertiary component is cooler and located inside the instability strip, accounting for both its pulsation content and its very weak contribution to the spectrum. This target will be reassessed after the 2017 observing run.

\subsection{KIC\,5724440 -- single}
The very fast rotation ($v\sin i = 240$\,km\,s$^{-1}$) made determination of the RVs difficult, with \citeauthor{lampensetal2017} quoting uncertainties of $\sim$10\,km\,s$^{-1}$ per measurement and finding no significant variability, except some features in the cross-correlated line profiles that may be pulsational. We also found no evidence for binarity via PM.

\subsection{KIC\,5965837 -- single}
\citeauthor{lampensetal2017} found that this star rotates very slowly ($v\sin i = 15$\,km\,s$^{-1}$) and is a $\rho$\,Pup star. Pulsational features were observed in the line profiles, but no binarity was suspected. We also found no binarity with PM.

\subsection{KIC\,6381306 -- triple}
The pulsation in this light curve is not favourable for PM because of both low frequencies and low amplitudes. We classified this object as single because there was no consistent behaviour from the different Fourier peaks. There is some weak evidence for a binary with $\sim$200-d period, but the Fourier peaks above 5\,d$^{-1}$ are very closely spaced and still give inconsistent results when 30-d segments are used for better frequency resolution. \citeauthor{lampensetal2017} found this to be a triple (SB3) system, with a 3.9-d pair and a 212-d tertiary. Here, RVs are much better suited to an orbital solution and PM offers no improvement.

\subsection{KIC\,6756386 and KIC\,6951642}
\citeauthor{lampensetal2017} grouped these stars together, as having pulsationally-induced line-profile variations and long-period RV variability of undetermined origin. We found both objects to be PM binaries.
\subsubsection{KIC\,6756386 -- binary}
KIC\,6756386 was already in our PM sample with a 134.4-d orbit of moderate eccentricity ($e=0.41$). The time delays have considerable scatter. The RV measurements refined the orbit, halving the uncertainties on some parameters. The remaining uncertainty is dominated by the 15\% error on $a_1 \sin i / c$. We inferred a primary mass of approximately 1.8\,M$_{\odot}$ from the spectroscopic $T_{\rm eff}$ and $\log g$ from \citeauthor{lampensetal2017}. The mass function of our refined orbit suggests a companion mass of $m_{\rm 2,min} = 0.37$\,M$_{\odot}$. The orbit is given in Table\:\ref{tab:appendix_c_orbits}.
\subsubsection{KIC\,6951642 -- long-period binary}
The time delays for 9 different modes show the same general character, indicating a binary system with a period considerably longer than the 4-yr \textit{Kepler} data set. The pulsation has quite low signal-to-noise for a $\delta$\,Sct star, and several Fourier peaks are packed into a small frequency region. The close spacings of those peaks cause some irregularities in the time delays that did not detract from a binary classification, but did hamper an orbital solution. The situation was scarcely improved when using 30-d sampling. For these reasons, we did not attempt a solution in our survey, but the availability of RVs prompted us to revisit this star. While neither the RVs nor the time delays are particularly useful data sets on their own, when we combined them we found a good orbital solution at $P=1867$\,d (Table\:\ref{tab:appendix_c_orbits}). We used the spectroscopic $T_{\rm eff}$ and $\log g$ from \citet{niemczuraetal2015} to infer a primary mass of $1.7\pm0.05$\,M$_{\odot}$. Then, from the binary mass function we calculated the minimum companion mass $m_{\rm 2,min} = 0.527\pm0.012$\,M$_{\odot}$.

\subsection{KIC\,6756481 and KIC\,7119530}
\citeauthor{lampensetal2017} grouped these stars together as rapid rotators with shell-like features in the cross-correlation function, indicated by a sharp but non-moving feature. They prefer the shell explanation, but speculate that these could be long-period binaries with two similar components. A PM analysis shows no evidence of binarity for either.
\subsubsection{KIC\,6756481 -- single}
The frequency spectrum is very dense and of low amplitude, which leads to rather noisy time delays, but there is no evidence for binarity with $a_1 \sin i / c > 25$\,s at any period.
\subsubsection{KIC\,7119530 -- single}
The time delays are variable but present rather weak and mutually inconsistent evidence for a long-period binary. The pulsation pattern is rather unusual for a $\delta$\,Sct star, and is almost certainly a $\gamma$\,Dor star with combination frequencies instead, like those described by \citet{kurtzetal2015}. The low frequency content likely includes r\:modes, too (Saio et al., submitted). Some of the modes are not resolved, which would explain the variable but irregular time delays.

\subsection{KIC\,7756853 -- triple}
\citeauthor{lampensetal2017} found this to be an SB2 system with a 99-d period. In our PM survey we saw no such binarity at that period, but did see a very long-period binary (too long for us to attempt a solution). We were therefore motivated to re-examine this system. We found no evidence of the 99-d binary in the time delays to very high precision, from which we conclude that neither of the 99-d pair is the pulsator. \citeauthor{lampensetal2017} determined the inner components to have temperatures of $\sim$9600 and $\sim$8400\,K, so both are probably beyond the blue edge of the $\delta$\,Sct instability strip, explaining their lack of p-mode pulsation. They are orbited by a $\delta$\,Sct star in an orbit of no less than a few thousand days. The fact that no discernible RV variation exists for the tertiary orbit, despite the substantial masses of the inner two stars, suggests the period must be long indeed.

\subsection{KIC\,7770282 -- single}
This star exhibits only two p\:modes, and at low signal-to-noise of around 9 and 4. The time delay noise therefore reaches several tens of seconds, so we were unable to place meaningful constraints on any companions. The star does have g\:modes of mmag amplitudes that are unsuited to a time-delay analysis (see \citealt{comptonetal2016}). These cause line profile variations that \citeauthor{lampensetal2017} observed in their spectra.

\subsection{KIC\,8975515 -- binary}
This binary was detected in our PM survey, but only one maximum and one minimum were observed in the time-delay curve due to the $>1000$-d period. We did not obtain a unique solution, but did derive a satisfactory one at 1900\,d. This is a lesson not to trust orbits longer than the data set length.

\citeauthor{lampensetal2017} reported an SB2 system with a rapidly rotating primary and slowly rotating secondary. Following private communication with one of us (SJM), they reported the preliminary period $P>1000$\,d. When we combined the time delays and both sets of RVs, we obtained a solution at $1090$\,d that fits all of the data well. There is a slight aperiodicity in the time delays that undoubtedly contributed to our originally overestimated period. In the combined data set, it can be seen that the time delay curve does indeed have two phases of maximum, but the irregularities are seen at one of them.

We calculated a mass ratio of $q=0.71\pm0.12$. The pulsator is the narrow-lined secondary component. We have been monitoring this system in our own RV campaign and will report full results at its conclusion.

\subsection{KIC\,9700679 -- binary or blend}
The KIC temperature of this object ($\sim$5000\,K) is much too low to have been included in our systematic study, but the Fourier transform of the light curve clearly shows low-amplitude p\:modes that can be used for a PM analysis. \citeauthor{lampensetal2017} have described it as a G2 giant in an SB1 system. The period of that binary cannot be inferred from their four RV measurements. We found no strong indication of binarity in the PM analysis, but a binary with $P < 20$\,d is not ruled out by either data set (though the radius of the G giant argues against such an orbit), nor can we rule out a binary with a period much longer than the \textit{Kepler} data set. The amplitudes of the oscillations in the 5--24.5\,d$^{-1}$ frequency interval are consistent with a heavily diluted $\delta$\,Sct light curve.

\subsection{KIC\,9775454 -- binary}
There are two dominant p\:modes in the Fourier transform of the light curve, and they are separated by 0.2\,d$^{-1}$. The time delays are variable and in excellent agreement, but the period, if strict, is long.

\begin{figure}
\begin{center}
\includegraphics[width=0.50\textwidth]{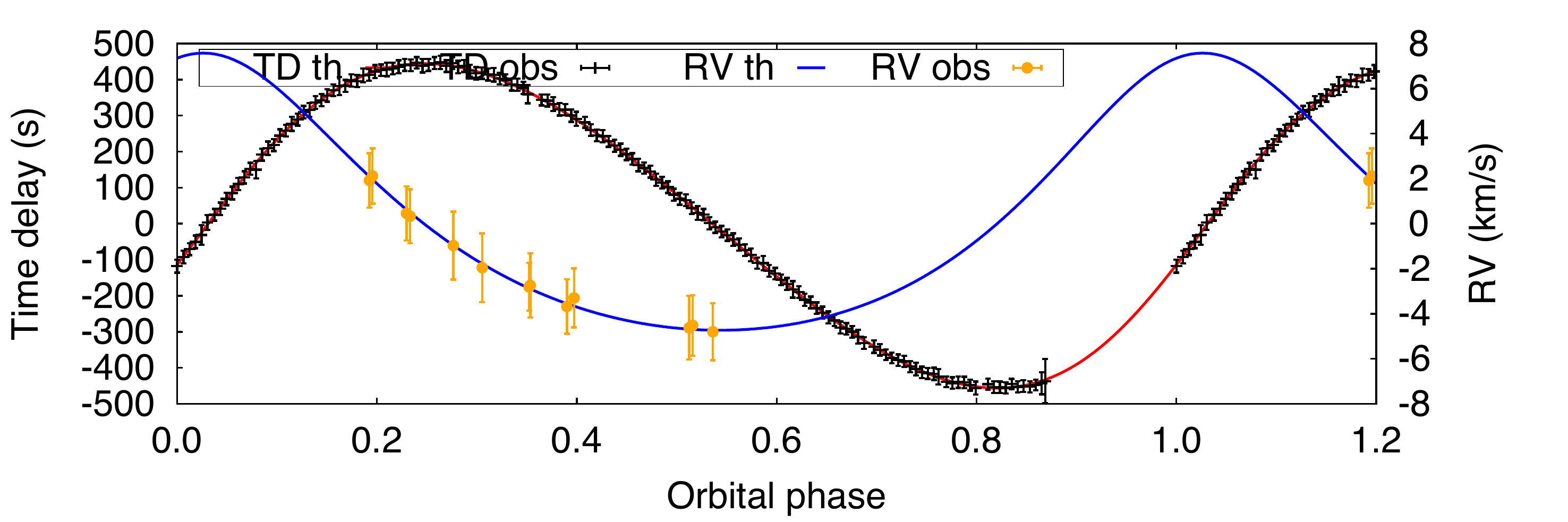}
\includegraphics[width=0.50\textwidth]{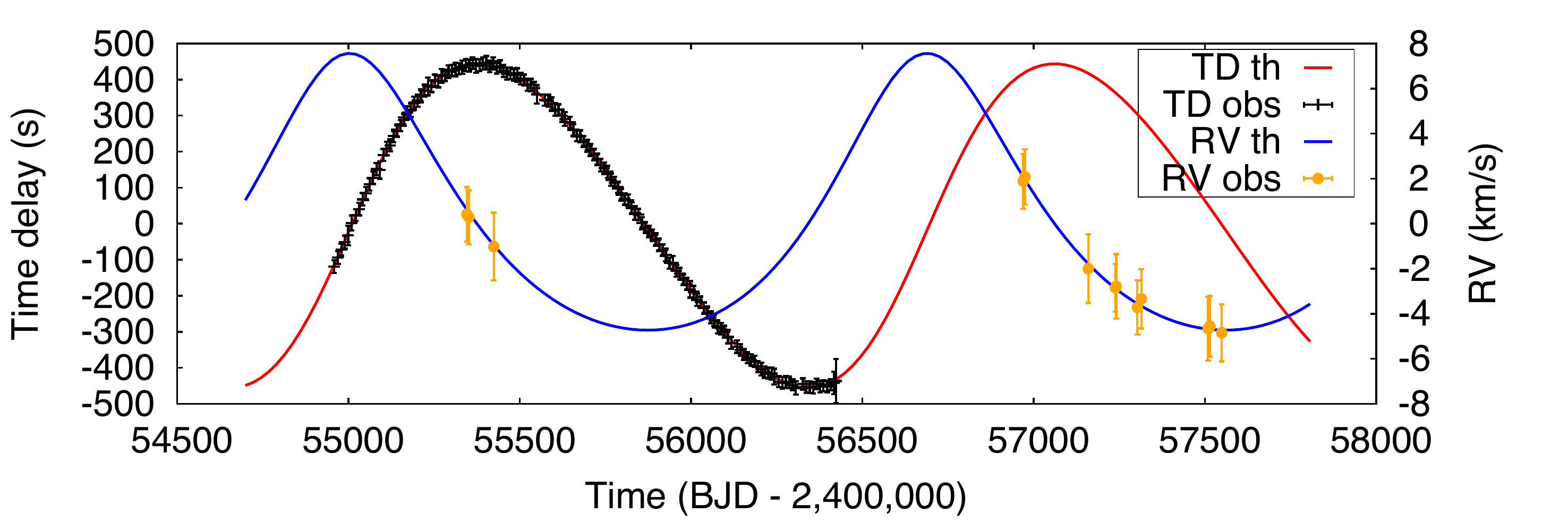}
\caption{The combined RV and time delay data set for KIC\,9775454, and the best fitting orbit.}
\label{fig:9775454}
\end{center}
\end{figure}

With the addition of RV measurements, the observed time span increases to 2600\,d and the solution is exquisite (Fig.\,\ref{fig:9775454}). It has $P = 1697\pm15$\,d and $e=0.242\pm0.018$ (Table\:\ref{tab:appendix_c_orbits}). From the spectroscopic atmospheric parameters published by \citeauthor{lampensetal2017}, we estimated a primary mass of 1.8\,M$_{\odot}$. The companion then has a minimum mass of $m_{\rm 2,min} = 0.60\pm0.02$\,M$_{\odot}$.

\subsection{KIC\,9790479 -- binary}
This binary system was missed by the PM survey, despite being detectable upon further inspection. The reason was a pair of modes with noisy time delays that obscured the periodic variation. In most cases, such binaries are still found because Fourier peaks with particularly noisy time delays are cleaned away in order to examine the more stable peaks. For some reason it did not happen in this case. This one system does not change the statistics we have presented.

\citeauthor{lampensetal2017} detected the SB1 nature and offered a possible solution of 230\,d. We confirmed the 230-d period with time delays alone, and refined the period to $230.5\pm0.7$\,d when combining the time delays and RVs (Table\:\ref{tab:appendix_c_orbits}). The tightly constrained mass function and our estimate of 2.0\,M$_{\odot}$ for the primary give $m_{\rm 2,min} = 0.25$\,M$_{\odot}$.

\subsection{KIC\,10537907 -- long-period binary}
The Fourier transform of the light curve has several p\:modes exceeding 0.5\,mmag in amplitude at or below $\sim$15\,d$^{-1}$. The modes show variable time delays but with no mutual consistency or apparent periodicity within the data set length. There is, however, tentative evidence for a binary with a much longer period. From their RV data, \citeauthor{lampensetal2017} noted a small shift in RV on a time-scale of 1500\,d, classifying it as a probable SB1.
After combining the data sets we obtained a solution at $P=2080\pm50$\,d, but the RV uncertainties are equal to the semi-amplitude of the RV variation ($K_1$). Nonetheless, future RV monitoring can refine the orbit.

It is noteworthy that the eccentricity is very small, at 0.01, for such a long-period binary. It has perhaps been tidally circularised during a mass-transfer episode. This idea is supported by the companion mass, $m_{\rm 2,min} = 0.4$\,M$_{\odot}$, being typical for a white dwarf.

\subsection{KIC\,10664975 -- single}
The time delays are variable but different modes showed no agreement and we found no clear periodicity. The RVs also show no significant variation, though \citeauthor{lampensetal2017} reported pulsational variation in the line profiles.

\subsection{KIC\,11180361 -- eclipsing binary}
In this work we have not investigated eclipsing binaries. These systems offer excellent tests of model physics through the inference of fundamental stellar parameters, but also demand a more involved analysis. We noticed the eclipses when closely inspecting the light curve. A solution is available from the \textit{Kepler} eclipsing binaries website\footnote{\url{http://keplerebs.villanova.edu/overview/?k=11180361}}, where the orbital period is given as 0.533\,d. We have revisited this target to analyse the time delays without fitting and removing the eclipses, which does make the analysis difficult. In addition, the target lies on the failed Module three, so data are missing for $\sim$90 consecutive days every 372.5-d \textit{Kepler} orbit. Although the Fourier transform of the light curve shows rich pulsation content, no evidence for a tertiary is seen. Given the aforementioned difficulties, this is not surprising. \citeauthor{lampensetal2017} did not detect any tertiary component in the RVs, either, though they have queried the accuracy of the orbital period of the eclipsing pair. We confirm the 0.533-d period.

\subsection{KIC\,11445913 -- binary}
PM shows this is a PB2 system. It is also an SB2 that \citeauthor{lampensetal2017} described as consisting of an early-F and K pair, both with low $v\sin i$. They could not find an orbital period from the RVs, but the time delays show it is at least as long as the 4-yr \textit{Kepler} data set.

\citeauthor{lampensetal2017} assumed $\log g = 4$ for both components, which may not be a good assumption for a co-eval pair of such different masses, and may explain why they were unable to achieve a satisfactory fit.

The unusual pulsation spectrum and the PB2/SB2 nature makes this system particularly worthy of continued RV monitoring, which we are doing.

\subsection{KIC\,11572666 -- blend?}
\citeauthor{lampensetal2017} offered a `preliminary solution' of 611\,d for this SB2 system, consisting of a broad and a narrow lined component. We found no evidence of binarity at that (or any other) period in the time delays. We did, however, note significant amplitude modulation in the light curve.

We took the raw RV data and performed a period search of the secondary's RVs, only. The period that \citeauthor{lampensetal2017} quoted does fit the RVs better than any other, but we suggest that more observations are required to confirm it. We cannot explain why the period is not evident in the time delays if it is real. Perhaps the pulsating star is a chance alignment.


\section{Lampens et al. `single' stars}
\label{app:single}

Here we continue the analysis from the previous section, focussing now on the stars that \citeauthor{lampensetal2017} classified as single.

\subsection{KIC\,3851151 -- long-period binary}
This is a very low amplitude pulsator whose highest p-mode peaks are only 20\,$\upmu$mag. The RVs show no clear variation within the quoted uncertainties, but when combined with the time delay observations it is evident that the small amount of variation present is consistent with a $\sim$2200-d orbit from the time delays. However, due to the long period and scant data, many possible orbital configurations remain. Nonetheless, the companion must have low mass or the orbit lies at low inclination -- we have estimated $m_{\rm 2,min} \sim 0.15$\,M$_{\odot}$.

\subsection{KIC\,4044353 -- binary}
\begin{figure}
\begin{center}
\includegraphics[width=0.50\textwidth]{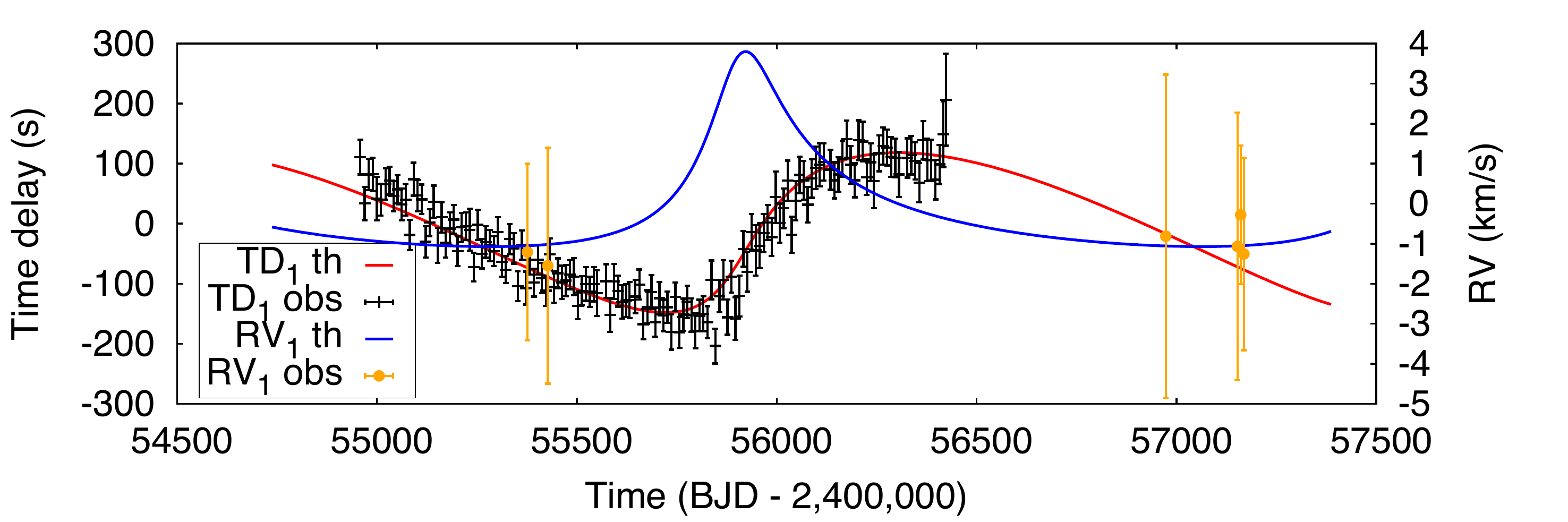}
\caption{The combined RV and time delay data set for KIC\,4044353, and the best fitting orbit. The long period and large RV uncertainties leave the orbit relatively unconstrained. Unfortunately, the RVs were obtained at approximately the same phase, one orbit apart.}
\label{fig:4044353}
\end{center}
\end{figure}

With PM we found an orbital period longer than the \textit{Kepler} data set, with a probable value around 1800\,d. The RV observations have been coincidentally obtained at the same orbital phase, one orbit apart (Fig.\,\ref{fig:4044353}). As with KIC\,3851151 above, the companion has low mass or the orbit lies at low inclination -- we have estimated $m_{\rm 2,min} \sim 0.2$\,M$_{\odot}$.

\subsection{KIC\,7668791 -- binary}
\begin{figure}
\begin{center}
\includegraphics[width=0.50\textwidth]{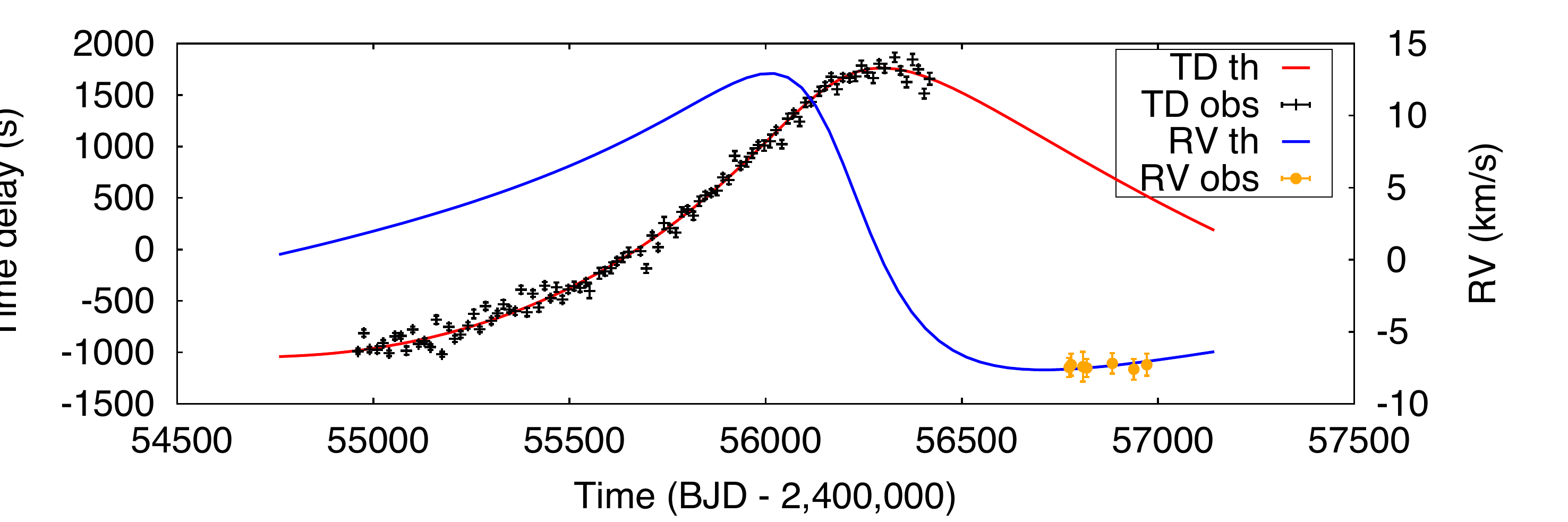}
\caption{The combined RV and time delay data set for KIC\,7668791, showing an incomplete orbital solution.}
\label{fig:7668791}
\end{center}
\end{figure}

The time delays show clear variability but the period is not well constrained by the \textit{Kepler} data. Unfortunately, the RVs are not well-spaced in time, either, and all fall near the same orbital phase (Fig.\,\ref{fig:7668791}). The attainable RV precision is better than 1\,km\,s$^{-1}$ and so RVs can confirm the orbit, but with the long orbital period one must be patient. The RVs are able to rule out a short-period orbit, implying this is just a binary and not a triple. The lower bound on the mass function implies $m_{\rm 2,min} \sim 1.1$\,M$_{\odot}$, while the primary mass is $\sim$2\,M$_{\odot}$. It is possible that the companion could be observable in the line profiles, but \citeauthor{lampensetal2017} did not report any indication of an SB2.

\subsection{KIC\,7827131 -- inconclusive}
This object has only one Quarter of \textit{Kepler} data available, so no meaningful time-delay analysis can be conducted.

\subsection{KIC\,8915335 -- inconclusive}
The time delays suggest a period of few thousand days, but are not conclusive.

\subsection{KIC\,9413057 -- inconclusive}
As for KIC\,8915335, the time delays suggest a period of few thousand days, but are not conclusive.

\subsection{Single stars}
The time delays and RV observations for the following KIC numbers show no consistent variability, implying they are single objects:
3097912,
3437940,
4281581,
4671225,
4989900,
5437206,
5473171,
6032730,
6289468,
6432054,
6587551,
6670742,
7748238,
7959867,
8738244,
9351622,
9509296,
9650390,
9764965,
9970568,
10264728,
11193046, and
11602449.


\end{document}